\documentclass[a4paper,11pt]{article}
\pdfoutput=1 % if your are submitting a pdflatex (i.e. if you have
\usepackage{lineno}
\usepackage{siunitx} 
\usepackage{enumitem}

\usepackage{jinstpub} % for details on the use of the package, please
\title{New perspectives on segmented crystal calorimeters for future colliders}

\author[a,1]{M.T.~Lucchini,\note{Corresponding author.}}
\author[a]{W.~Chung,}
\author[b]{S.C.~Eno,}
\author[b]{Y.~Lai,}
\author[c]{L.~Lucchini,}
\author[a]{M.~Nguyen,} %Minh-Thi Nguyen
\author[a]{C.G.~Tully}

\affiliation[a]{Princeton University, Princeton, New Jersey, USA}
\affiliation[b]{University of Maryland, College Park, Maryland, USA}
\affiliation[c]{Fondazione Bruno Kessler, Trento, Italy}

\emailAdd{marco.lucchini@cern.ch}

\abstract{
Crystal calorimeters have a long history of pushing the frontier on high-resolution electromagnetic (EM) calorimetry for photons and electrons. 
We explore in this paper major innovations in collider detector performance that can be achieved with crystal calorimetry when longitudinal segmentation and dual-readout capabilities are combined with a new high EM resolution approach to Particle Flow in multi-jet events, such as $e^+e^+\rightarrow HZ$ events in all-hadronic final-states at Higgs factories. 
We demonstrate a new technique for pre-processing $\pi^0$ momenta through combinatoric di-photon pairing in advance of applying jet algorithms. This procedure significantly reduces $\pi^0$ photon splitting across jets in multi-jet events. The correct photon-to-jet assignment efficiency improves by a factor of about 3 when the EM resolution is improved from 15 to $3\%/\sqrt{E}$. 
In addition, the technique of bremsstrahlung photon recovery significantly improves electron momentum measurements.
A high EM resolution calorimeter increases the Z boson recoil mass resolution in Higgstrahlung events for decays into electron pairs to 80\% of that for muon pairs.
We present the design and optimization of a highly segmented crystal detector concept that achieves the required energy resolution of $3\%/\sqrt{E}$, and a time resolution better than 30 ps providing exceptional particle identification capabilities.
We demonstrate that, contrary to previous detector designs that suffered from large neutral hadron resolution degradation from one interaction length of crystals in front of a sampling hadron calorimeter, the implementation of dual-readout on crystals permits to achieve a resolution better than $30\%/\sqrt{E}\oplus 2\%$ for neutral hadrons. 
Our studies find that the integration of crystal calorimetry into future Higgs factory collider detectors can open new perspectives by yielding the highest level of combined EM and neutral hadron resolution in the PFA paradigm.
}


\begin{document}

\maketitle
\flushbottom

\section{Introduction}
\label{sec:intro}
The compelling physics motivations for future collider experiments -- that will follow the High Luminosity LHC era -- inspire requirements on their accelerator infrastructure and detector performance.
Precise measurements of the Higgs boson properties, along with those of the mediators of the weak interaction, the W and Z bosons, will provide critical tests of the underlying fundamental physics principles of the Standard Model (SM) and are vital in the exploration of new physics beyond the SM (BSM). 
Such goals represent major milestones for future electron-positron collider proposals, such as the Circular Electron Positron Collider (CEPC) \cite{CEPC_CDR_Vol1, CEPC_CDR_Vol2} in China, the Future Circular Collider (FCC-ee) \cite{FCC_CDR} at CERN, and the International Linear Collider (ILC) \cite{ILC_TDR}.
An electron-positron ($e^+e^-$) Higgs factory has also been recognized by the 2020 Update of the European Strategy for Particle Physics as the highest-priority next collider \cite{European:2720131}.

To fully exploit the potential of future colliders, the use of complementary and diverse detector technologies is the best approach to efficiently achieve the performance requirements necessary for different physics goals.
In this context, thanks to major technological progress in the last decades, crystal calorimetry can offer unique advantages in the design of a new concept of collider detector that merges the features of homogeneous calorimetry with the particle-flow approach \cite{THOMSON200925, BUSKULIC1995481}.

The combination of high granularity with excellent energy resolution and intrinsic dual-readout capabilities may be an extremely powerful tool to further enhance the performance of Particle-Flow Algorithm (PFA) on which future collider detectors will have to rely.
Beside the obvious gain in jet energy resolution from a precise measurement of the neutral particles within the jet, as summarized in Section~\ref{sec:pfa_drivers}, we discuss two examples where a EM resolution at the level of $3\%/\sqrt{E}$ yields significant performance gains with respect to a EM resolution in the range of $15-30\%/\sqrt{E}$.
Two examples are discussed: the recovery of photons from bremsstrahlung to improve electron momentum measurement and the clustering of photons into the mother $\pi^{0}$'s to improve jet reconstruction, as presented in Section~\ref{sec:highlights}.
In particular, we demonstrate a new technique for pre-processing $\pi^0$ momenta through combinatoric di-photon pairing in advance of applying jet algorithms. This procedure significantly reduces the effective angular spread and the sharing of photons from $\pi^0$ decay between reconstructed jets in multi-jet events. 

We then present, in Section~\ref{sec:calo_overview}, a calorimeter concept consisting of a two-layer minimum ionizing particle (MIP) timing detector based on thin crystals combined with a segmented crystal electromagnetic calorimeter that features excellent particle identification capabilities.
Homogeneous crystal calorimetry is not limited by sampling fluctuation and can thus provide an energy resolution to photons and electrons at the level of $3\%/ \sqrt{E}$.
We also demonstrate that contrary to previous detector designs that suffered from large neutral hadron resolution degradation from one interaction length of crystals in front of a sampling hadron calorimeter, such as the L3 BGO \cite{L3_ECAL_results} and CMS PbWO$_4$ \cite{CMS_ECAL_TDR}, the implementation of dual-readout on crystals can maintain exceptional neutral hadron performance of better than $30\%/ \sqrt{E} \oplus 2\%$, comparable to that of a fiber-based dual-readout calorimeter \cite{DRO_calorimetry_Ferrari2109}.
Furthermore, this performance is maintained in a configuration where a thin solenoid is inserted between the electromagnetic (ECAL) and hadronic (HCAL) calorimeter compartments.
We show that this configuration yields excellent electromagnetic (EM) resolution for low energy photons and allows for increased flexibility in the optimization of the ECAL and HCAL segments in terms of cost and performance.

\section{Key calorimeter parameters for Particle Flow algorithms}\label{sec:pfa_drivers}
The vast majority (97\%) of the SM Higgstrahlung signal at a $e^+e^-$ collider with center of mass energy around 250 GeV has jets in the final state.
About one-third of the events have only two jets, while the rest have final states with 4 or 6 jets and need color-singlet identification, i.e. grouping the hadronic final-state particles into color-singlets (Z, W, H, ...) \cite{CEPC_CDR_Vol2}.
For instance, a typical benchmark requirement of future collider experiments is to achieve a jet energy resolution at the level of 3\% at 50 GeV, enabling the separation of W and Z dijet decays with a $2.5-3\sigma$ confidence level.

An important figure of merit for collider detectors performance is the resolution of the jet energy measurement.
However, due to the rather complex nature of such objects, the jet energy resolution relies on the combined performance of all subdetectors (mostly tracker and calorimeters) and on the algorithm used to cluster the jets.

\begin{figure}[!tbp]
\centering % \begin{center}/\end{center} takes some additional vertical space
\includegraphics[width=0.495\textwidth]{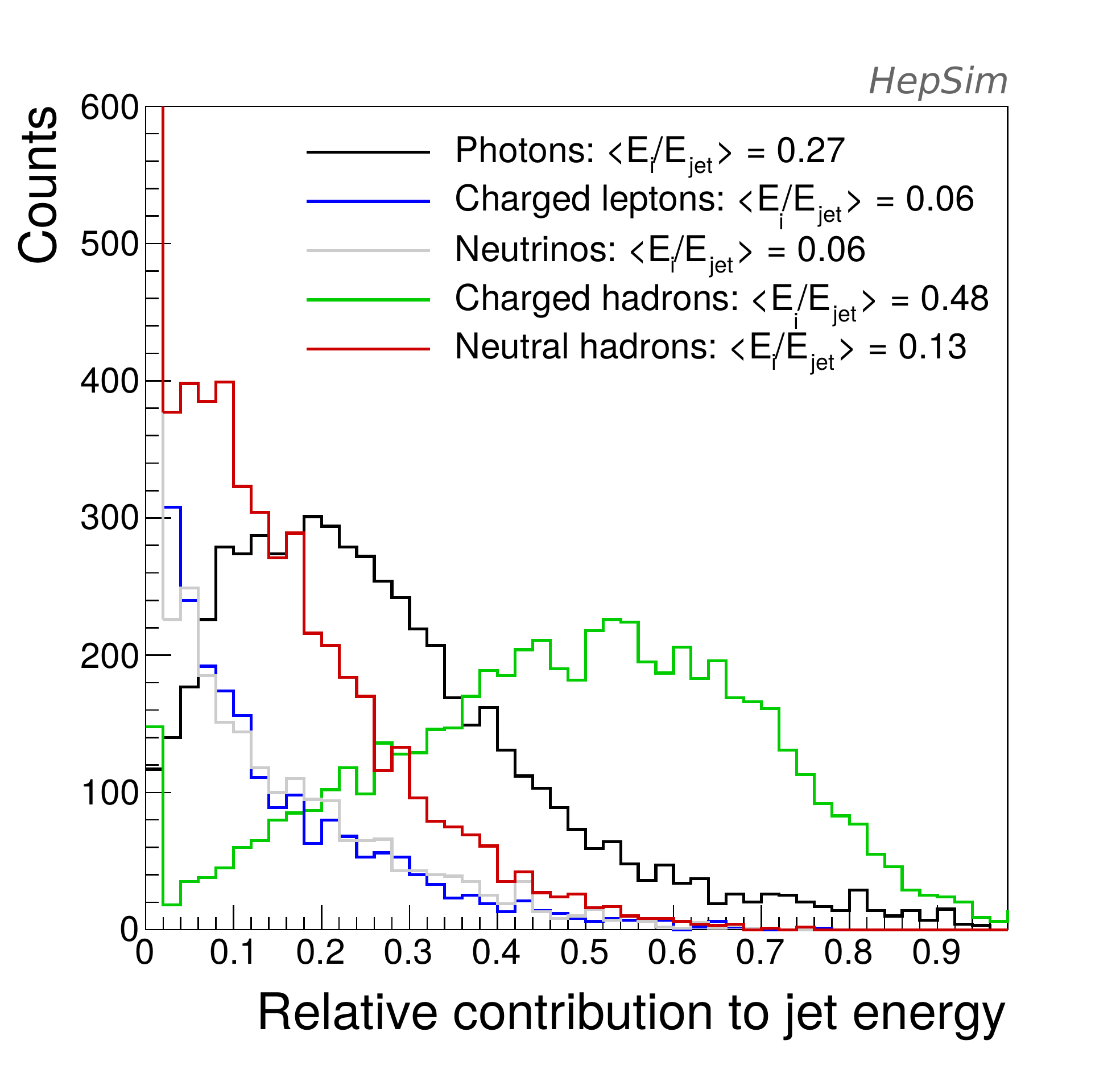}
\includegraphics[width=0.495\textwidth]{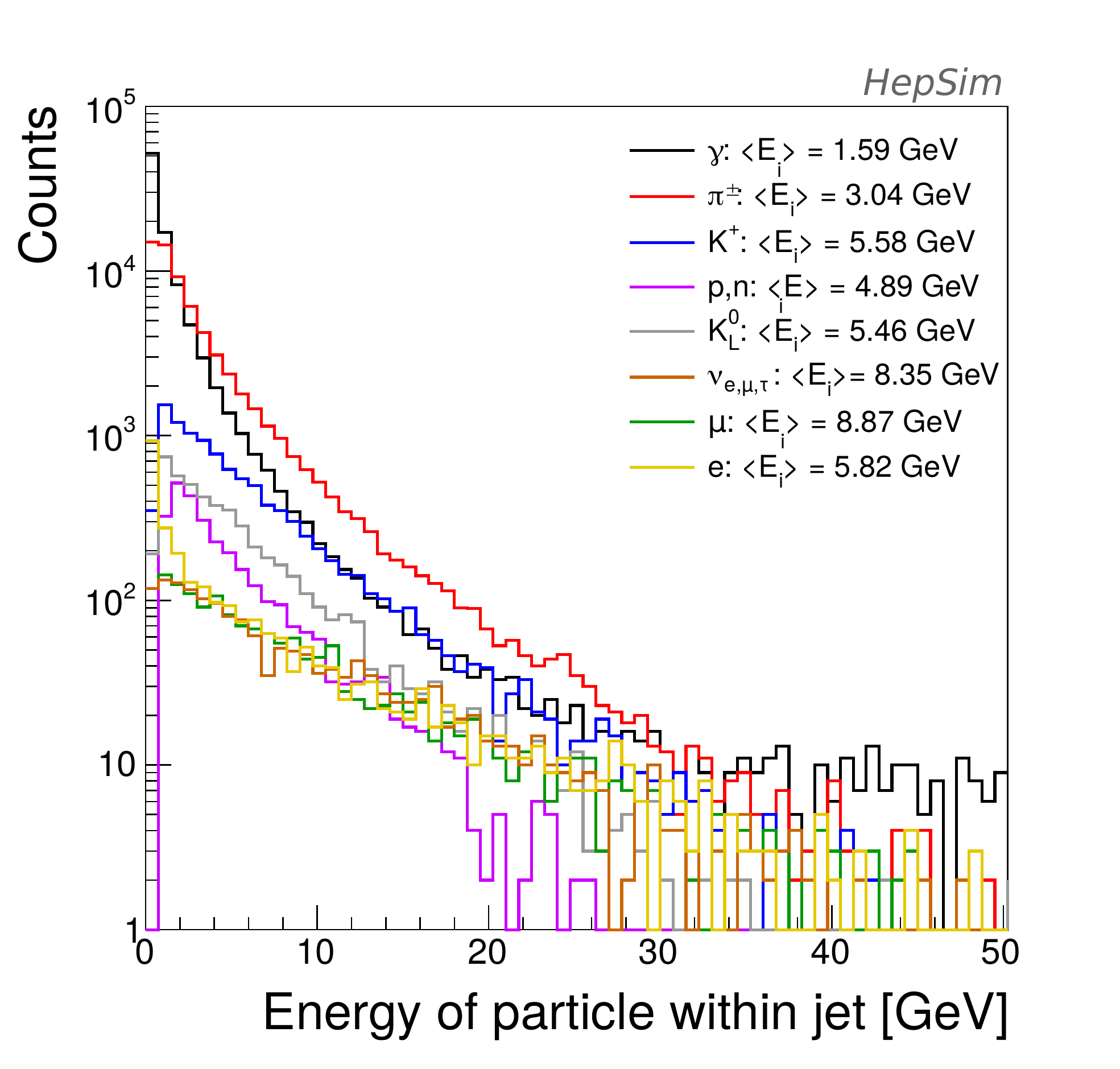}
\caption{\label{fig:jet_fractional_energy} Left: relative contribution of different particles (photons, neutrinos, leptons, neutral hadrons, charged hadrons) to the jet energy, defined as the ratio of the energy of a given particle $E_i$ and the energy of the jet, $E_{jet}$, to which the particle belongs to. Right: energy distribution of different particle types clustered within jet.}
\end{figure}

To briefly illustrate this aspect, Figure~\ref{fig:jet_fractional_energy} shows the relative contribution to jet energy from different particles: photons, neutrinos, neutral hadrons, charged leptons, and charged hadrons. The results are shown for a sample of $Z\rightarrow b\overline{b}$ events from $e^+e^-$ collisions at 250 GeV from the HepSim repository \cite{HepSim_repository}. Data are analyzed using the HepSim package, and Monte Carlo truth jets are reconstructed with the Durham algorithm \cite{Durham} (using jet resolution parameter, $y_{cut} = 0.05$) and required to have a minimum transverse momentum, $p_{\rm T}$, of 10~GeV. %

%The software used for jet reconstruction... comment on the sample and jet cuts, etc.

After the decay of short-lived particles, more than half of the jet energy is from charged hadrons whose energy can be precisely measured with the tracker.
Roughly 13\% consists of long-lived neutral hadrons (e.g. $n$, $\bar{n}$ and $K^0_L$), and about 27\% is made of photons. Such neutral components of the jet have to be measured by the calorimeters with good precision. This is particularly true since the majority of photons and hadrons clustered within a jet have relatively low energy as shown in Figure~\ref{fig:jet_fractional_energy}, and thus the stochastic term of the calorimeter energy resolution has to satisfy rather tight requirements.
In particular, the neutral hadron component is dominated by neutrons and long-lived neutral kaons, $K^{0}_{L}$, with mean energy of about 5 GeV, while the photons have a mean energy below 2 GeV.

We estimated the corresponding impact of either hadronic or electromagnetic particle energy resolution to the jet resolution by smearing the momenta of the Monte Carlo truth particles used by the jet clustering algorithm, for different levels of calorimeter energy resolution.
Since, as discussed in \cite{THOMSON200925}, the energy distributions of jets reconstructed with the Pandora algorithm can feature non-Gaussian tails, we quantify the resolution as the {\it effective sigma}, $\sigma_{eff, X}$, i.e. half of the smallest interval containing a fraction X of the unbinned event distribution, with $X=0.68$ (or 0.90 to show the impact of tails).

\begin{figure}[!tbp]
\centering
\includegraphics[width=0.49\textwidth]{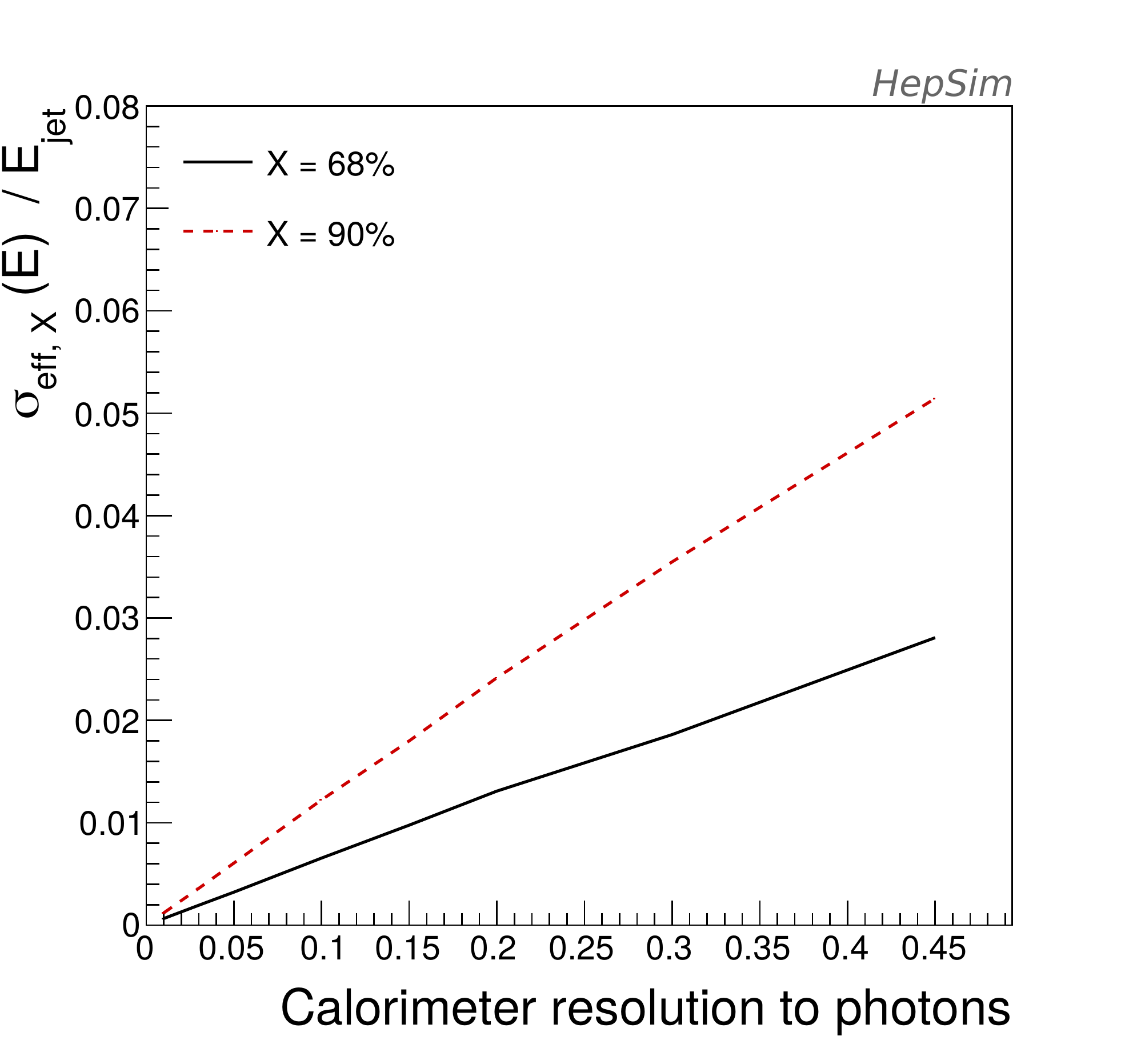}
\includegraphics[width=0.49\textwidth]{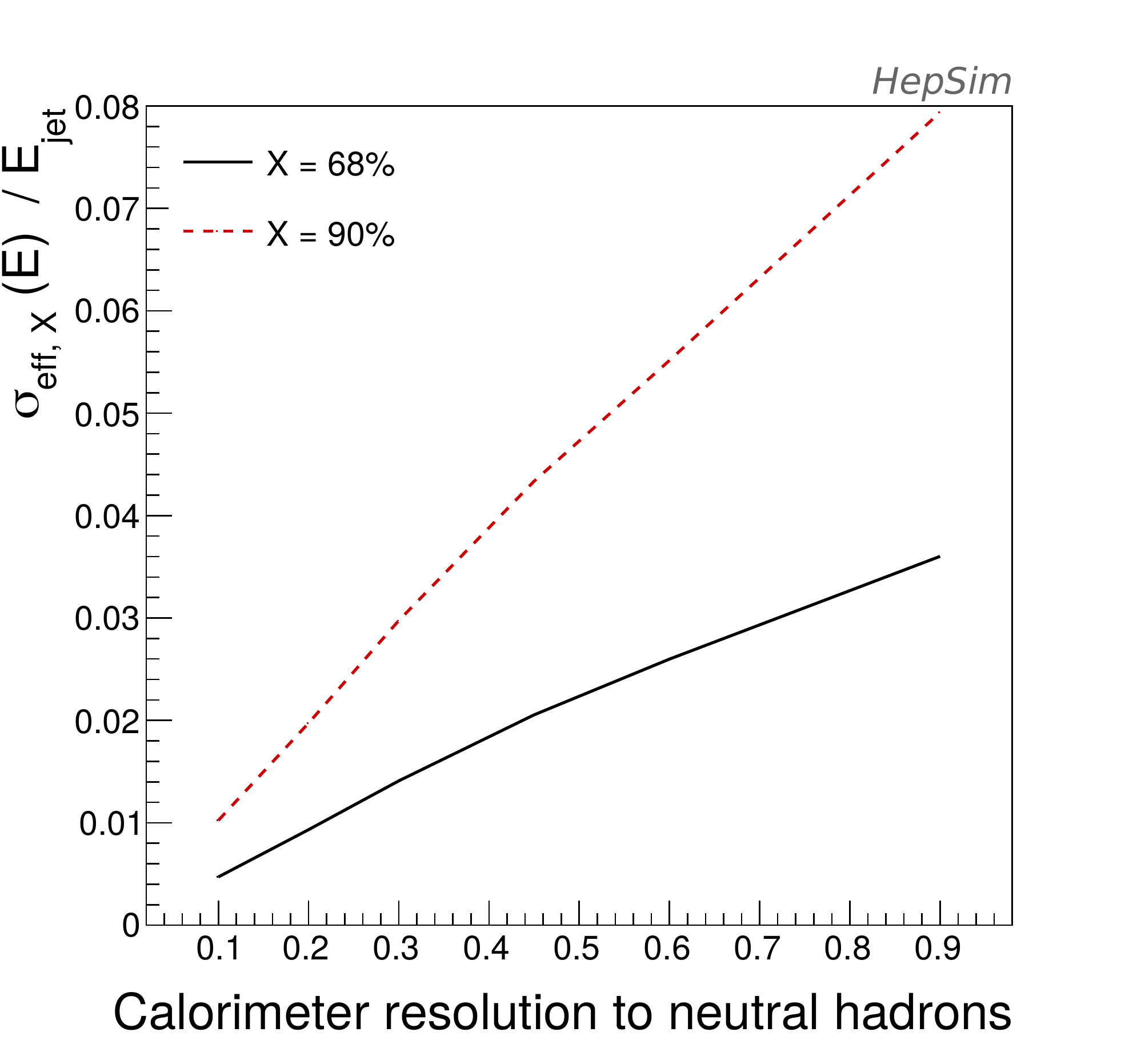}
\caption{\label{fig:calo_to_jet_res_contr} Contribution to the total jet energy resolution from the EM (left) and hadronic (right) energy resolution in HZ$\rightarrow$ $q\bar{q}\nu\bar{\nu}$ events, quantified as $\sigma_{eff, X}$, i.e. half of the smallest interval containing a fraction X of the unbinned event distribution, with $X=0.68$ (or 0.90). }
\end{figure}

Figure~\ref{fig:calo_to_jet_res_contr} shows that to maintain the contribution from the calorimeters to the jet energy resolution below $3\%$ at 50~GeV, it is in general required to measure the photons with an EM resolution better than $20\%/\sqrt{E}$ ($\sim1.5\%$ to a 50 GeV jet) and the neutral hadron component should be measured with a resolution better than $45\%/\sqrt{E}$ ($\sim2.2\%$ to a 50 GeV jet).
Beside an excellent energy resolution, other key calorimeter parameters that are relevant for an optimal performance of the Particle Flow algorithms are:
\begin{itemize}
\itemsep0em 
\item a small Moli\`ere radius, $R_{M}$ (radius of the cylinder containing 95\% of the EM transverse shower development), to provide better separation of EM calorimeter clusters by keeping the shower lateral development as narrow as possible;
\item a transverse segmentation similar to the Moli\`ere radius (at shower maximum) or smaller (at the early stage of shower development);
\item some level of longitudinal segmentation to exploit different shower profiles at different depths and enhance particle separation and identification capabilities.
\end{itemize}

In particular, it has been shown that the PFA performance for a Si-W sampling calorimeter is similar for a number of equally partitioned EM longitudinal layers between 4 and 30, and degrades when the number of longitudinal layers is smaller than 4 \cite{ManqiPrivate}.
A precise measurement of time for both charged particles (before the calorimeters) and for calorimeter clusters can further improve the performance of PFA, reduce effects of pileup and jets overlap and improve the hadronic calorimeter performance.

%\newpage

\section{Performance gains specific to high-resolution EM calorimetry}\label{sec:highlights}
%\textcolor{red}{What are the benefits from a EM resolution better than 15\%/$\sqrt{E}$?}
%\newline

In Section~\ref{sec:pfa_drivers} we briefly discussed the key aspects driving the performance of particle-flow algorithms.
Beyond the advantages of a high-resolution calorimeter in improving jet energy resolution by precisely measuring the neutral component, there are additional performance gains that become available with an energy resolution for EM particles at the level of 3-5\%/$\sqrt{E}$.
We report in Sections~\ref{sec:brem} and \ref{sec:pi0clust} two examples: the possibility to recover the electron energy loss due to bremsstrahlung radiation thus improving the momentum measurement provided by the tracker and the possibility to pair up photons originating from $\pi^0$ decays thus enhancing the performance of jet clustering algorithms.

In particular, the pairing of photons from $\pi^0$ decays has the potential to improve the precision of particle flow by reducing the splitting of $\pi^0$ photons across jets. This is particularly relevant to achieve a precise separation of the Higgs particle decay products from the Z boson decay products used to estimate the recoil mass of the system.
In a recoil mass analysis, when looking at rare Higgs decays, which may even be photons and missing energy, the physics of the decay is more important that the resolution of the jet energies - especially since the Z boson has a finite width of 2.5 GeV, the Higgs boson mass is known to a precision of 1 GeV and center-of-mass constraints on the momenta and energies from the beams are available.
For this reason, in Section~\ref{sec:pi0clust}, we focus on the improvements achieved, using the $\pi^0$ photon pairing algorithm, in the correct association of a photon to a jet rather than on improvements to the jet energy resolution (previously demonstrated in \cite{pi0clustering}).

\subsection{Recovery of photons from bremsstrahlung}\label{sec:brem}
At typical collider detectors, a strong magnetic field is present, and a tracker with a material budget equivalent to $0.1-1.0$ radiation lengths ($X_{0}$) is located before the calorimeters. In such conditions, the measurement of the electron energy in the electromagnetic calorimeter is challenged by the electrons losing energy in the form of bremsstrahlung radiation while traversing the silicon layers of the tracker.

\begin{figure}[!tbp]
\centering
\includegraphics[width=0.495\textwidth]{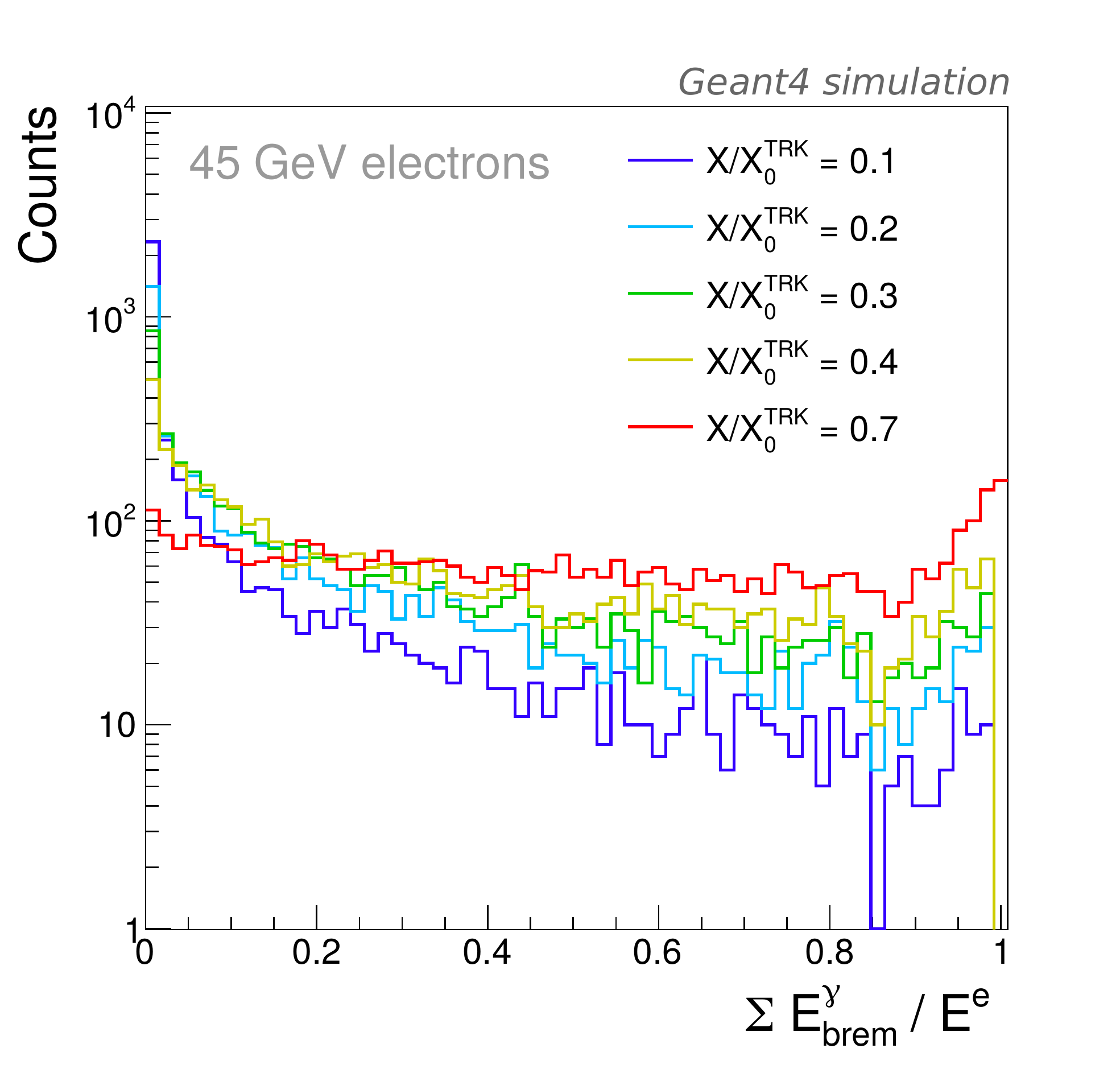}
\includegraphics[width=0.495\textwidth]{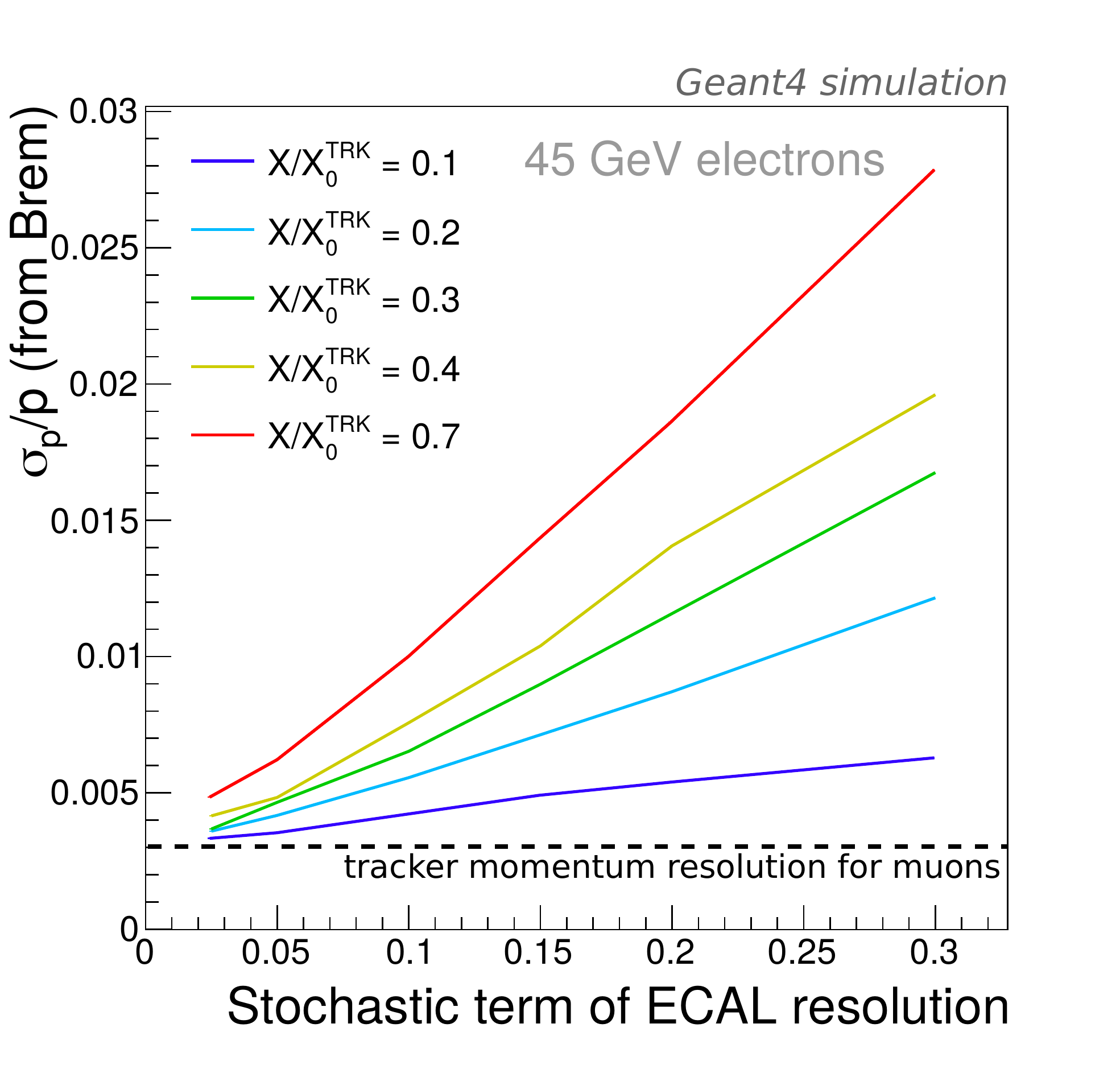}
\caption{\label{fig:brem} Left: fraction of the total energy lost by 45 GeV electrons through bremsstrahlung radiation within the tracker volume (before reaching the calorimeter) for different scenarios of tracker material budget. Right: contribution to the resolution of the electron momentum at vertex due to bremsstrahlung assuming the energy of photons emitted within the tracker volume are measured by calorimeter with a certain stochastic term of energy resolution.}
\end{figure}

Clustering algorithms, which identify energy deposits by bremsstrahlung photons in the ECAL, are essential for recovering such energy losses and improving the momentum measurement from the tracker, especially for energies in the 7-30 GeV range, as demonstrated in \cite{Brem_CMS} for the CMS ECAL. Above 30~GeV the precision on the electron energy is mainly dominated by the resolution of the ECAL.
A standalone \textsc{Geant4} simulation \cite{AGOSTINELLI2003250} was used to study this effect as a function of the material budget from the tracker. A simplified tracker geometry, consisting of seven silicon layers and radial dimension of 1.9~m in a 3T magnetic field, was used as discussed in more detail in Section~\ref{sec:overview_calo}.
The amount of electron energy lost through bremsstrahlung within the tracker volume (before reaching the calorimeter) has a strong dependence on the thickness of the tracker in radiation lengths, as shown in Figure~\ref{fig:brem} for 45 GeV electrons.
Assuming a clustering algorithm, such as those currently used by the CMS and ATLAS reconstruction software, is used, the degradation of the electron momentum resolution will depend on the ECAL capability to precisely measure the energy of bremsstrahlung photons.
By adding to the electron momentum measured at the ECAL all the truth-level energies of the photons ($\Sigma E_{brem}^{\gamma}$), after applying a Gaussian smearing defined by the ECAL resolution, the contribution to the electron momentum resolution at vertex due to bremsstrahlung can be estimated.
The result is shown in the right plot of Figure~\ref{fig:brem} for 45 GeV electrons. A calorimeter with energy resolution at the level of $3\%/\sqrt{E}$ can reduce the contribution of bremsstrahlung radiation to the electron momentum measurement below 0.4\% for a tracker material budget of 0.4~$X_0$, compared to a contribution of about 2\% for a calorimeter with resolution of about $30\%/\sqrt{E}$.

\begin{figure}[!tbp]
\centering
\includegraphics[width=0.495\textwidth]{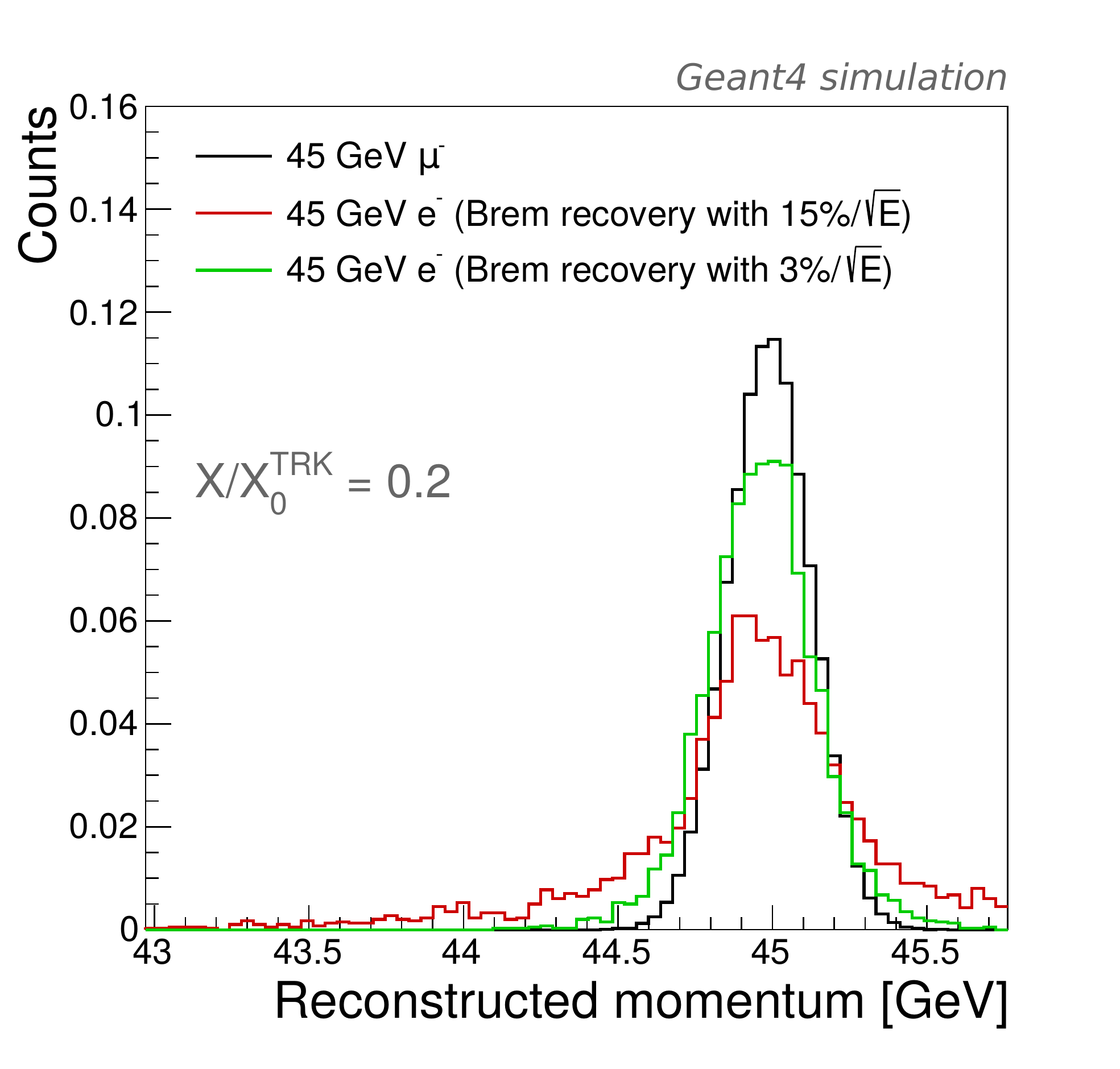}
\includegraphics[width=0.495\textwidth]{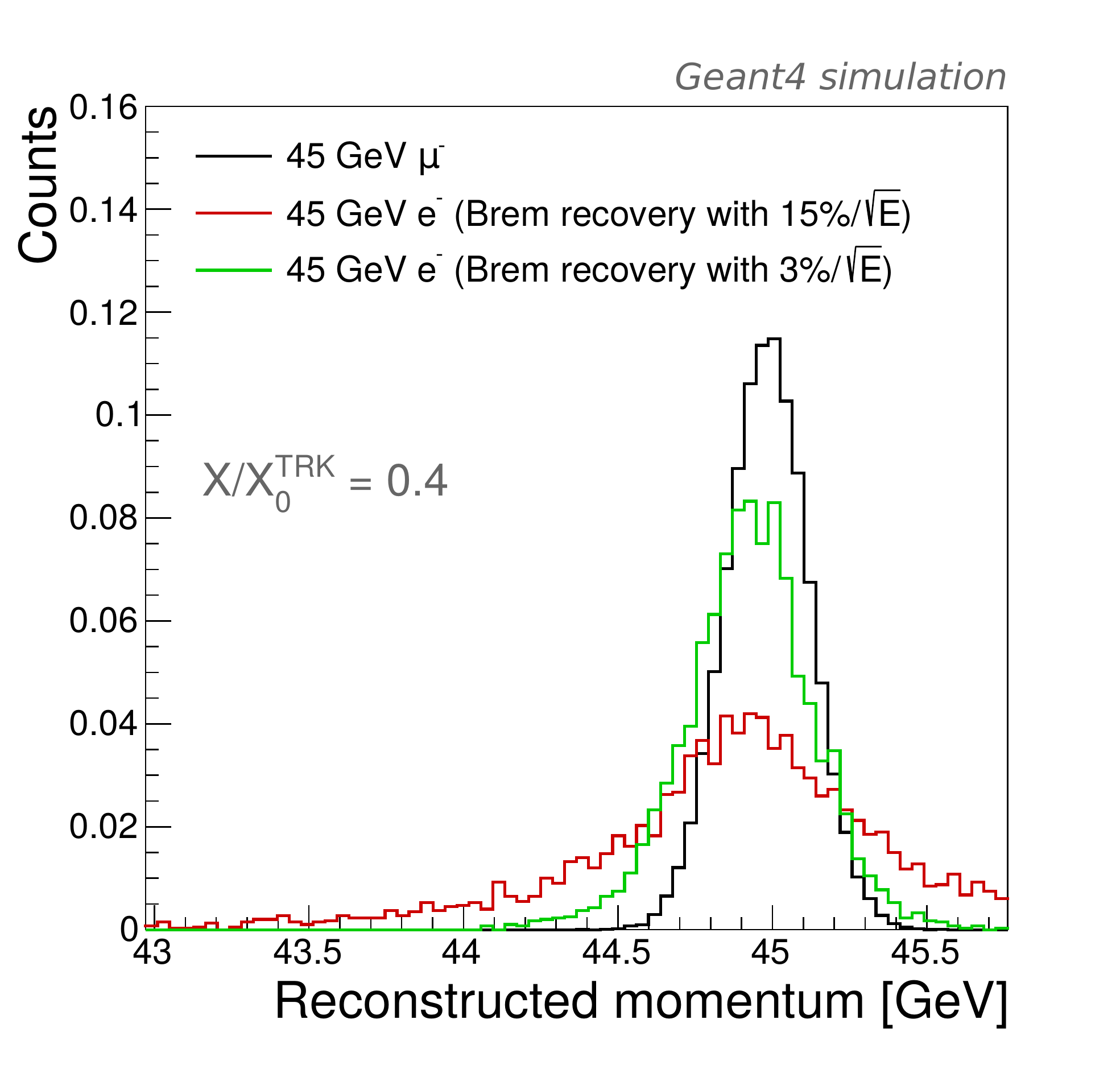}
\caption{\label{fig:brem_distr} Distributions of the reconstructed momentum for muons (assuming a tracker momentum resolution of 0.3\%) and for electrons assuming recovery of Bremsstralhung photons with an energy resolution of $15\%/\sqrt{E}$ and $3\%/\sqrt{E}$. Two scenarios are shown: assuming a tracker material budget equivalent to $0.2~X_0$ (left) and $0.4~X_0$ (right).}
\end{figure}

For the physics program of future $e^{+}e^{-}$ colliders, it is crucial to reconstruct the Higgs boson in Higgstrahlung events ($e^+e^-\rightarrow ZH$) by measuring the mass of the system recoiling against the Z boson.
Both the mass and the width of the Higgs boson can be inferred by the distribution of the recoil mass, as it has a peak corresponding to the Higgs boson mass. The highest resolution in the recoil mass peak can be achieved with the Z boson decaying to a pair of leptons ($\mu^{+}\mu^{-}$ or $e^{+}e^{-}$).
As shown in Figure~11.3 of \cite{CEPC_CDR_Vol2}, compared to the $Z \rightarrow \mu^+\mu^-$ decay, the analysis of the $Z \rightarrow e^+e^-$ decay is affected by additional background contributions due to Bhabha scattering and single boson production. In addition, the measurement of the electron momentum is limited by bremsstrahlung radiation from the electron passing through the beampipe and tracker material that leads to a more pronounced high-mass tail in the signal distribution.
Starting with a track momentum measurement of 0.3\% for 45 GeV muons (dashed black line in Figure~\ref{fig:brem}), as a benchmark, and quantifying the bremsstrahlung recovery performance shown in Figure~\ref{fig:brem} for a tracker material budget of $X/X_0^{\rm TRK} \leq 0.4$, an EM resolution of order $3\%/\sqrt{E}$ would allow the precise measurement of the energy of photons from bremsstrahlung. 
The final electron momentum resolution is 0.375\% for 45 GeV electrons compared to 0.3\% for muons.  The resolution of the recoil mass signal from $e^+e^-$ pairs (with typical energies around 45 GeV) would thus similarly improve achieving about 80\% of that for muon pairs.

\subsection{Clustering of $\pi^{0}$ photons in multi-jet events}\label{sec:pi0clust}

Another possibility enabled by high-resolution EM calorimeter is the possibility of reconstructing $\pi^{0}$'s decaying into pairs of photons, which is an additional tool to further enhance the performance of jet reconstruction algorithms in particular for event topologies featuring 4 or 6 jets.
The potential of such a tool has been investigated using the HepSim software package and repository and smearing the momenta of Monte Carlo truth particles according to different levels of energy resolution, following the procedure described in the next paragraphs.

\begin{figure}[!tbp]
\centering % \begin{center}/\end{center} takes some additional vertical space
\includegraphics[width=0.495\textwidth]{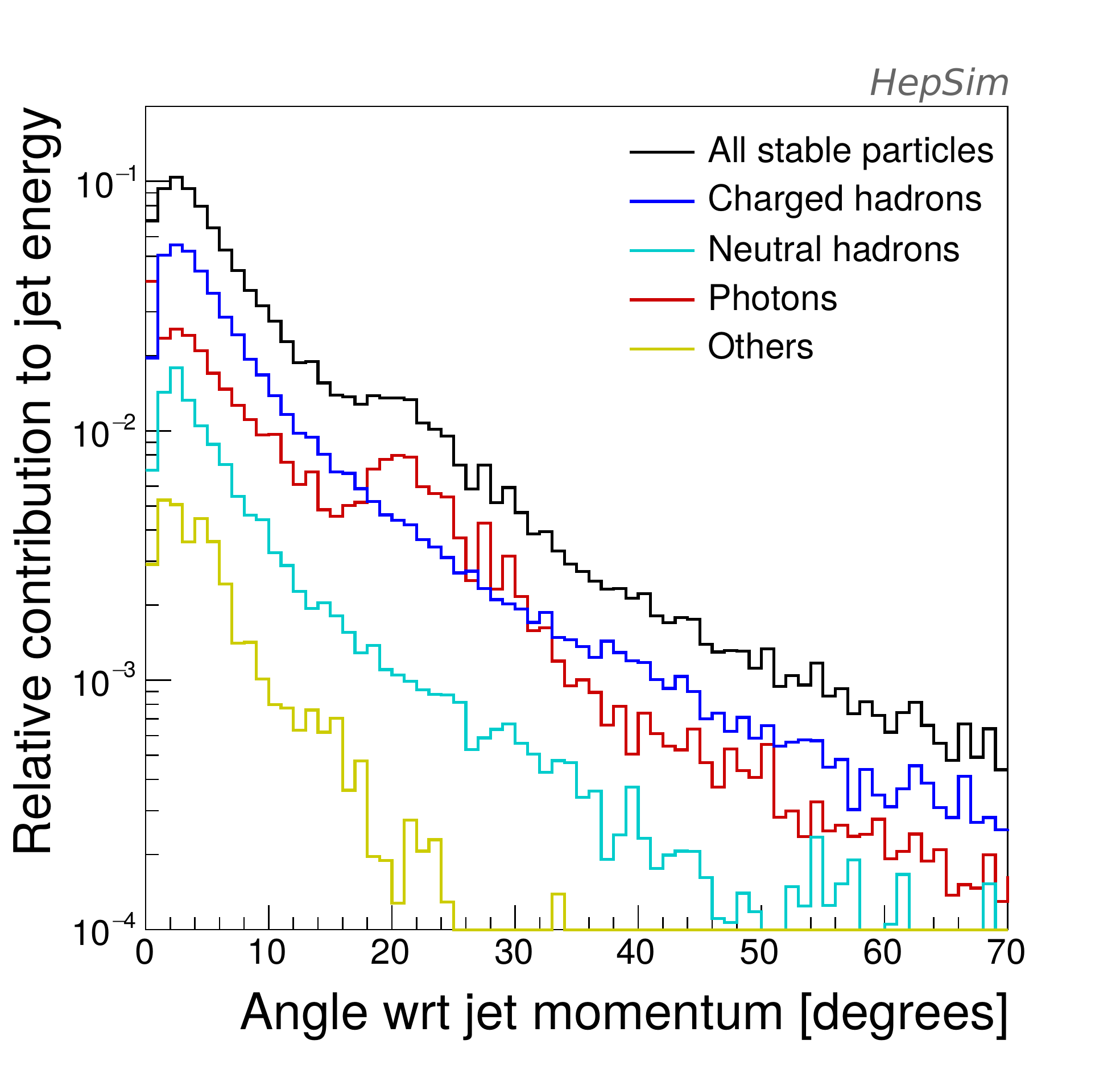}
\includegraphics[width=0.495\textwidth]{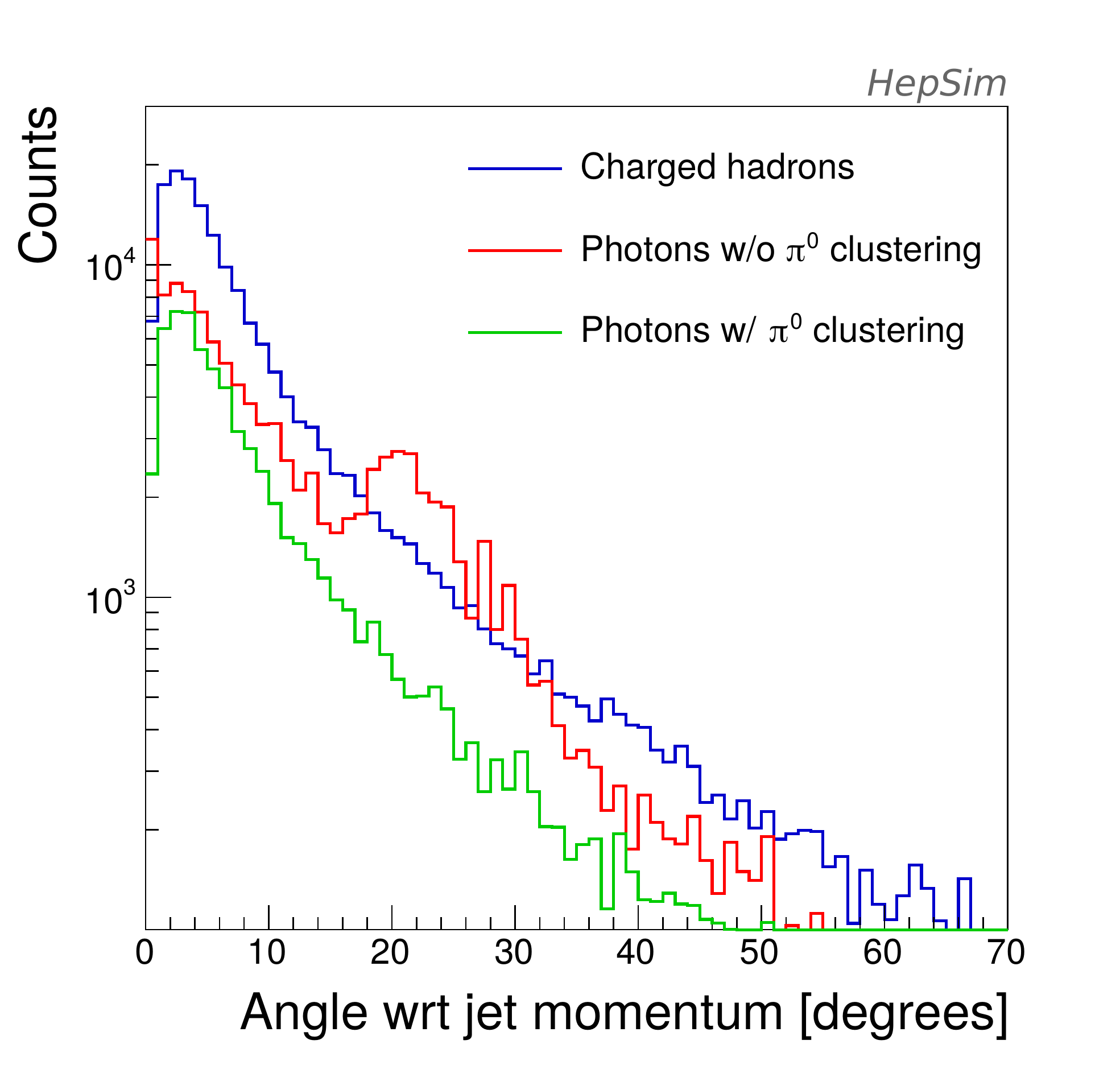}
\caption{\label{fig:jet_angular_composition} Left: relative energy contribution at a given angle with respect to jet momentum for different particle types within a jet. Right: comparison of photon angular distribution within jet with the distributions of charged and neutral pions. Jets from the decay of the Z boson to $b\bar{b}$ and $p_{\rm T}$ between 35 and 45 GeV are used. In both figures the contribution from photons originating from a $\pi^{0}$ decay, emitted preferably at 10-20 degrees angle with respect to the reconstructed jet momentum, is visible.}
\end{figure}

\subparagraph{Angular distribution of particles within jets}
About 27\% of the jet energy is made of photons, of which about 90\% originates from the decay of $\pi^{0}$-mesons within the tracker volume and about 7\% from the decay of $\eta$-mesons.
These photons are emitted with an average angle of about 20 degrees with respect to the direction of the mother $\pi^{0}$ particle and thus cause an additional widening of the particles within a jet.
Figure~\ref{fig:jet_angular_composition} shows, for different types of particle within a jet, their angular aperture, defined as the angle between the particle momentum, measured at the vertex, with respect to the direction of the reconstructed jet.
If photons were first clustered into corresponding $\pi^{0}$'s, the angular distribution of the clustered $\pi^{0}$'s would then follow the same distribution as the charged hadrons (mainly $\pi^{\pm}$'s).
It should be noted that low momentum charged hadrons are substantially deflected by the magnetic field and would not be clustered efficiently within a jet based only on their calorimetric energy despositions. The PFA approach makes use of the tracker measurement to reconstruct the momentum of the charged hadron at vertex, therefore eliminating the confusion that would otherwise be created by the bending effect. The angular distribution of charged hadrons shown in Figure~\ref{fig:jet_angular_composition} is thus calculated using the particle momentum at vertex.
%[add a picture of an event display?]

\begin{figure}[!tbp]
\centering % \begin{center}/\end{center} takes some additional vertical space
\includegraphics[width=0.495\textwidth]{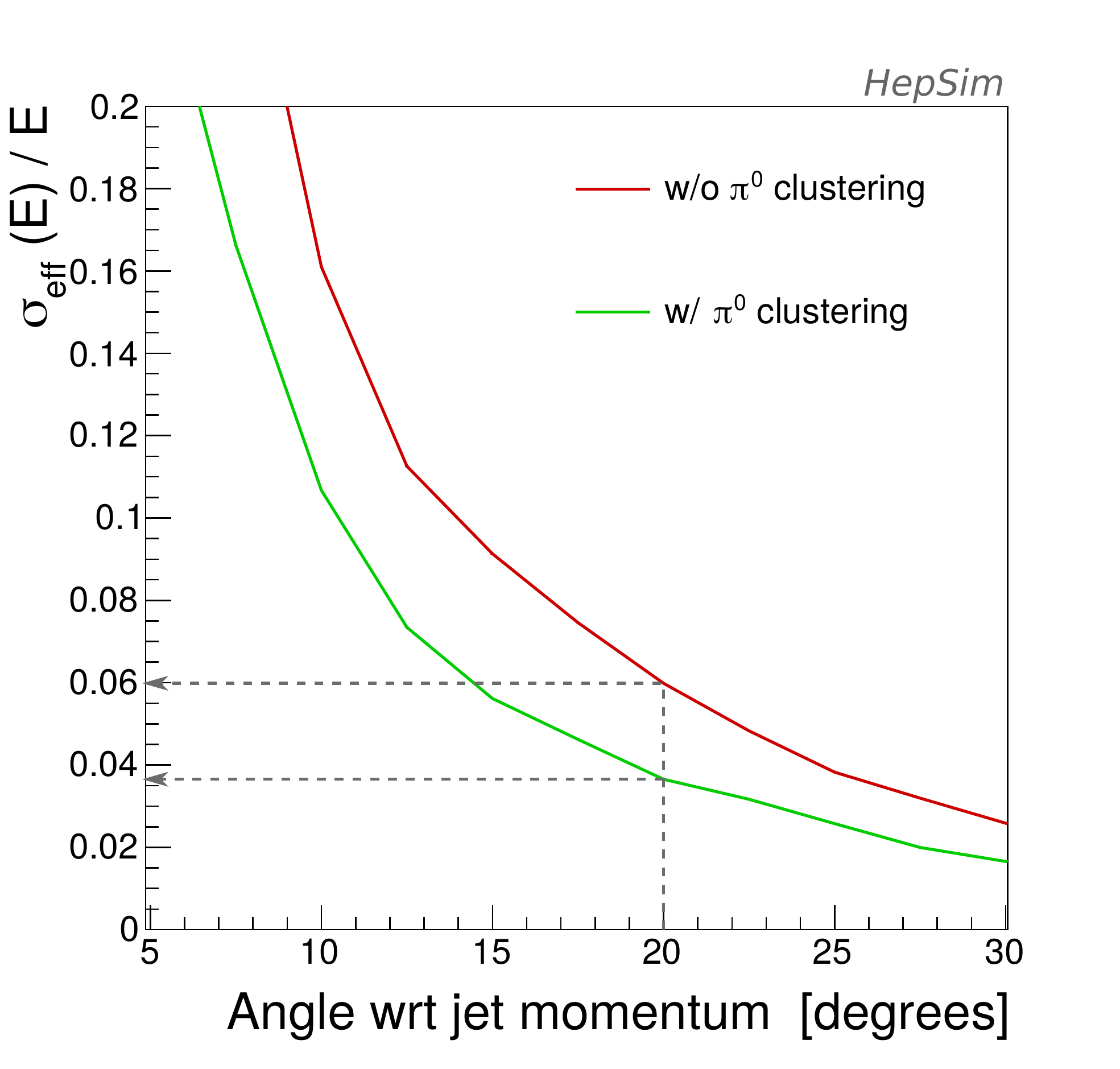}
\includegraphics[width=0.495\textwidth]{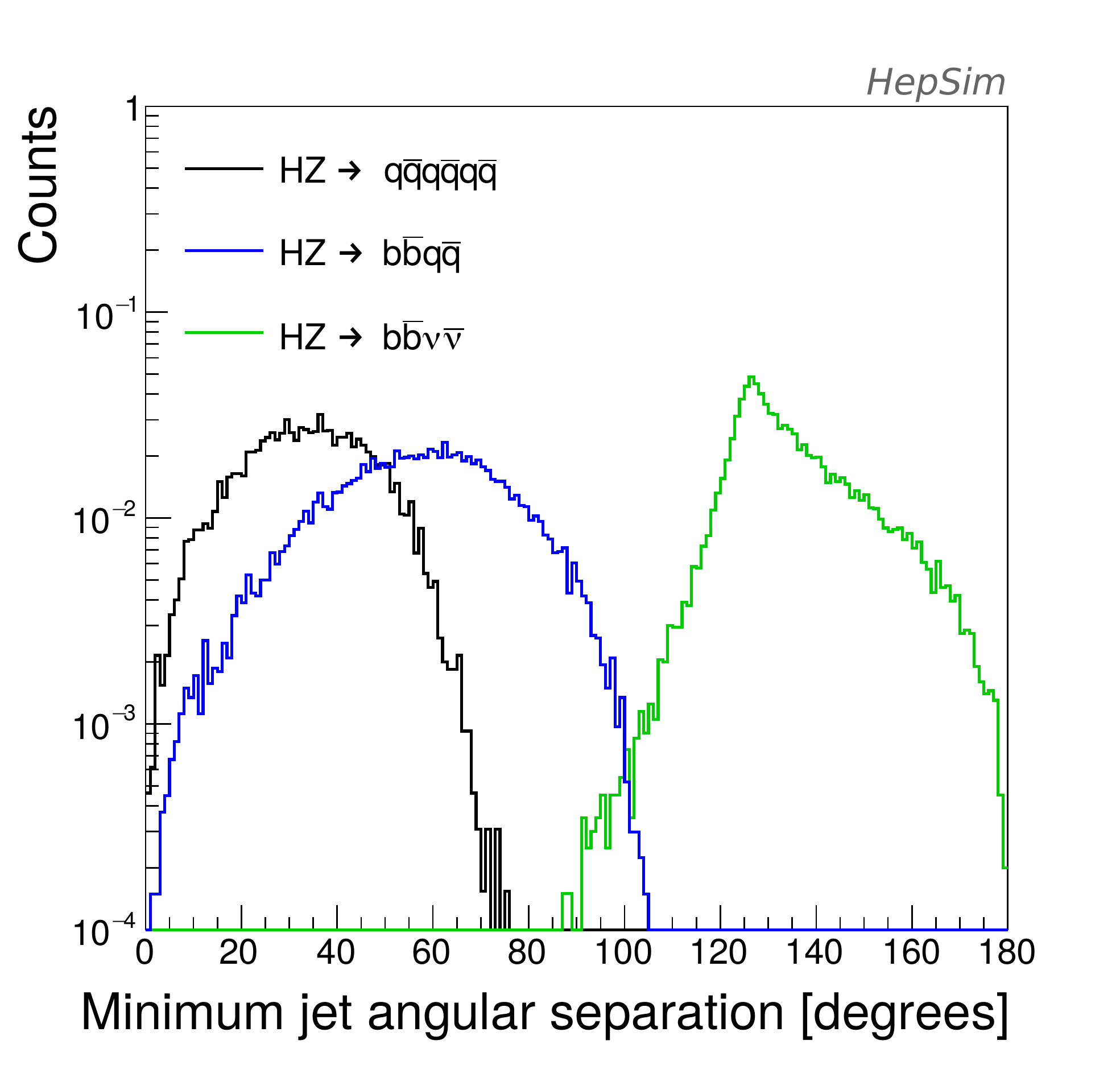}
 \caption{\label{fig:jet_res_angular_cut_andMinSep} Left: contribution from exclusion of particles outside a given angular aperture to the jet energy resolution. Jets have a typical angular aperture of about 10 degrees, however the largest opening angle of a particle with respect to the reconstructed jet momentum varies in a wide range depending on the jet algorithm used. Right: minimum angular separation between jets in multi-jets event from different Higgs boson decay processes featuring different numbers of jets (2,4,6 jets).}
\end{figure}

For jets constructed, for simplicity, using a simple cone algorithm, the loss of particles outside the cone affects the reconstructed jet energy resolution. The resulting resolution as a function of cone size is shown in the left plot of Figure~\ref{fig:jet_res_angular_cut_andMinSep}.
The pairing of photons into $\pi^{0}$'s would reduce the impact of a cut on the cone angle on the jet energy resolution. For instance, excluding from the jet particles at an angle larger than 20 degrees with respect to the jet momentum would impact the jet energy resolution with a 3.6\% contribution (with photon pairing) instead of 6\% (without photon pairing).

%, such that if one were to exclude from the clustering particles outside a 20 degrees angle cone, the impact on the jet energy resolution from the rejected photons would be 6\% instead of 4\% as for the case of charged hadrons.

While for the events with 2-jet topologies this is not critical, in events with 4 or 6 jet topologies the average minimum angular separation between jets is much smaller, peaking around 50 and 30 degrees respectively. In such event topologies, a substantial fraction of the events have jets closer than 30 degrees, as shown in the right plot of Figure~\ref{fig:jet_res_angular_cut_andMinSep}.
This means that roughly one-third of the multi-jet Higgs physics is impacted by the low energy photon annulus.

%The minimum angular separation between jets peaks at the edge of the parameter R used for jet clustering with the anti-k$_T$ algorithm \cite{Cacciari_AntiKT} (R=0.4 for the case shown in the figure).

%In addition.... gluon radiation jets will tend to have a relatively smalle angle with respect to other jets as in 6q events. Gluon jets tend to be emitted at smalla ngles relative to the emitting quark jet in 4q events. as discussed in \cite{J.Mans https://arxiv.org/pdf/hep-ex/0204029.pdf}.

\subparagraph{Performance of the $\pi^{0}$ clustering algorithm}

In event topologies featuring 4 or 6 jets, the probability that photons from $\pi^{0}$'s get scrambled across jets increases with respect to 2-jets events, thus degrading the jet resolution relative to the hadron-based ($\pi^{\pm}$, $\pi^0$, etc.) jet energy.
We therefore evaluate a technique for pre-processing $\pi^0$ momenta through combinatoric di-photon combinations in advance of applying jet algorithms. Such a technique is particularly powerful with a high-resolution EM calorimeter. We found that this procedure significantly reduces the effective angular spread and particle sharing across jets of photons from $\pi^0$ decay in multi-jet events.

The average number of photons with energy above 100 MeV involved in multi-jet event topologies grows from about 20 for 2-jet events to about 40(50) for 4(6)-jet events. 
The number of potential solutions when considering all candidate $\pi^0$'s (all possible photon pairs) can thus be very large. As proposed in \cite{pi0clustering}, a natural way to approach such a problem is to use methods from graph theory.
In particular, to identify the set of photon pairs that better matches their parent pions, we build a graph in which each node (vertex) is a photon. 
Each edge between nodes thus represents a potential pairing of photons.
We then assign to each edge a weight, $w_{ij}$, defined as:
\begin{equation}
w_{ij} = 1 - \frac{\chi^2_{ij}}{\chi^2_{max}}
\end{equation}
where $\chi^2_{ij}$ is the chi-squared between the invariant mass of the photon pair, $M_{\gamma_i\gamma_j}$, and the $\pi^0$ mass:
\begin{equation}
\chi^2_{ij} = \frac{(M_{\gamma_i\gamma_j}-M_{\pi^{0}})^2}{M_{\pi^{0}}}
\end{equation}
and $\chi^2_{max}$ is the maximum chi-squared among all edges of the event.
The solution of the problem is the one that maximizes the sum of the edge weights, $\sum w_{ij}$, while using each node (photon) at most once.
%To account for the fraction of photons ($\sim 10\%$) that do not originate from $\pi^0$ (\emph{spare} photons) and thus should not be paired, we exploit the method suggested in \cite{10.1007/3-540-44691-5_3} that involves the creation of a new duplicate graph and allows for photons to go unmatched.
An implementation of the Edmonds' algorithm \cite{edmonds_1965} called Blossom V \cite{Galil} was used, as provided by the \textsc{NetworkX} Python package \cite{NetworkX}, which can solve the graph in polynomial time. 
The algorithm was tested on samples from the HepSim repository featuring respectively 2, 4 and 6 jets event topologies:
\begin{itemize}
\itemsep0em 
\item $HZ\rightarrow b\bar{b}\nu\bar{\nu}$, 
\item $HZ\rightarrow b\bar{b}q\bar{q}$,
\item $HZ\rightarrow q\bar{q}q\bar{q}q\bar{q}$ .
\end{itemize}

Different assumptions of the calorimeter resolution to EM particles were compared by smearing the energy of Monte Carlo truth particles with a stochastic term parameterized by A$/\sqrt{E}$, with A in the range from 0 to 30\%.

To reduce computational time and increase the matching efficiency, a set of constraints can be imposed on the edges.
Such constraints represent an underlying \emph{graph structure} which defines which edges can be considered by the algorithm in the optimization process. 
We consider the following variables to discriminate between true and fake photon pairs:
\begin{itemize}
\itemsep0em 
\item the invariant mass of the two photons, $M_{\gamma_1 \gamma_2} = 4E_1E_2 \sin^2(\alpha/2)$ (with $\alpha$ being the angle between the two photons);
\item the angular distribution of the leading photon momentum in the rest frame of the reconstructed $\pi^0$, with respect to the $\pi^0$ momentum in the lab frame: $\cos [\theta (\gamma^{1}_{RF}, \pi^{0}_{LB}]$;
\item the Lorentz boost of the reconstructed $\pi^0$: $E_{\pi^{0}} / M_{\pi^{0}}$.
\end{itemize}

\begin{figure}[!tbp]
\centering 
\includegraphics[width=0.495\textwidth]{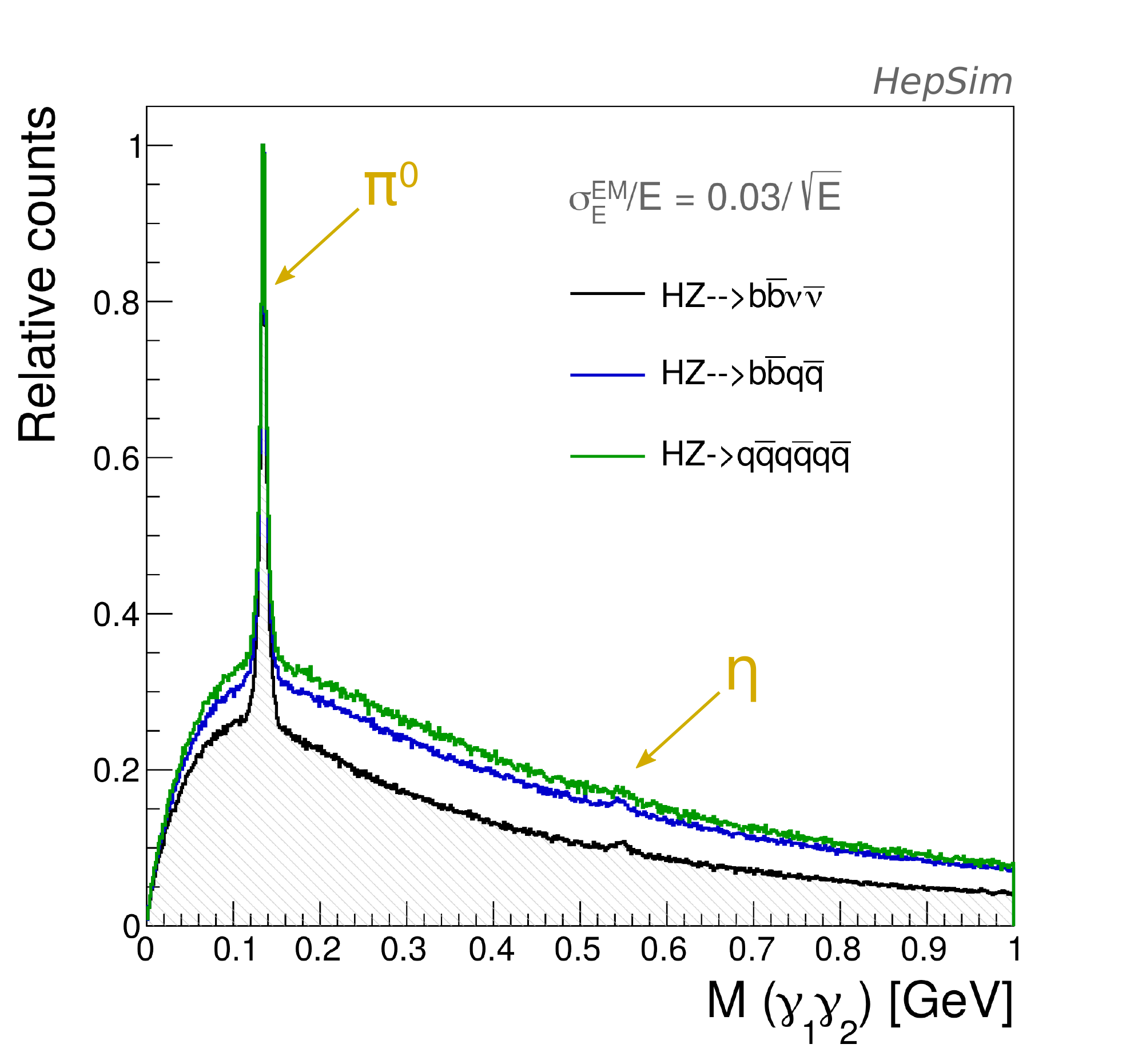}
\includegraphics[width=0.495\textwidth]{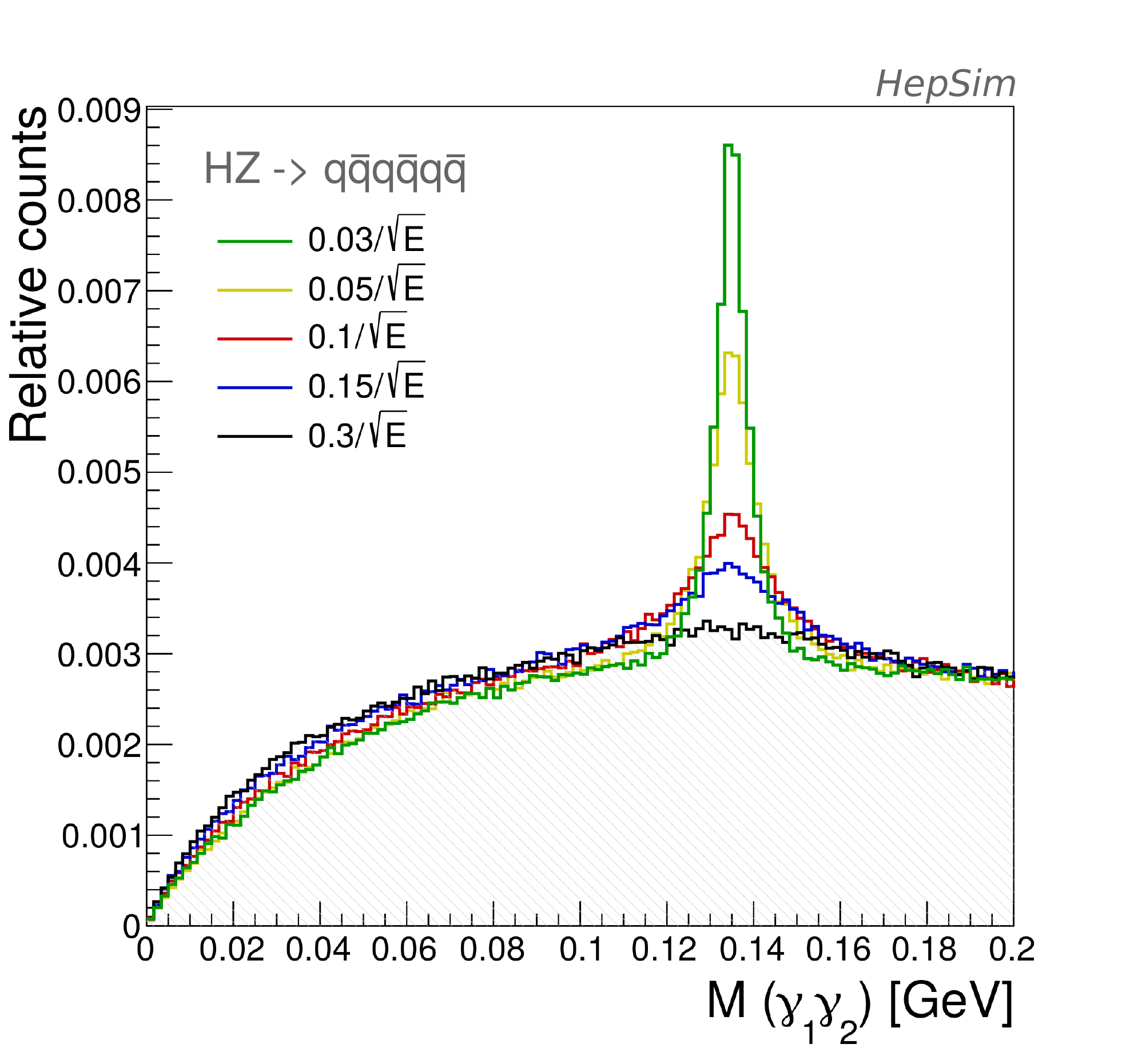}
\caption{\label{fig:diphoton_masses}  Left: distributions (normalized to the maximum) of the combinatoric di-photon invariant mass for different event topologies assuming a calorimeter with 3\%$/\sqrt{E}$ energy resolution to photons. Right: distributions (normalized to the integral) of the combinatoric di-photon invariant mass for $HZ \rightarrow q\bar{q}q\bar{q}q\bar{q}$ events varying the EM energy resolution of the calorimeter.}
\end{figure}

Depending on the EM resolution of the calorimeter, the peak corresponding to $\pi^0$ photons will be more or less resolved on top of the combinatorial background of the invariant mass from possible photon pairs. This is shown in Figure~\ref{fig:diphoton_masses} for different event topologies and different EM energy resolutions.
In particular, we apply a constraint on the invariant mass of the photon pair belonging to a given edge to be within $\pm3\sigma^{\pi^0}_{EM}$ around the true $\pi^0$ mass (where $\sigma^{\pi^0}_{EM}$ is the resolution of the $\pi^0$ peak for a given calorimeter EM resolution), which thus includes 99.73\% of the signal events. The fraction of background will depend on $\sigma^{\pi^0}_{EM}$. For a poor EM resolution, the number of photon pairs from combinatorial background included in such a window grows very quickly as the acceptance window on the diphoton invariant mass becomes wider. The algorithm computational time and the fraction of wrong photon pairs matched thus increases to a level where the probability of a correct photon pairing is smaller than 50\% as shown in Figure~\ref{fig:graph_performance}.

\begin{figure}[!tbp]
\centering
\includegraphics[width=0.495\textwidth]{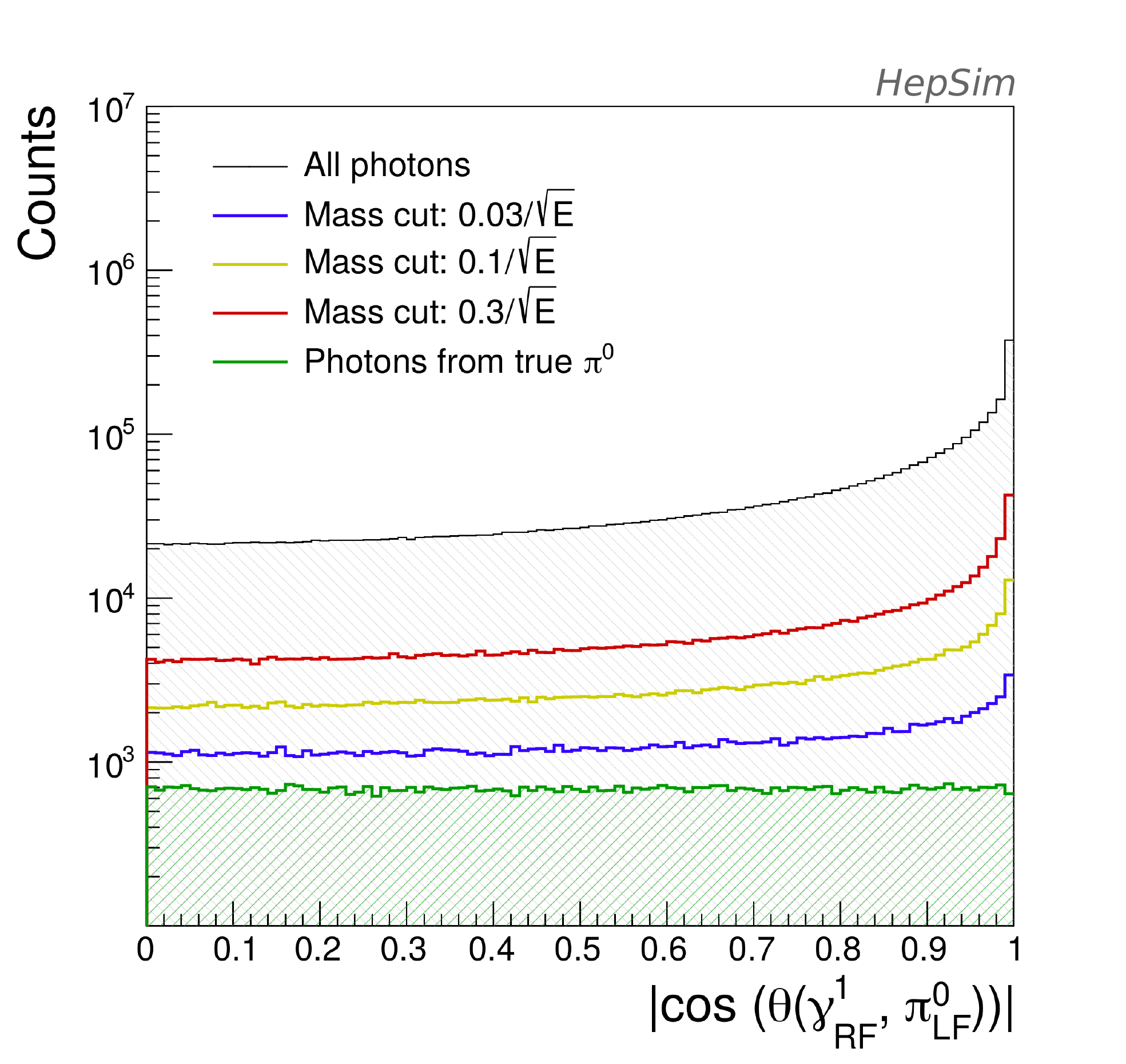}
\includegraphics[width=0.495\textwidth]{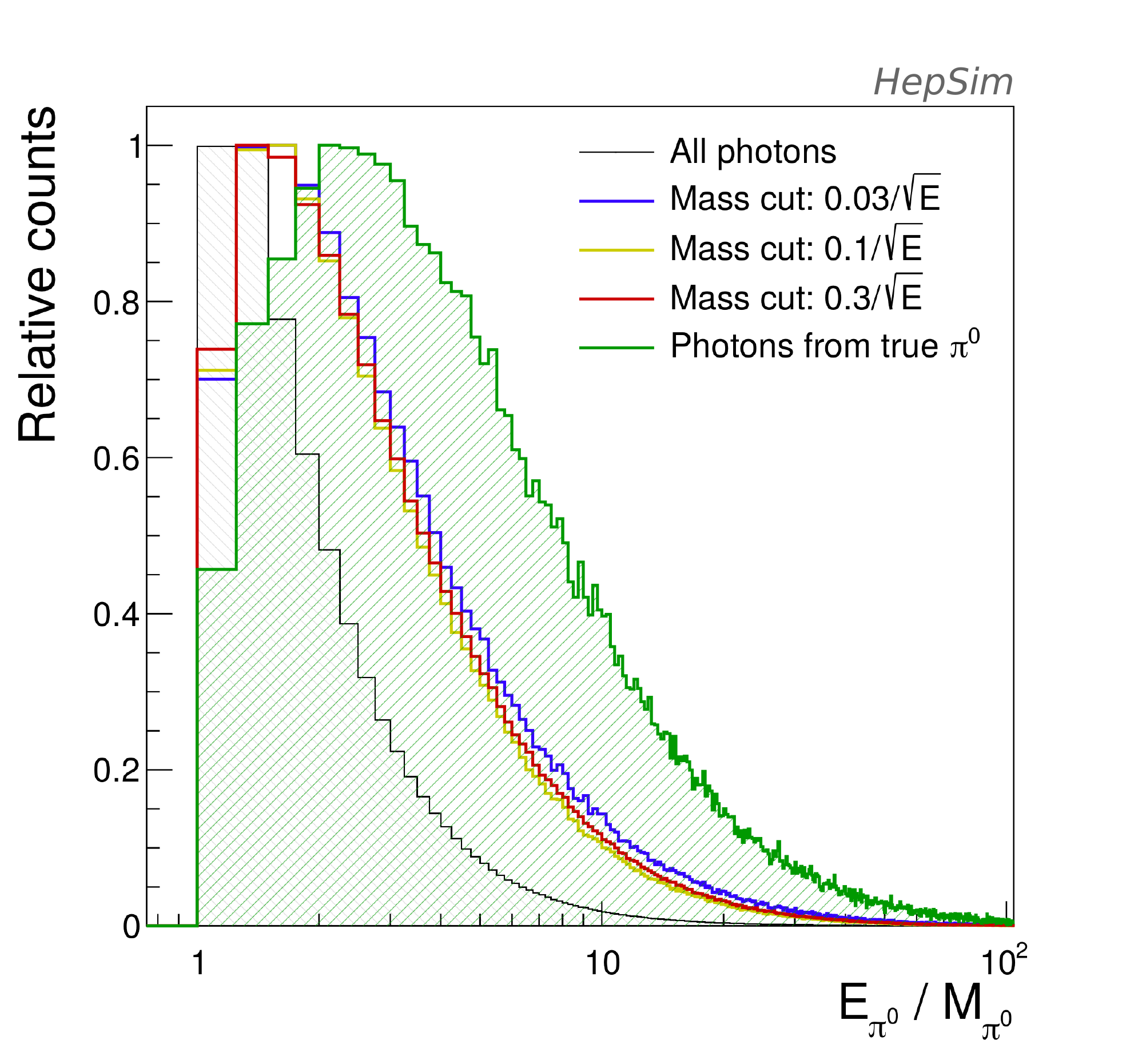}
\caption{\label{fig:algo_cuts} Distribution of $\cos [\theta (\gamma^{1}_{RF}, \pi^{0}_{LB})]$ (left) and Lorentz boost factor (right) for all photon pairs in the event (black), for all photon pairs that pass the selection cut based on the di-photon invariant mass and for photon pairs from real $\pi^{0}$'s (green) for $HZ \rightarrow  q\bar{q}q\bar{q}q\bar{q}$ events.}
\end{figure}

\begin{figure}[!tbp]
\centering
\includegraphics[width=0.495\textwidth]{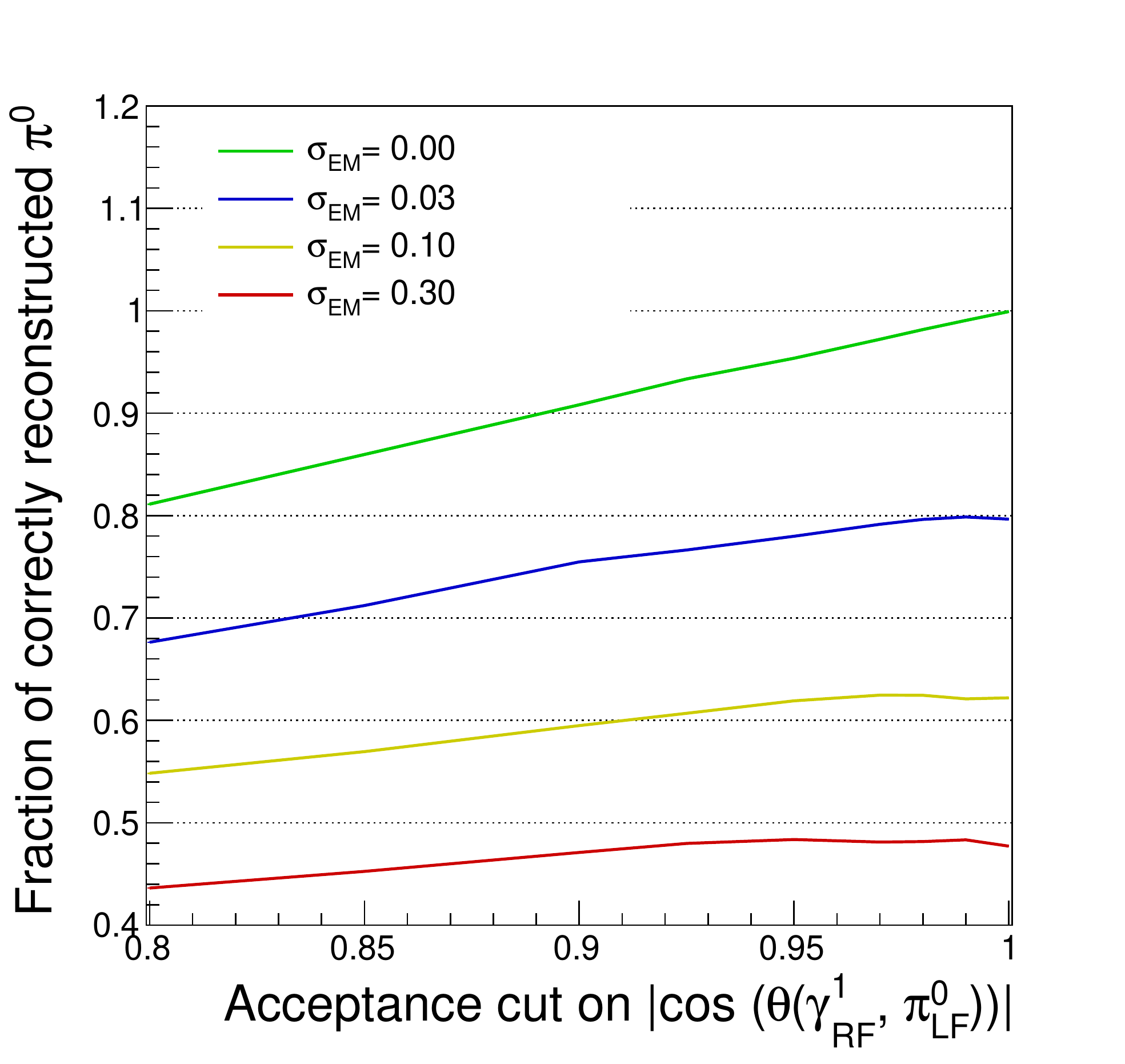}
\includegraphics[width=0.495\textwidth]{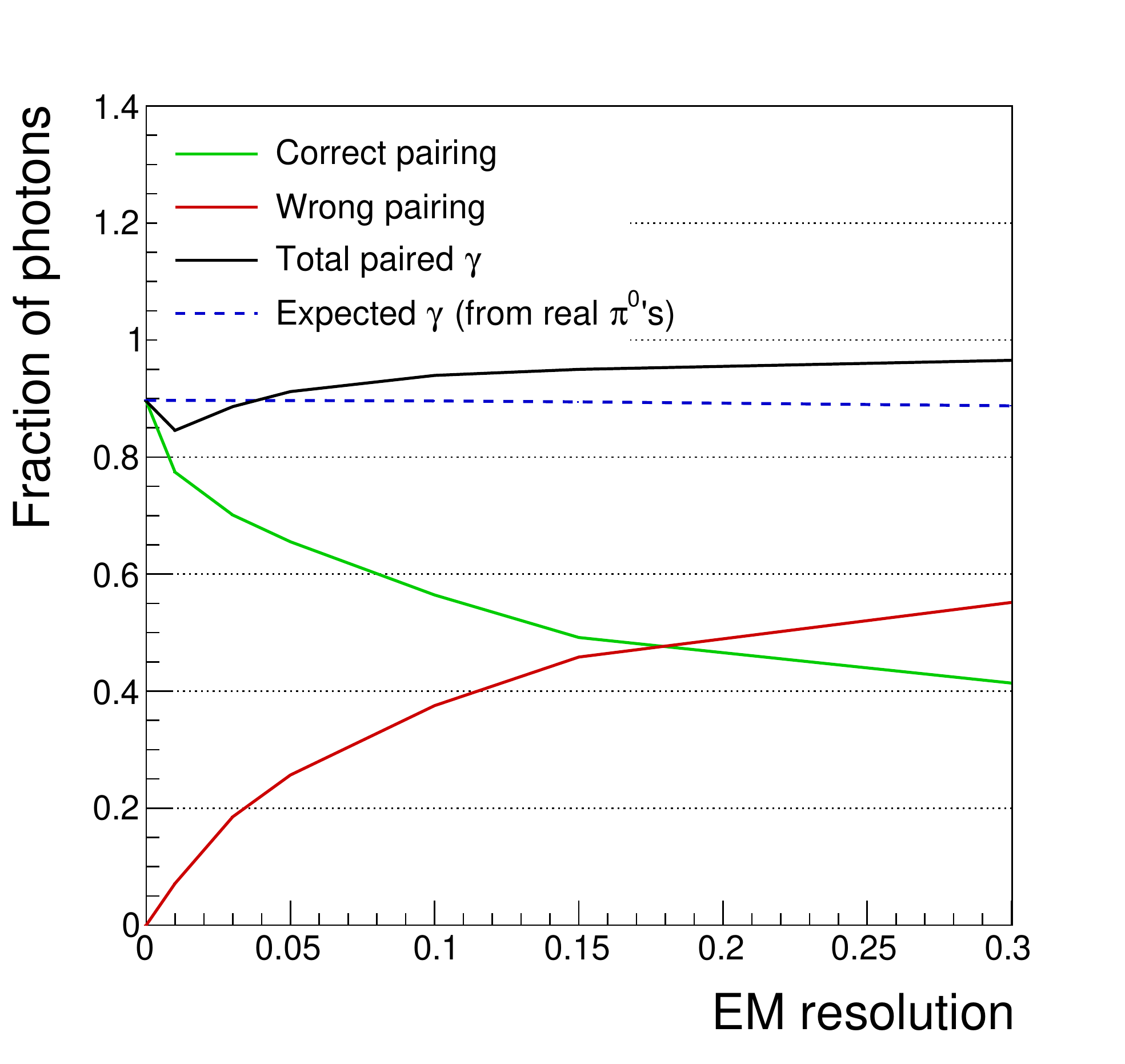}
\caption{\label{fig:graph_performance} Left: fraction of $\pi^0$'s correctly reconstructed from photon pairing for different EM resolution and varying the acceptance cut on the angular distribution of photons in the rest frame wrt to the momentum of the clustered $\pi^0$. Right: average fraction of photons within an event which are correctly paired into a $\pi^0$ (green) and incorrectly clustered as a $\pi^0$ (red) compared with the expected fraction of photons from $\pi^0$ (blue dashed line) and to the total fraction of photons that are paired by the clustering algorithm (black).}
\end{figure}

To reduce the contribution from the combinatorial background, additional cuts can be applied on the Lorentz boost factor ($E_{\pi^{0}} / M_{\pi^{0}}$) and on the angle between the leading photon momentum (highest energy in the pair) in the rest frame with respect to the reconstructed $\pi^{0}$'s momentum in the lab frame, $\cos [\theta (\gamma^{1}_{RF}, \pi^{0}_{LB})]$. 
The distributions of such variables are shown in Figure~\ref{fig:algo_cuts} for the combinatorics of all photons (background), with and without a cut on their invariant mass, and for just photon pairs originating from the same mother $\pi^0$.
In the rest frame, the two $\pi^0$ photons are emitted back-to-back and isotropically. This yields an angular distribution that is flat for photons from real $\pi^0$'s, while it is strongly peaked around one for the combinatoric background.
Similarly, a difference between the Lorentz boost distributions from real and background photon pairs is visible. In particular, the real $\pi^{0}$'s have on average a larger boost (peaking around 2.4 GeV) with respect to the combinatorial background which peaks at 1.2-1.5 GeV.
Additional cuts on these variables to further constrain the graph structure have a direct impact on the signal efficiency, i.e. the fraction of photon pairs correctly clustered decreases.
For poor EM resolutions, however, the simultaneous decrease of incorrectly clustered photon pairs can be beneficial for the photon-to-jet assignment efficiency to the limit where no photons are clustered and the choice of whether to cluster a photon within a jet is entirely left to the jet clustering algorithm (no gain and no degradation from the $\pi^0$ clustering algorithm).

As shown in Figure~\ref{fig:graph_performance}, the performance of the $\pi^{0}$ clustering algorithm strongly depends on the calorimeter EM resolution. 
The impact of imposing an additional cut on the photon angular distribution (from Figure~\ref{fig:algo_cuts}) to reduce the combinatorial background contribution is shown in the left plot of the figure. For energy resolutions at the level of $3\%/\sqrt{E}$ or better, there is no gain from this additional cut. For poorer energy resolutions the requirement $\cos [\theta (\gamma^{1}_{RF}, \pi^{0}_{LB})]<0.96-0.98$ slightly improves the fraction of $\pi^{0}$'s that are correctly reconstructed.
For perfect energy resolution, all $\pi^{0}$'s are correctly reconstructed and no photons are incorrectly clustered. For poorer energy resolutions, the algorithm tends to cluster all photons, including the ones that do not originate from a real $\pi^0$ decay.
A very high efficiency of about 80\% in correctly pairing photons from $\pi^{0}$ is achieved for a calorimeter resolution of $3\%/\sqrt{E}$ while for a calorimeter EM resolution of $30\%/\sqrt{E}$ less than 50\% of the $\pi^{0}$ are correctly reconstructed. In addition, the fraction of photons that are incorrectly clustered as a $\pi^{0}$ increases from less than 19\% up to about 55\%. For a calorimeter resolution worse than $18\%/\sqrt{E}$, the majority of photons are incorrectly clustered into $\pi^0$'s.

\subparagraph{Improvements in correct photon-to-jet assignment}
We then explored the impact of the photon clustering algorithm in improving the efficiency of correctly assigning a photon to the corresponding jet. 
We do this by comparing the particles clustered within a jet using the Jade algorithm for three different cases: 
\begin{enumerate}[itemsep=1pt]
\item {\bf Standard}: the jet clustering algorithm utilizes all photons present in the event (without any $\pi^0$ clustering);
\item {\bf Ideal}: the jet clustering algorithm utilizes true $\pi^0$'s based on Monte Carlo truth (instead of photons from $\pi^0$ decays) and with \emph{spare} photons (not originated from a $\pi^0$ decay);
\item {\bf Pre-clustering of $\pi^{0}$'s}: the jet clustering algorithm utilizes $\pi^0$'s clustered using the algorithm presented above and \emph{spare} photons (which have not been clustered into $\pi^0$'s by the algorithm);
\end{enumerate}

For each jet in the event, we count how many photons are correctly assigned to the corresponding jet, i.e. the jet to which they get assigned in the ideal case (2) (that uses real $\pi^0$'s from Monte Carlo truth).
We then compare the fraction of photons that are correctly assigned to a jet with and without the use of the $\pi^0$ clustering algorithm: cases (3) and (1) respectively.

\begin{figure}[!tbp]
\centering
\includegraphics[width=0.495\textwidth]{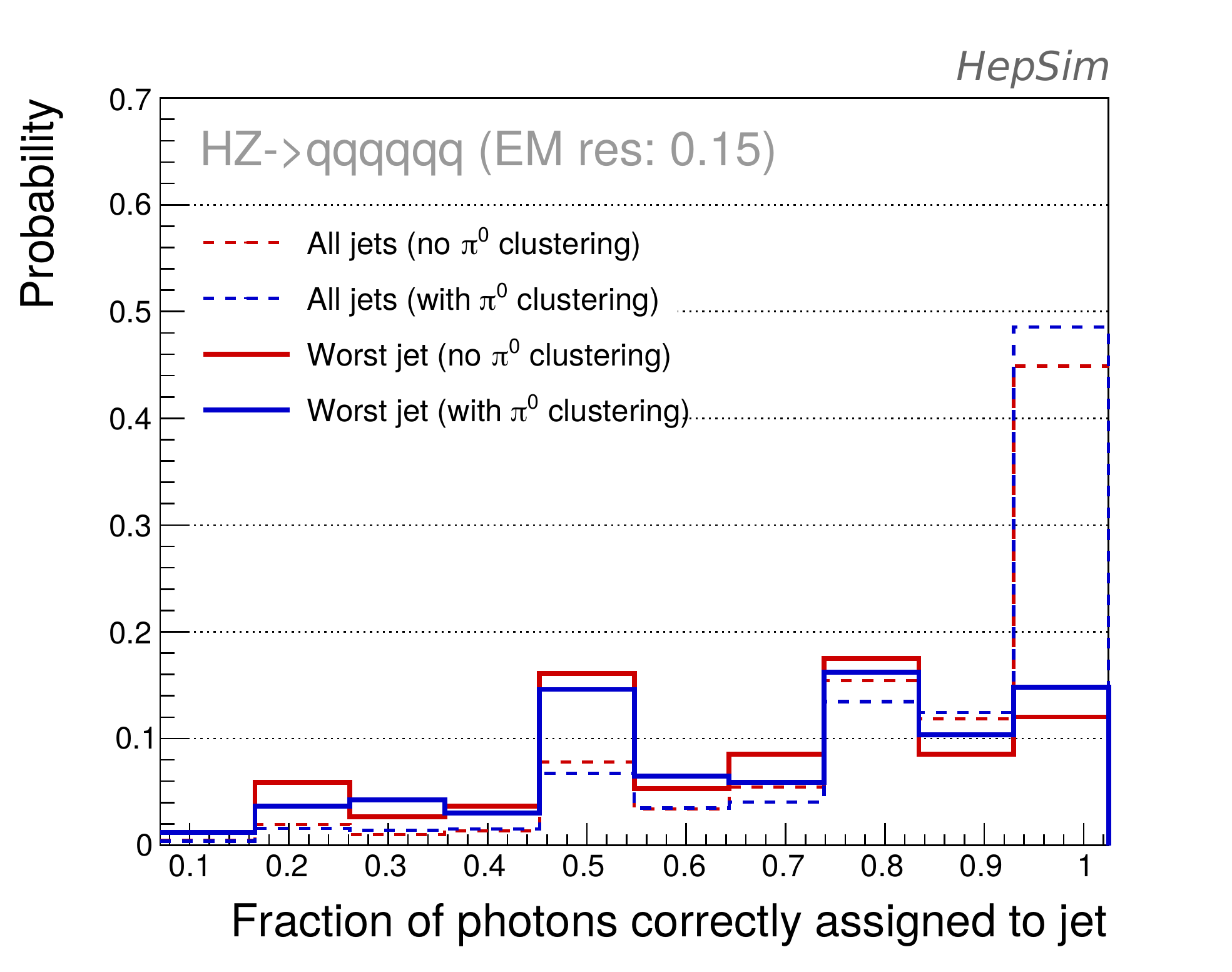}
\includegraphics[width=0.495\textwidth]{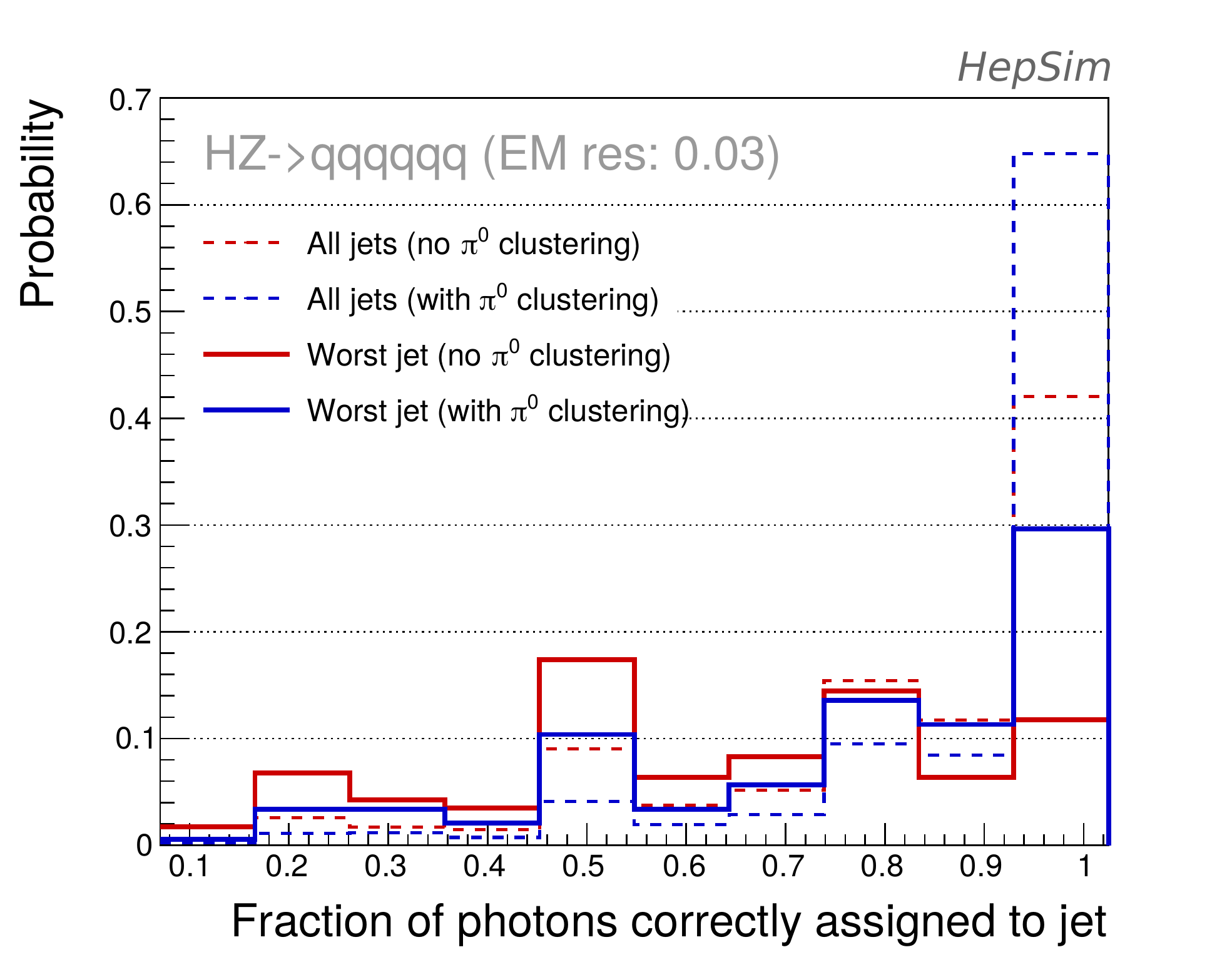}
\caption{\label{fig:algo_performance_distr} Distribution of the fraction of photons that are correctly assigned to a jet in $HZ\rightarrow q\bar{q}q\bar{q}q\bar{q}$ events with and without the use of the $\pi^0$ clustering algorithm for all jets and the worst jet in the event and for a EM resolution of $15\%/\sqrt{E}$ (left) and $3\%/\sqrt{E}$ (right).}
\end{figure}

\begin{figure}[!tbp]
\centering
\includegraphics[width=0.495\textwidth]{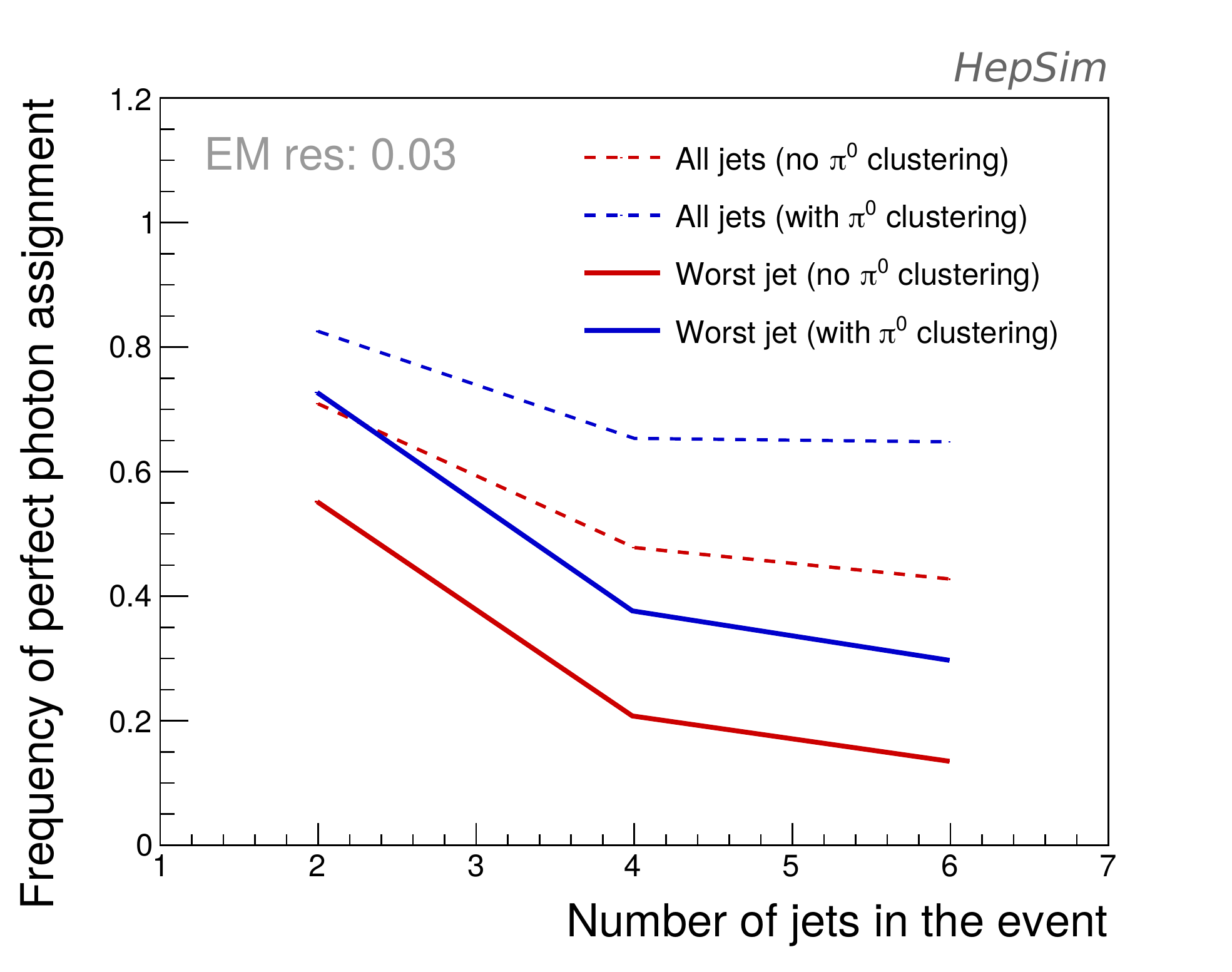}
\includegraphics[width=0.495\textwidth]{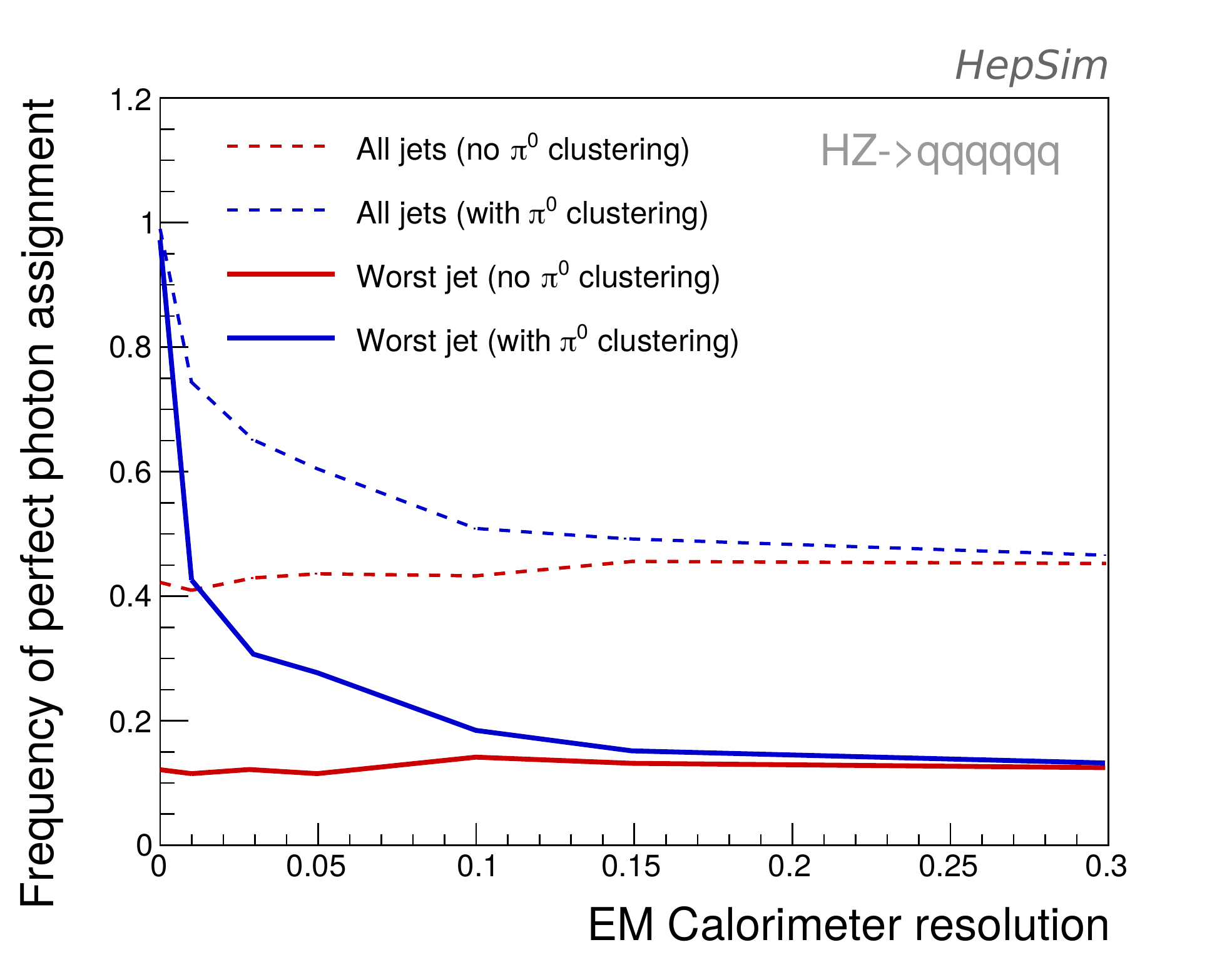}\\
\caption{\label{fig:algo_performance} Frequency of events where photons are perfectly assigned to the corresponding jet as a function of the number of jets in the event, assuming a calorimeter resolution of $3\%/\sqrt{E}$ (left), and as a function of calorimeter EM resolution in the case of the $HZ\rightarrow q\bar{q}q\bar{q}q\bar{q}$ sample (right).}
\end{figure}

For a perfect EM energy resolution (i.e. no smearing of the photon energy applied), the $\pi^0$ clustering algorithm yields substantial improvements in the photon-to-jet assignment. The gain is, as expected, more pronounced for the event topologies with 4 or 6 jets, where standard algorithms are more affected by particles scrambled across jets.
We then compare the performance of the algorithm assuming different energy resolutions for the calorimeter by applying a smearing to the photon energy in the $3\%-30\%/\sqrt{E}$ range. 
As shown in Figure~\ref{fig:algo_performance_distr} and \ref{fig:algo_performance}, the algorithm performance degrades quickly once the calorimeter resolution is larger than 3-5\%$/\sqrt{E}$.
As discussed earlier, this is strongly connected to the increasing fraction of fake $\pi^0$'s clustered by the algorithm, which actually deteriorates as a confusion term the performance of the jet clustering algorithm.

\section{Optimization of a hybrid segmented dual-readout calorimeter}\label{sec:calo_overview}

In Section~\ref{sec:pfa_drivers}, some of the key criteria for the optimization of a calorimeter design to achieve the jet energy resolution goal when using the particle flow algorithm have been introduced. 
In addition, it was shown in Section~\ref{sec:highlights}, that an electromagnetic calorimeter capable of measuring photons with an energy resolution of $3-5\%/\sqrt{E}$ enables novel approaches for improving particle reconstruction. 

Based on such considerations, we propose in the following a segmented calorimeter design that combines excellent timing capabilities and energy resolution by exploiting the potential of scintillating crystals and new photodetectors. We then show how such a system can be effectively integrated with hadron calorimeters that feature dual-readout capabilities to achieve excellent energy resolution for neutral hadrons.

\subsection{Overview of the calorimeter layout}\label{sec:overview_calo}
\begin{figure}[!tbp]
\centering % \begin{center}/\end{center} takes some additional vertical space
\includegraphics[width=0.99\textwidth]{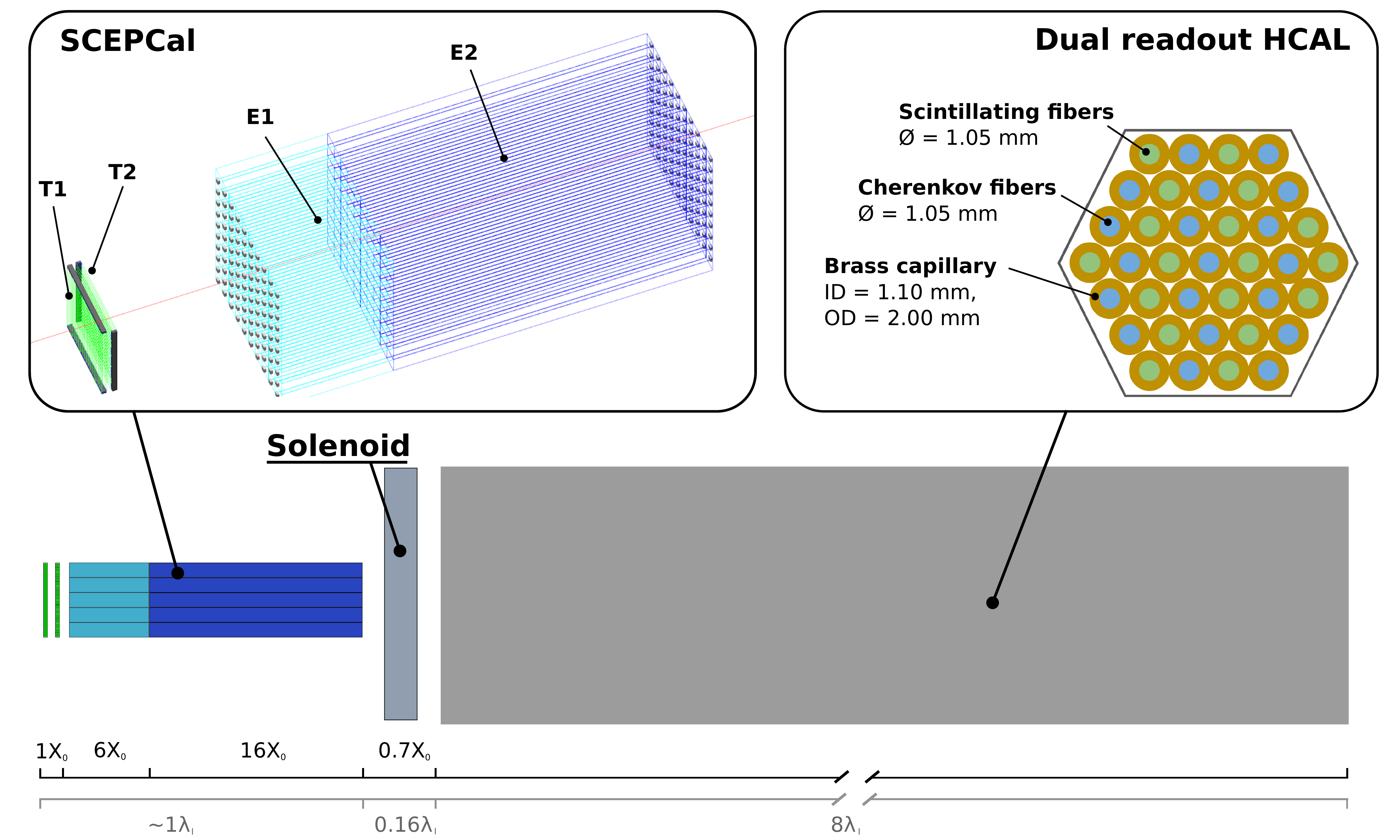}
\caption{\label{fig:overall_layout} Overview of a hybrid segmented calorimeter layout featuring 4 front segments which exploit scintillating crystals for detection of EM showers followed by an ultrathin-bore solenoid and a hadron calorimeter based on scintillating and quartz fibers.}
\end{figure}

The proposed calorimeter concept consists of two highly segmented timing layers, a 2-layer crystal ECAL with dual readout, followed by an ultrathin-bore solenoid and a hadron calorimeter with dual-readout capabilities. A sketch of the overall calorimeter system, implemented in a \textsc{Geant4} simulation, is shown in Figure~\ref{fig:overall_layout} and discussed in the following.
A tracker layout inspired by the one proposed for the CEPC detectors is used, made of seven layers, for an integrated material budget of about 0.1~$X_0$ and radial dimensions of about 1.9~m.

The Segmented Crystal Electromagnetic Precision Calorimeter (SCEPCal) consists of two thin layers (T1,T2) of a fast and bright scintillator (e.g. LYSO:Ce crystals) providing time tagging of MIPs with a precision of 20 ps ({\it timing} part), followed by a two segments (E1,E2) of a dense crystal with dual-readout capabilities (e.g. PbWO$_4$, BGO or BSO) for precise measurements of electromagnetic showers ({\it ECAL} part).
Space for cooling, readout and other services is allocated in the front and the back of both the timing and ECAL part, simulated as an Aluminum layer of 3 mm thickness. In such a configuration, no dead material is present between the two ECAL segments, thus avoiding any disruption in the sampling fraction in the location where the maximum of the electromagnetic showers occurs.

An ultrathin-bore superconducting solenoid, similar to the one designed for the so-called IDEA detector \cite{UltraThin_Solenoid} is placed between the SCEPCal and the HCAL section. In this way, the solenoid radius and cost are contained, while its material budget (although limited to less than 0.7~$X_0$) does not affect the SCEPCal energy resolution for electromagnetic showers.

Several designs have been proposed that implement the dual-readout calorimeter concept based on the use of both scintillators and quartz (Cherenkov sensitive), with possibly a third neutron-sensitive material (e.g. hydrogen-rich). The active material could be in the shape of fibers embedded in an absorber groove and extending in the longitudinal direction or in the shape of tiles forming active layers interleaved with absorber layers.
In both cases, a key parameter for an optimal performance is the sampling fraction, i.e. the ratio of the fraction of the shower energy deposited in the active material  to that in the passive material and the dimension or frequency of active sampling, as discussed in Section~\ref{sec:hcal_opt}.
%We the arrangement of fibers within the absorber and the absorber structure (material, air gaps, etc.).

In this work, we use a geometry similar to that proposed by the IDEA collaboration \cite{TB_IDEA_layout} in which sensitive fibers of 1~mm diameter are inserted into brass tubes with outer (inner) diameter of 2.2 (1.1) mm. This option provides a fill factor of about 90\% with a sampling fraction of about 2\% and has potential advantages for the practical implementation of such a detector. 
For a given sampling fraction, the fiber geometry uniformly samples the shower over its entire longitudinal development and represents thus a practical way to reduce the number of channels while keeping a high transverse segmentation.

The hybrid calorimeter layout shown in Figure~\ref{fig:overall_layout} features, in total, 5 longitudinal segments. In particular, the 4 longitudinal layers of the SCEPCal (T1,T2,E1,E2) bring an advantage for PFA with respect to a monolithic HCAL block with no longitudinal segmentation. 
It should be noted that some level of longitudinal segmentation in the HCAL could be achieved virtually by implementing, for instance, double-ended readout of the fibers and exploiting the difference in the light signal arrival time between the front and rear readout to infer the average longitudinal location of the shower.

In this work, we focus on the optimization of the SCEPCal section and a first demonstration of how it can be integrated with a dual-readout HCAL. While the optimal resolution for neutral hadrons is achieved using a dual-readout HCAL, the SCEPCal could be similarly integrated with other more conventional HCAL designs with longitudinal segmentation and no dual-readout. We also discuss some of the parameters that have been studied to evaluate possible performance optimization with respect to cost and power budget. Some of the considered factors include the length of the crystal, the transverse and longitudinal segmentation and the active area of the photodetectors.

The photodetectors of choice for the entire calorimeter system are Silicon Photomultipliers (SiPMs), a very compact and robust solution immune to magnetic field effects. Developments over the last decade have been such that the dimension of the cells (SPADs = single avalanche diodes) constituting the SiPM can be produced as small as \SI{5}{\um} \cite{FBK_RGB_UHD}. Standard cell pitches of $10-15$~\SI{}{\um} provide a large dynamic range spanning from single photon counting to more than $10^{5}$ photoelectrons for devices of few mm$^2$ active area. Improvements in the cell layout and electric field design for small cell size devices are now part of industrial production processes at many manufacturers and can provide a large photon detection efficiency (up to 40-50\% for 15~\SI{}{\um} cell size) in a wavelength range extending from the Near-UV (300~nm) \cite{FBK_NUV_SiPMs} or Vacuum-UV (175~nm) \cite{FBK_VUV_SiPMs} to the infrared (900~nm) \cite{FBK_RGB_SiPMs}.
SiPMs are a flexible technology that can be manufactured in different sizes and its key parameters can be optimized for each calorimeter section to achieve the desired performance within contained cost.
In addition, the large gain of SiPMs (of order 10$^5$-10$^6$) allows simpler, less expensive readout electronics and lowers the noise contribution to the calorimeter performance to a negligible level.

\begin{figure}[!tbp]
\centering % \begin{center}/\end{center} takes some additional vertical space
\includegraphics[width=0.99\textwidth]{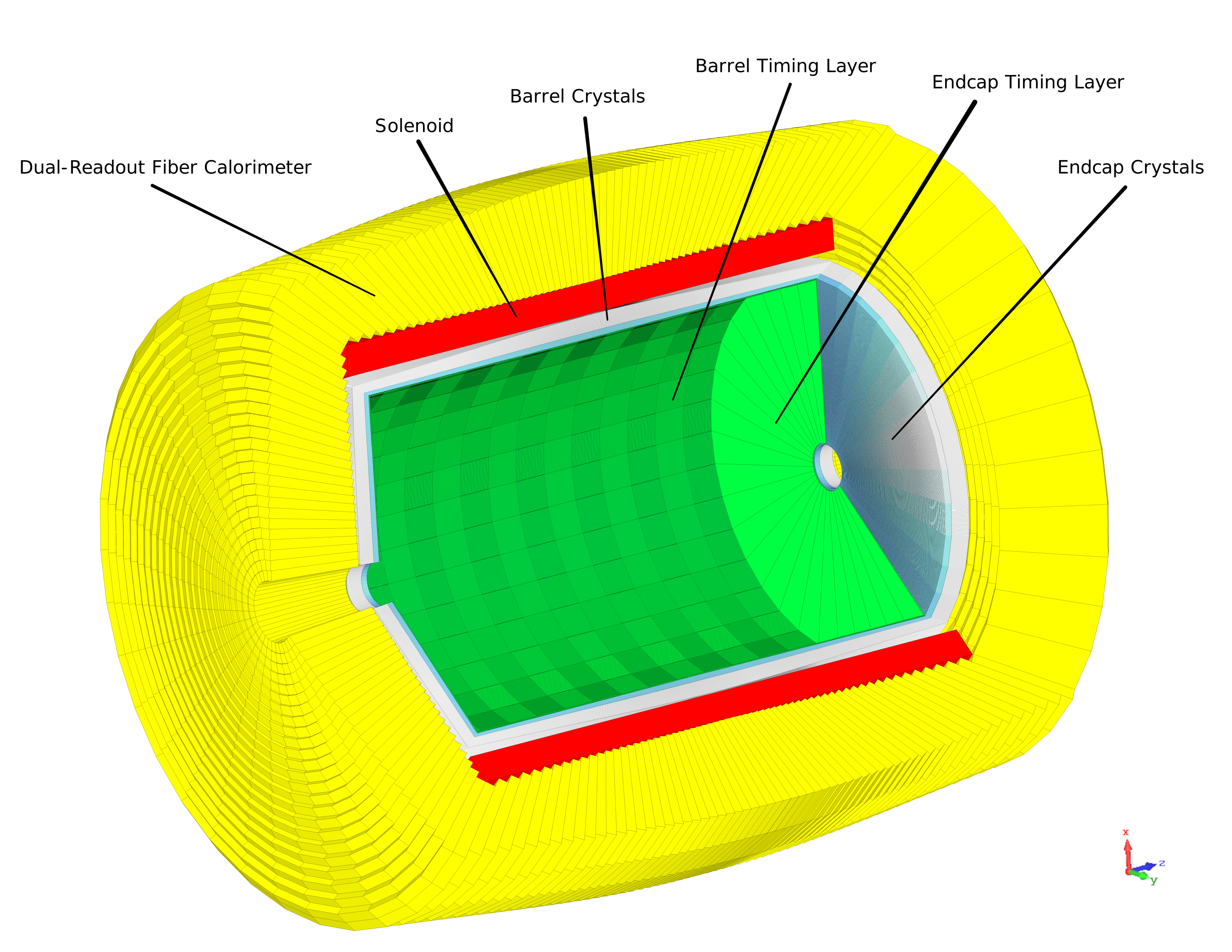}
\caption{\label{fig:dd4hep_layout} Implementation of the hybrid calorimeter system described in Figure~\ref{fig:overall_layout} in a $4\pi$ detector geometry. The layers of the detector from the inner one to the outer one are: crystal timing layers T1 and T2 (green), crystal ECAL layers E1 (light blue) and E2 (white), solenoid (red), dual-readout fiber calorimeter (yellow).}
\end{figure}

The details and highlights of each longitudinal calorimetric compartment are discussed in the respective sections below. A possible implementation of such a combined hybrid calorimeter system in a $4\pi$ detector geometry is shown in Figure~\ref{fig:dd4hep_layout}.

\subsection{Timing layers}\label{sec:Timing_part}
The potential of exploiting time as a fourth dimension in the event reconstruction at colliders has a long history. However, recent technological developments on several fronts have enabled time resolutions that are orders of magnitudes better than the previous generation of collider detectors. This new possibility triggered the interest of several groups in the High Energy Physics (HEP) community and of many industries involved in the production of silicon detectors and scintillators, thus further boosting developments in this direction.

Proof-of-principle that 30~ps time resolution for MIP tagging is achievable was demonstrated in test beam for a wide range of sensors: e.g. low gain and deep depleted avalanche diodes \cite{CARTIGLIA2015141, CENTISVIGNALI2020162405}, microchannel plates \cite{BRIANZA2015216}, micropattern gas detectors \cite{Titov:2013hmq} and scintillating crystals coupled to SiPMs \cite{BENAGLIA201630, LUCCHINI20171}. Although excellent time resolution has been measured on single small devices, the path towards the instrumentation of large area detectors present additional challenges.

In the particular case of collider detectors, cost, power consumption and radiation tolerance set additional stringent constraints on current state-of-the-art timing sensors. Among the technologies mentioned above, the one consisting of a scintillating crystal coupled to a SiPM represents a flexible option with intrinsic advantages towards the scalability to large areas and has, for instance, been selected for the instrumentation of the CMS barrel MIP Timing Detector \cite{CMS_MTD_TDR}. A similar approach is also used for the PANDA TOF detector \cite{PANDA_TOF} where a plastic scintillator, in the shape of elongated plates, is used.

Based on similar concepts, the first part of the proposed SCEPCal detector consists of two thin layers (T1 and T2) with the capability to measure single MIPs with a time resolution of about 20 ps.
Each layer is made of inorganic scintillator square fibers close to each other. The cross section of the fibers is about $3\times 3$~mm$^2$ while the length can be tuned in the range from 5 to 10 cm with minimal impact on the light collection efficiency and timing performance. 
We use in the following 10 cm long fibers to cover the equivalent of a module of $10\times10$ $1\times1$~cm$^2$ crystals in the ECAL that is described in Section~\ref{sec:ECAL_part}. 
Each fiber is readout on both sides by a SiPM, and the combination of the time stamps from both photodetectors is then used to extract the time information. The thickness of the fibers in each layer is about 3~mm, corresponding to about 0.4~$X_0$ per layer.

The layers are rotated by 90 degrees with respect to each other in a way to consitute a x-y grid with a granularity of few millimeters, thus providing a powerful tool for PFA algorithms.
The use of two layers improves the system time resolution by a factor of $1/\sqrt{N}$ (with $N=2$) with respect to a single layer, and enhances the particle identification capability of the SCEPCal as described in Section~\ref{sec:scepcal_pid}.

The potential uses of MIP tagging with a time resolution at the level of tens of picoseconds can span from pileup mitigation in high rate colliders (e.g. high luminosity hadron colliders like the HL-LHC or in high background as expected in a muon collider) to the opening of new possibilities for identification of charged hadrons through time-of-flight down to low energies and for searches of long-lived particles \cite{CMS_MTD_TDR}.

\subsection{Segmented crystal ECAL features}\label{sec:ECAL_part}
Homogeneous crystal calorimeters for high energy physics applications can provide the best energy resolution for photons and electrons, which is driven by the extremely low stochastic term achieved with a sampling fraction of EM showers close to unity. The addition of the higher granularity crystal segmentation (both transversally and longitudinally) is a natural direction for future crystal calorimetry as it can open new ways for improving particle reconstruction in the PFA paradigm.

There have been many developments in the field of crystal technology in the last few decades, including the discovery of very bright inorganic scintillators that exploit band-gap engineering to enhance their timing properties, e.g. garnet scintillators (e.g. YAG, LuAG, GAGG) \cite{Nikl_luag, LUCCHINI2016176}.
In parallel, new crystal production techniques tailored to provide higher quality and more granular crystals at a contained cost have also been developed (e.g. MPD and transparent ceramics) \cite{LECOQ2016130,C8CE01781F}.

For the design optimization of a highly granular calorimeter capable of exploiting the PFA potential for event reconstruction, the Moli\`ere radius ($R_M$) and radiation length ($X_0$), which defines respectively the size of the transverse and longitudinal development of EM showers, are key parameters driving the crystal choice.
Conversely, the crystal light yield tends to be a less crucial factor since the energy deposits typical of homogeneous crystal calorimeters for high energy physics applications are in the GeV range, about three order of magnitudes larger than other commercial applications such as Positron Emission Tomography.
The performance of crystals with very low light yield (e.g. PbWO$_4$) for homogeneous calorimetry is only marginally affected by photostatistics fluctuations, and thus other parameters related to the physics of high energy EM showers play a more crucial role.
Another recurring parameter which becomes relevant when large detector volumes need to be instrumented is the cost of the crystal, which is affected mainly by its raw material and the melting temperature required to grow the ingots \cite{LECOQ2016130}.

A comparison of a few scintillators, namely lead tungstate (PbWO$_4$), Bismuth Germanate (BGO), Bismuth Silicon Oxide (BSO) and Cesium Iodide (CsI), is given in Table~\ref{tab:crystal_specs}, to provide an example of the range of parameters that crystal technology can span. More examples can be found in \cite{Renyuan_crystals}.
For the specific goal of developing a segmented crystal ECAL for $e^+e^-$ colliders, where radiation levels are several order of magnitudes smaller than those of hadron colliders, we consider PbWO$_4$ and BSO as two of the best candidates, based on the criteria discussed above.
\begin{table}[!tbp]
\centering
\caption{\label{tab:crystal_specs} Comparison of some of the key crystal properties for HEP applications. From left to right: crystal name, density, interaction length, radiation length, Moli\`ere radius, light yield relative to that of PbWO$_4$, scintillation decay time, photon time density and estimated cost for mass production.}
\smallskip
\begin{tabular}{|l|c|c|c|c|c|c|c|c|}
\hline
Crystal & $\rho$ 		& $\lambda_{I}$ & $X_0$ & $R_M$ &  LY/LY$_0$ & $\tau_{D}$ & Photon density & Est. cost \\
	    & [g/cm$^3$]	& [cm]			&	[cm]	   &	[cm]	  & [a.u.]		&	[ns]	 & [photons/ns] & [\$/cm$^{2}$/$X_0$]\\
\hline
\hline
PbWO$_4$ & 8.3 & 20.9 & 0.89 & 2.00 & 1   & 10   & 0.10 & 7.1\\
BGO      & 7.1 & 22.7 & 1.12 & 2.23 & 70  & 300  & 0.23 & 7.8\\
BSO      & 6.8 & 23.4 & 1.15 & 2.33 & 14  & 100  & 0.14 & 7.8\\
CsI 	 & 4.5 & 39.3 & 1.86 & 3.57 & 550 & 1220 & 0.45 & 8.0\\
\hline
\end{tabular}
\end{table}

%Crystals can also be exploited for sampling calorimetry to address specific challenges (e.g. radiation tolerance). In some cases, the combination of a low density crystal with a dense absorber like tungsten can provide an effective way to reduce the Moli\`ere radius, increase granularity and reduce crystal volume at the cost, however, of an increased complexity of the system (e.g. `Shashlik'-like \cite{Becker_2015} or `Spaghetti'-like calorimeters \cite{Pizzichemi_2020}).
%Some mention of recent technological progress on the crystal technology and novel applications (e.g. LHCb SPACAL?

\subparagraph{Segmentation}
Typical electromagnetic calorimeters proposed for future colliders are based on thin active silicon pads within a tungsten sandwich, Si-W, and feature a large number of longitudinal layers ranging between 20 and 40. This is necessary to achieve a sufficient sampling fraction of the shower and maintain the stochastic term of energy resolution around 20-30\%. It was recently shown \cite{ManqiPrivate}, that a reduction of the number of longitudinal layers from 30 to 4 has only a minor impact on the BMR (Higgs mass resolution with full hadronic final state + standard cleaning) at 240 GeV, once the effect of sampling fraction is factored out. The need of such a granular longitudinal segmentation in Si-W detector is thus mainly motivated by achieving a sufficient sampling fraction to keep the contribution from stochastic shower fluctuations below $30\%/\sqrt{E}$.

\begin{figure}[!tbp]
\centering % \begin{center}/\end{center} takes some additional vertical space
\includegraphics[width=0.32\textwidth]{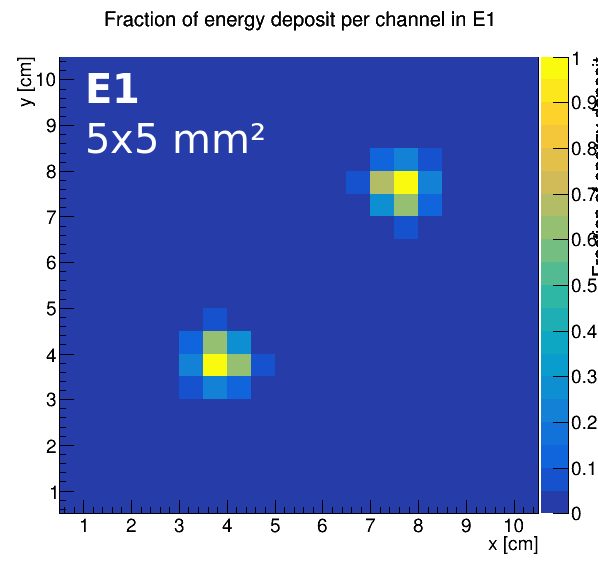}
\includegraphics[width=0.32\textwidth]{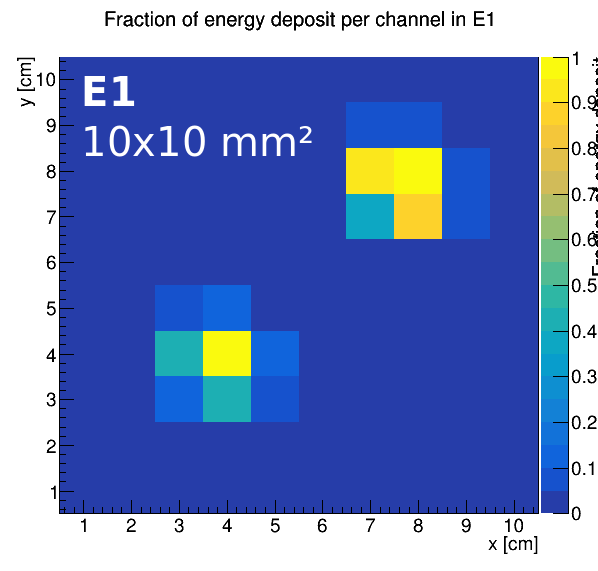}
\includegraphics[width=0.32\textwidth]{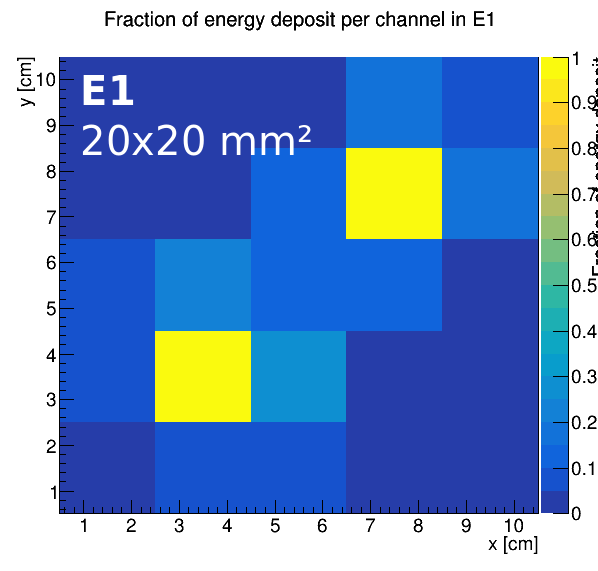}\\
\includegraphics[width=0.32\textwidth]{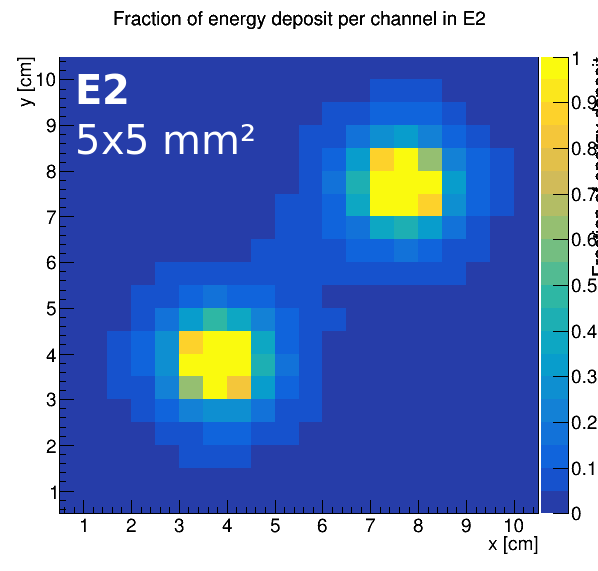}
\includegraphics[width=0.32\textwidth]{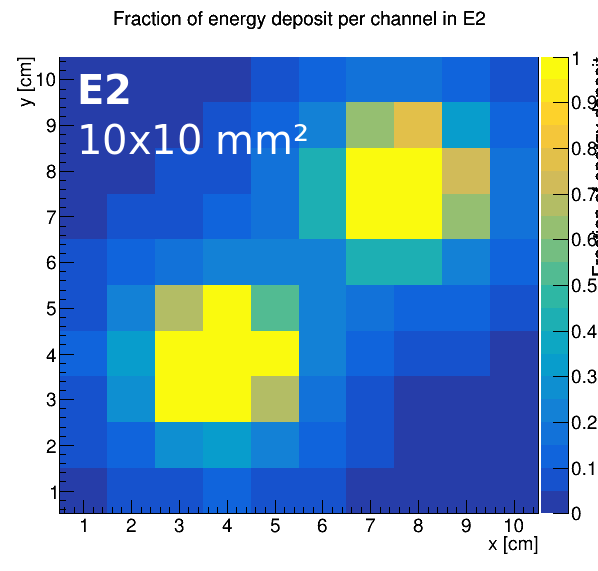}
\includegraphics[width=0.32\textwidth]{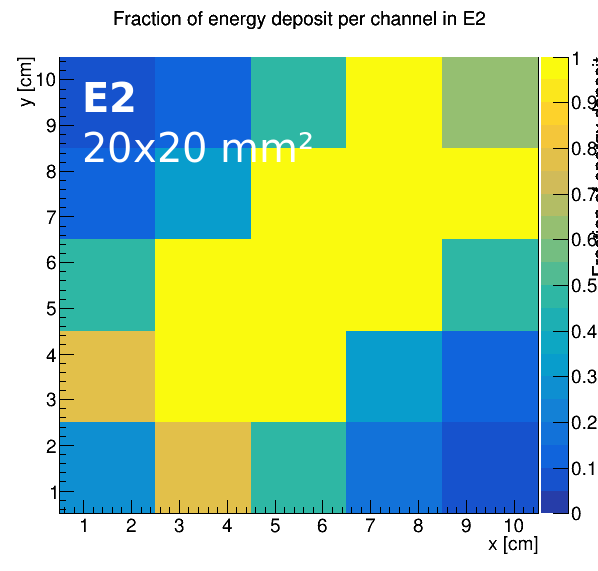}
\caption{\label{fig:ecal_t_granularity} Transverse separation of two photons emitted with an angle of about 3 degrees, in the front and rear layer of the crystal ECAL (with PbWO crystals), for different scenarios of transverse segmentation ($5\times5$~mm$^2$, $10\times10$~mm$^2$, $20\times20$~mm$^2$).}
\end{figure}

The SCEPCal features a total of 4 longitudinal layers. The two front timing layers provide mostly particle identification capabilities based on high sensitivity to MIPs. The two ECAL layers measure with precision the electromagnetic showers by avoiding any dead material between the two segments where the shower maximum occurs. The front segment enables better shower separation, crucial for PFA, since the effective shower radius in the first segment is only about half the Moli\`ere radius as can be observed from Figure~\ref{fig:ecal_t_granularity}.

The transverse granularity of a SCEPCal readout element is set to less than half of the crystal Moli\`ere radius, thus about 1~cm for PbWO$_4$. This provides a containment of about 65\% of the shower within one cell (front+rear segment) for particles impacting in the center of the crystal. Increasing further the transverse segmentation would have marginal gain for PFA algorithms since the capability of shower separation would remain limited by the intrinsic shower radius \cite{THOMSON200925}.

For comparison with a (Si-W) ECAL, it should also be noted that while very dense absorber materials feature small $R_{M}$ (e.g. $\leq 1$~cm for Pb and W), when the longitudinal layers, made of silicon with their associated readout electronics and air gaps, are included, the effective shower radius increases substantially.
For instance, the Moli\`ere radius of the Si-W ECAL foreseen for the CMS Phase 2 Upgrade of the endcap calorimeter \cite{HGCAL_TDR} is about 2~cm, practically equivalent to that of PbWO$_4$ crystals.

\subparagraph{Energy resolution}

\begin{figure}[!tbp]
\centering 
\includegraphics[width=0.495\textwidth]{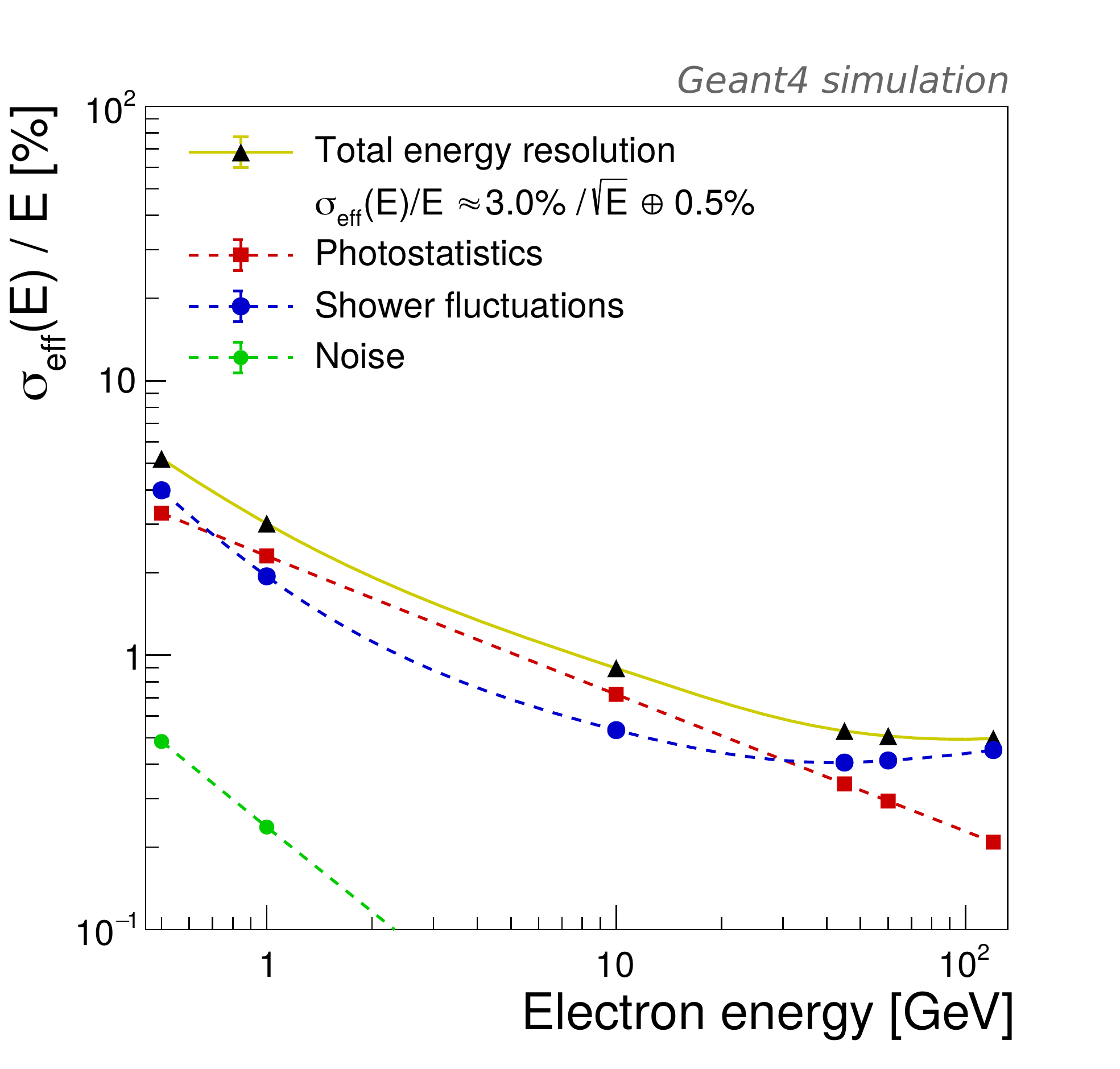}
\includegraphics[width=0.495\textwidth]{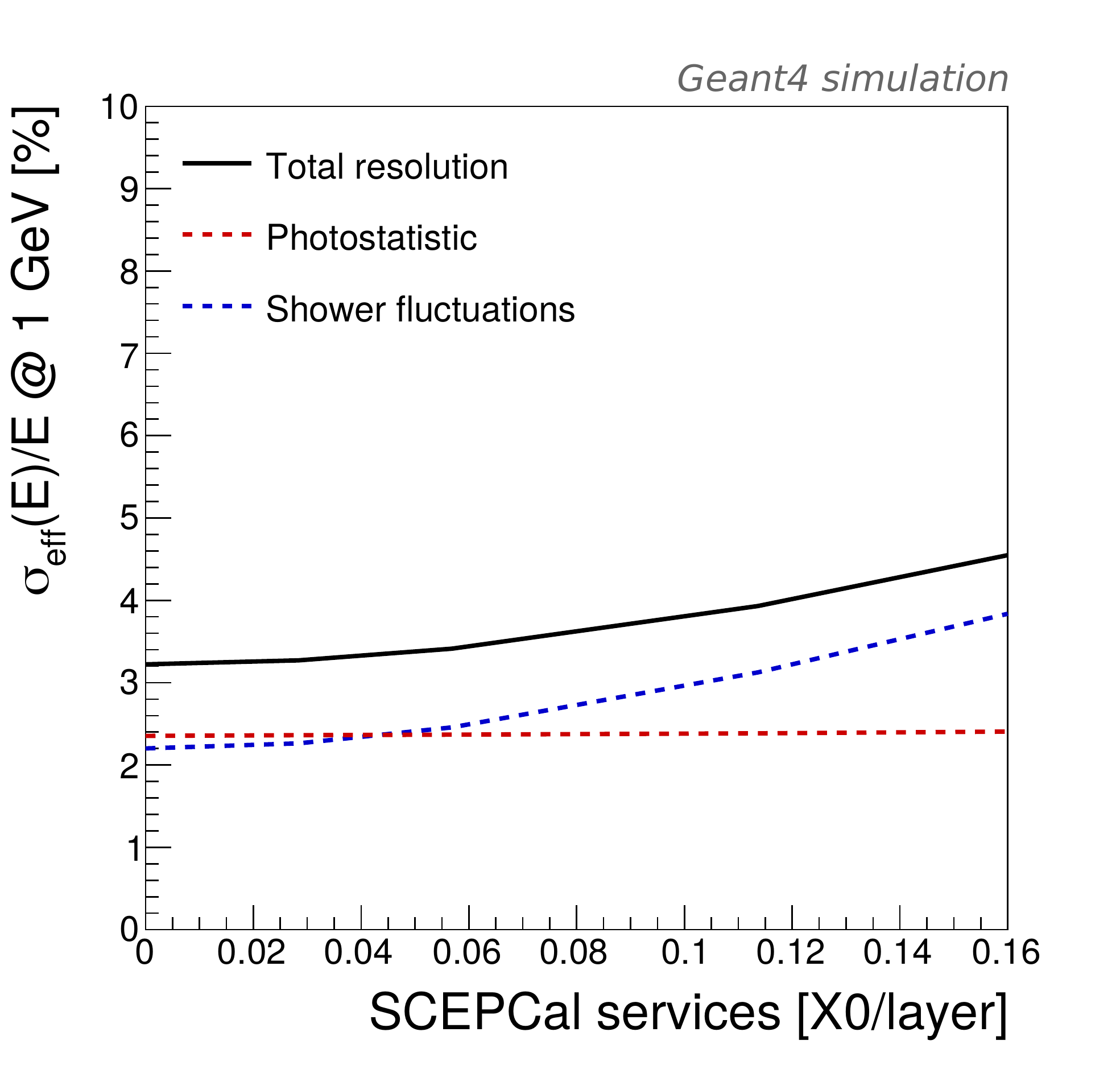}
\caption{\label{fig:ecal_eres} Left: energy resolution to electrons as a function of the electron energy with the assumption of a 2.4\% contribution from photostatistics, a tracker material budget of 0.1~$X_0$, and about 0.1~$X_0$ of dead material between layers. Right: impact of dead material between layers on energy resolution. }
\end{figure}

We propose a design of the SCEPCal calorimeter based on PbWO$_4$ crystals that can achieve an energy resolution to EM particles of:
\begin{equation}
\frac{\sigma_{E}}{E} = \frac{3\%}{\sqrt{E}} \oplus  \frac{0.2\%}{E} \oplus 0.5\%
\end{equation} 
The left plot in Figure~\ref{fig:ecal_eres} shows the overall energy resolution of the SCEPCal module for electrons of energy in the 1-120 GeV range obtained with the Geant4 simulation.
At energies above 10 GeV, the contribution to the energy resolution from shower fluctuations is below 0.5\% with a negligible increase at energies above 100 GeV due to longitudinal shower leakage from the rear side of the crystals. The dominant term, at low energies, is the stochastic one, and is ultimately limited by two effects:
\begin{itemize}
\itemsep0em 
\item Poisson fluctuations in the number of photoelectrons detected (photostatistics);
\item fluctuations in the fraction of electron energy deposited in the upstream material (tracker, services).
\end{itemize}

We studied both the impact of tracker material and the impact of the dead material due to services and readout for the two timing layers (T1,T2) and the front ECAL layer (E1).
For a typical tracker design proposed for future $e^+e^-$ colliders, the material budget is equivalent to about 0.1~$X_{0}$ (\cite{CEPC_CDR_Vol2,Tracker_forIDEA}) and has an impact of less than 0.01 in quadrature on the stochastic term.
The material budget for services and readout is typically dominated by the Aluminum plates required for the cooling system, which can be about 3 mm thick, i.e. 0.03~$X_0$ for each plate.
For a fixed tracker material budget of 0.1~$X_0$, by varying the material budget for each of the three readout layers in front of the ECAL segments (T1,T2,E1), we concluded that if the radiation length per layer is kept below 0.05~$X_0$ the stochastic term of the energy resolution is below 2.4\% as shown in Figure~\ref{fig:ecal_eres}.

\begin{figure}[!tbp]
\centering % \begin{center}/\end{center} takes some additional vertical space
\includegraphics[width=0.495\textwidth]{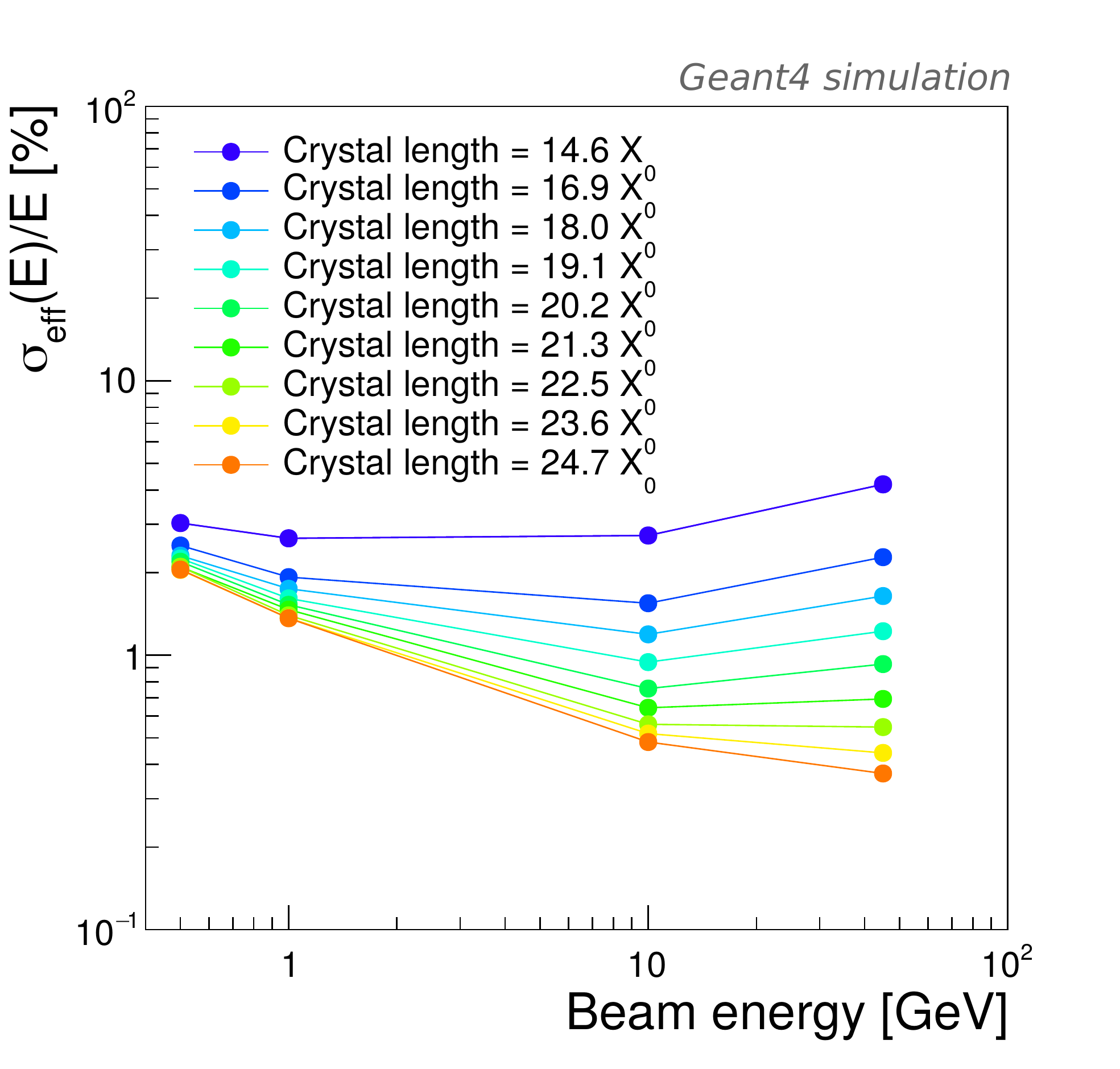}
\includegraphics[width=0.495\textwidth]{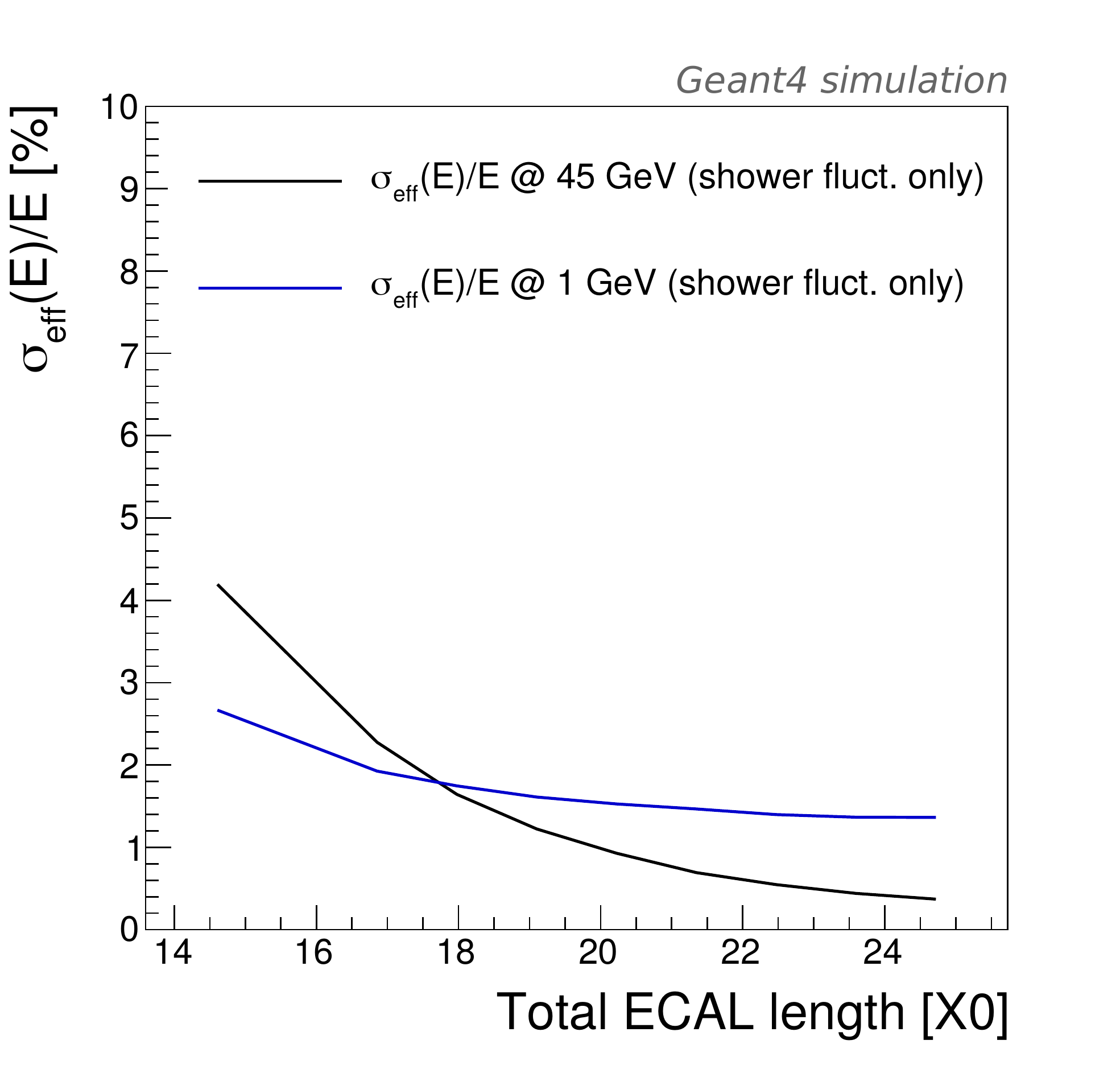}
\caption{\label{fig:ecal_eres_vs_length} Left: SCEPCal energy resolution to electrons in the energy range from 0.5 to 45 GeV for different total radiation lengths of the crystal (E1+E2). Right: evolution of the energy resolution of the SCEPCal at 1 (approximation of the stochastic term) and 45 GeV as a function of the total crystal length in units of radiation length $X_0$.}
\end{figure}

The other source of energy fluctuations (contributing to both the stochastic and the constant term) is due to partial longitudinal containment of the shower. The impact of the SCEPCal length on the energy resolution has been studied, and the results, reported in Figure~\ref{fig:ecal_eres_vs_length}, can be used for a performance-cost optimization of the detector. A total crystal length (front+rear segments) of more than 21~$X_0$ is required to keep the constant term below 1\% and the contribution to the stochastic term below 2.0\%. The performance drop for a shorter crystal calorimeter, down to 19~$X_0$, mainly impacts energetic photons and electrons with up to a 2\% constant term.

The photostatistics term is driven by the Poisson fluctuations due to a finite number of photoelectrons detected:
\begin{equation}
\text{N}_{\rm phe} = \text{LY} \cdot \text{LCE} \cdot \text{PDE}
\end{equation}
in which LY is the number of scintillation photons produced by the crystal per MeV of deposited energy, LCE is the light collection efficiency, defined as the fraction of light that reaches the SiPM and PDE is the photon detection efficiency of the SiPM.
For PbWO$_4$, a light yield of about 200 photons per MeV is assumed (supported by extensive measurements performed on CMS ECAL crystals \cite{Adams_2016}) and a PDE of 20\% weighed over the PbWO$_4$ emission spectrum (peaking at 420~nm), which can be achieved with 10~\SI{}{\um} cell size SiPMs.
The light collection efficiency is affected by the quality of the crystal surfaces (polishing), geometry of the crystal (length, width, tapering), wrapping material and optical coupling to the SiPM (refractive index of the glue and of the SiPM entrance window).

The light collection efficiency increases for shorter crystals due to less self-absorption of the scintillation light, and increases linearly with the fraction of the crystal end face covered by the photodetector. For the SCEPCal crystal segments, which are shorter and narrower than typical dimensions used in CMS PbWO$_4$ or L3 BGO calorimeters, the LCE varies between 2 and 20\% for SiPMs of active area between 10 and 100~mm$^2$.
Supported by the \textsc{Geant4} ray-tracing simulation results discussed in Section~\ref{sec:scepcal_dro} (see Figure~\ref{fig:scepcal_lce_optimization}), we assume an LCE of 5\% for a $5\times5$~mm$^2$ SiPM, which thus yields about 2000 photoelectrons per GeV and contributes to the stochastic term of the energy resolution with about 2\% in quadrature.
%\textcolor{red}{check consistency with Section~\ref{sec:scepcal_dro}}.

The constant term of such a detector due to shower sampling fluctuations is of the order of 0.5\%, small compared to other contributions such as the channel inter-calibration procedure, which typically bring this term up to about 1\%.

In the proposed design the noise term can be neglected since SiPMs (proposed for the readout) have a large gain of the order of 10$^5$ which makes the electronic noise fluctuations negligible with respect to the signal. 
Similarly, the dark count rate of the SiPM is typically below 1~MHz/mm$^2$, yielding less than 200~MHz even considering the extreme case where both the front and rear SiPMs have an active area that covers the entire crystal end face ($200$~mm$^2$). For a signal integration gate of 100 ns (10 times the PbWO$_4$ decay time), this corresponds to less than 5 photoelectrons noise and contributes to the energy resolution with about $0.2\%/E$, which is entirely negligible.

SiPMs with active area of 25~mm$^2$ and cell size of 10~\SI{}{\um} consist of more than $N_{tot}=2.5\times10^5$ cells and can cover the entire dynamic range from 1 to 100 GeV. A further extension of the dynamic range by about a factor of two could be achieved by exploiting SiPMs with cell size of 7.5~\SI{}{\um}. Non-linearities in the SiPM response due to the limited number of cells is a well-known effect and can be corrected for \cite{ACERBI201916}. The real number of photons impacting on the SiPM, $N_{ph}$, can be estimated based on the measured signal (i.e. the number of SiPM cells, $N_{fired}$, that actually fired) according to the equation: $N_{fired} = N_{tot}\cdot(1-\exp(-\frac{N_{ph}\cdot PDE}{N_{tot}}))$.

\subparagraph{Particle identification}\label{sec:scepcal_pid}

\begin{figure}[!tbp]
\centering % \begin{center}/\end{center} takes some additional vertical space
\includegraphics[width=0.495\textwidth]{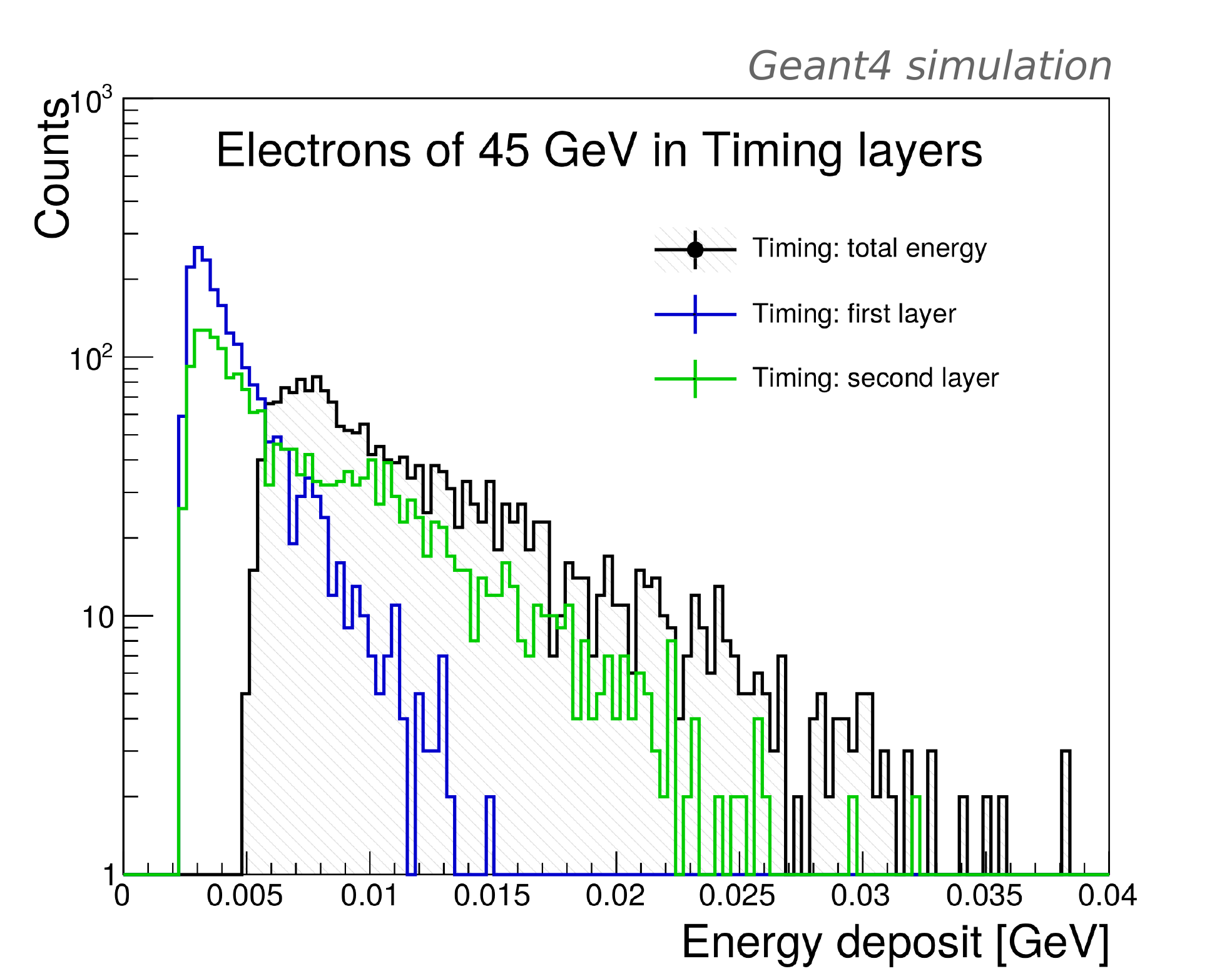}
\includegraphics[width=0.495\textwidth]{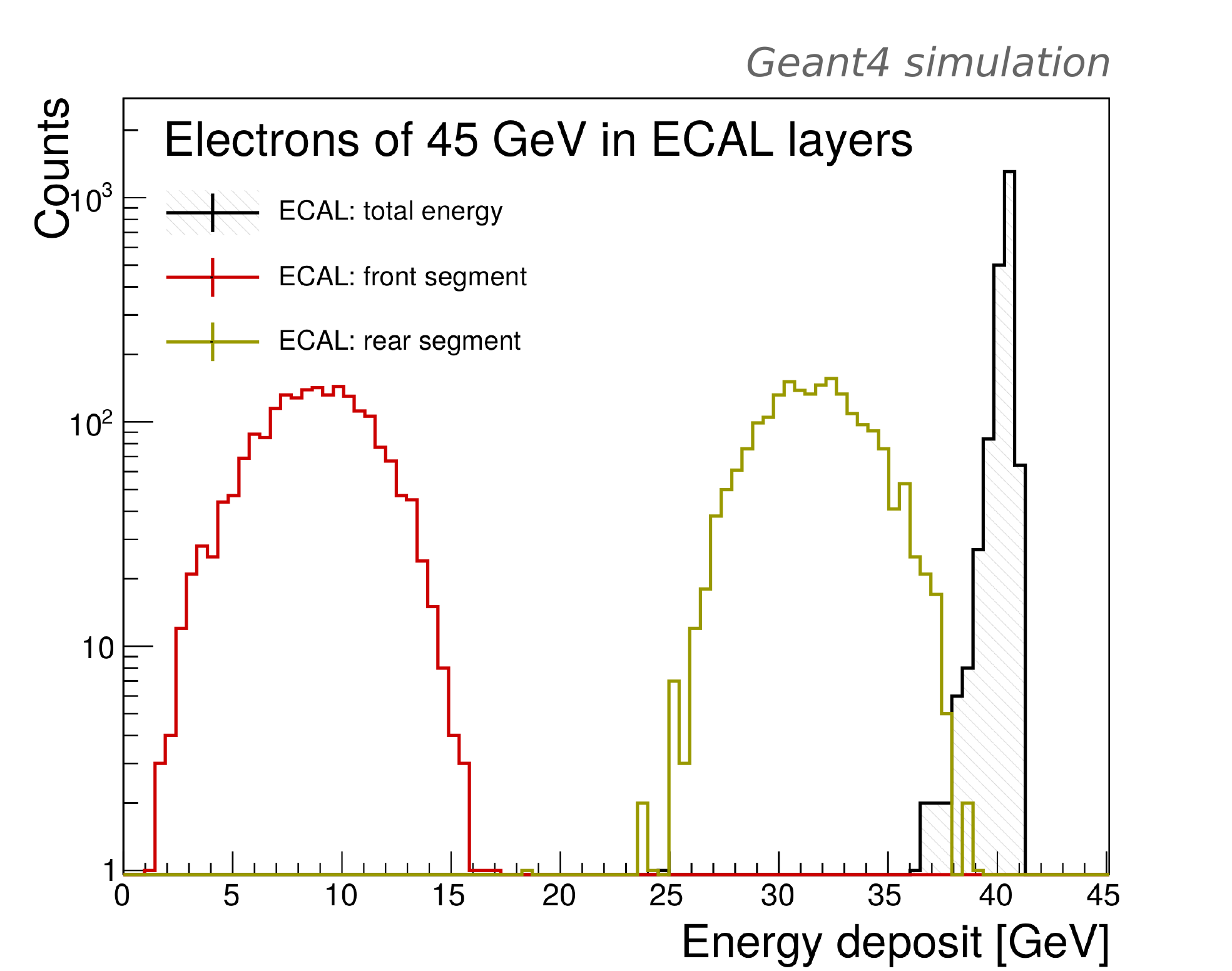}\\
\includegraphics[width=0.495\textwidth]{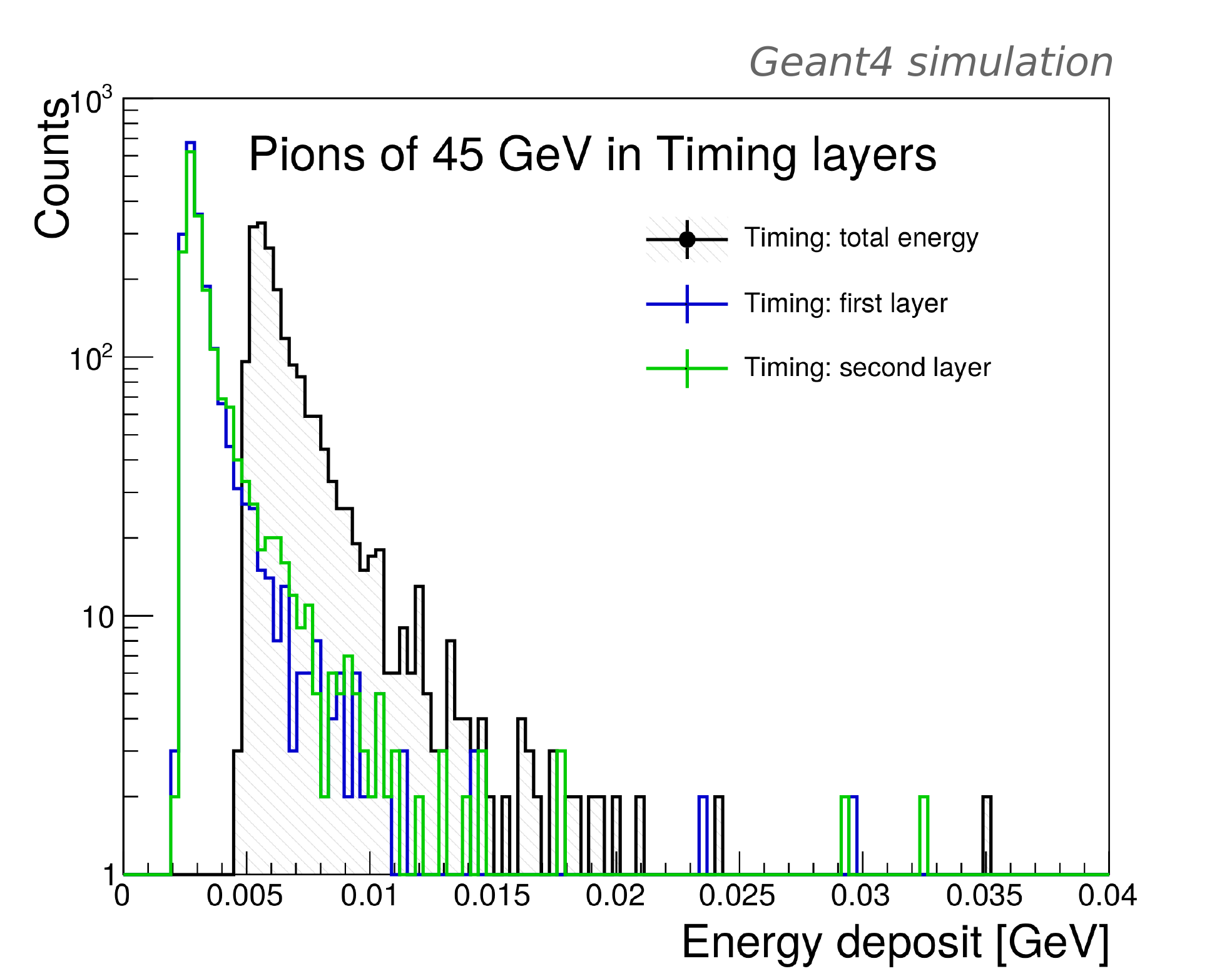}
\includegraphics[width=0.495\textwidth]{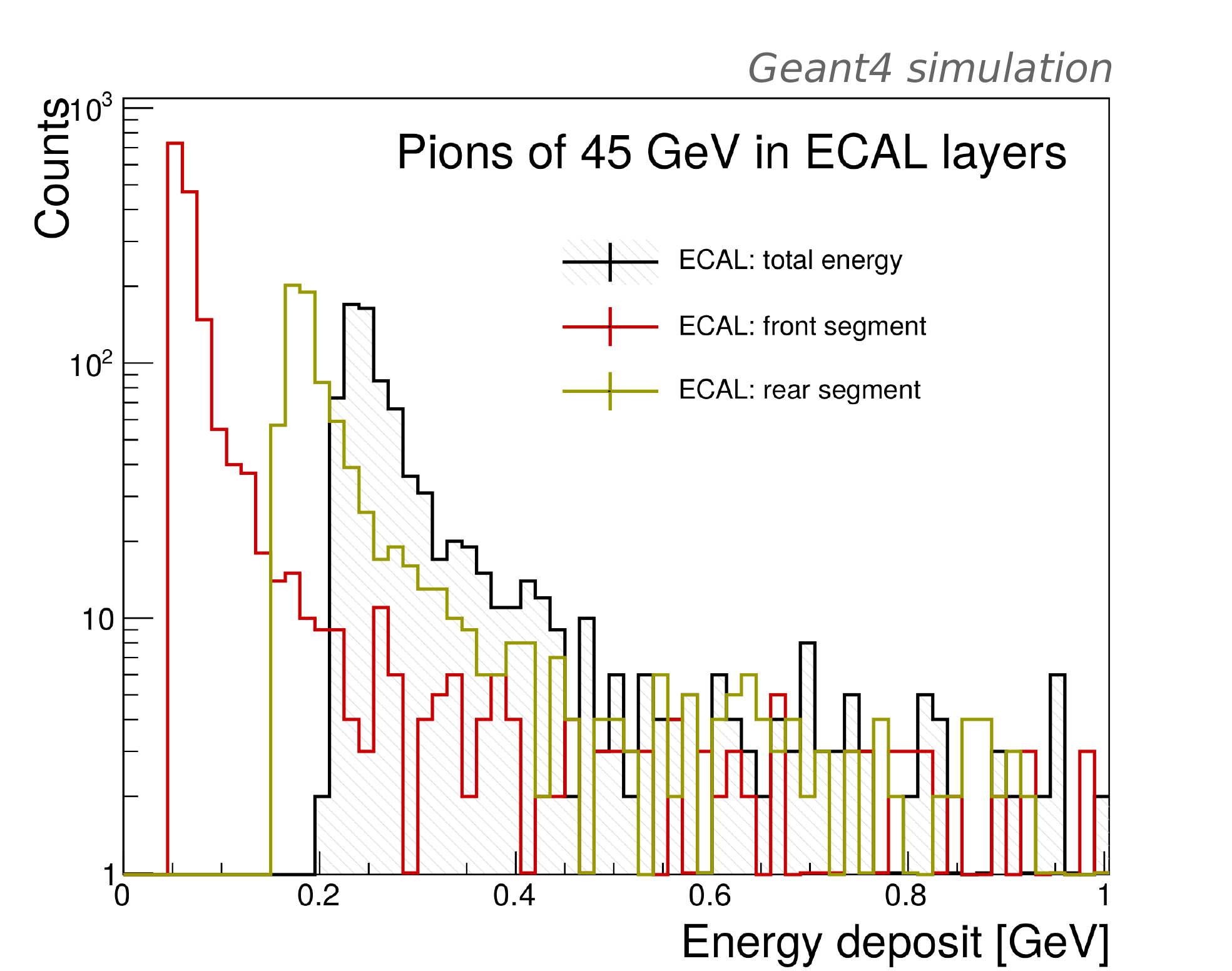}
\caption{\label{fig:ecal_energy_deposits} Energy deposits in different crystal layers (timing for the left plots, ECAL for the right plots) for electrons (top) and pion (bottom) of 45 GeV momentum.}
\end{figure}

The identification of heavy hadrons based on their time-of-flight, is enabled by the excellent time resolution of 20~ps from the timing layers.
This is particularly relevant for particle momenta around 2 GeV for which the energy loss per track length by charged particles (dE/dx) approaches its minimum and thus discrimination between pions and kaons is typically not possible.
A clean discrimination between electrons and pions is also possible based on shower profile. The energy deposits in the 4 SCEPCal longitudinal layers for electrons and pions of 45 GeV energy are shown in Figure~\ref{fig:ecal_energy_deposits}. The different patterns of energy depositions represents a powerful tool for particle discrimination. 

\begin{figure}[!tbp]
\centering % \begin{center}/\end{center} takes some additional vertical space
\includegraphics[width=0.495\textwidth]{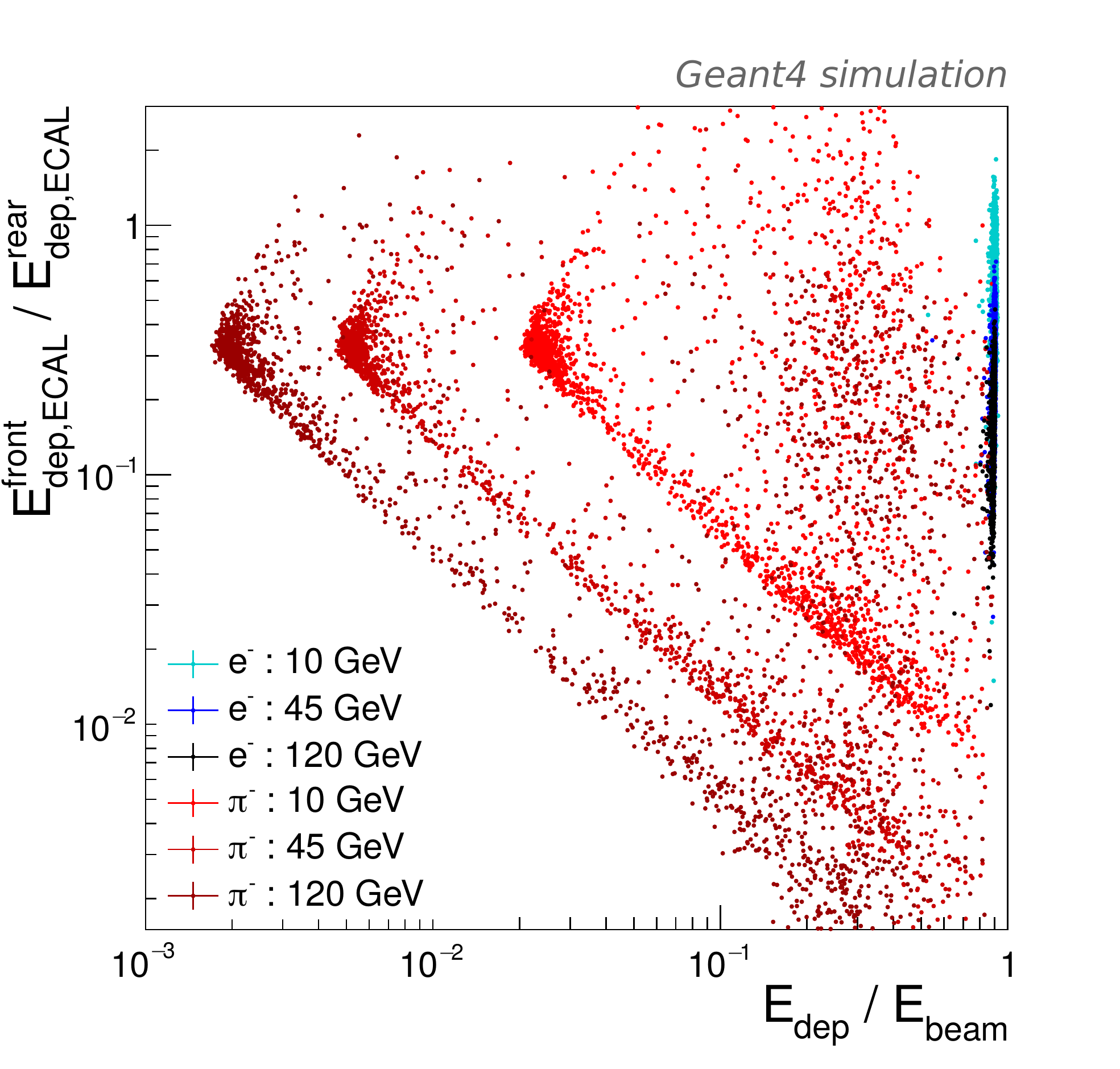}
\includegraphics[width=0.495\textwidth]{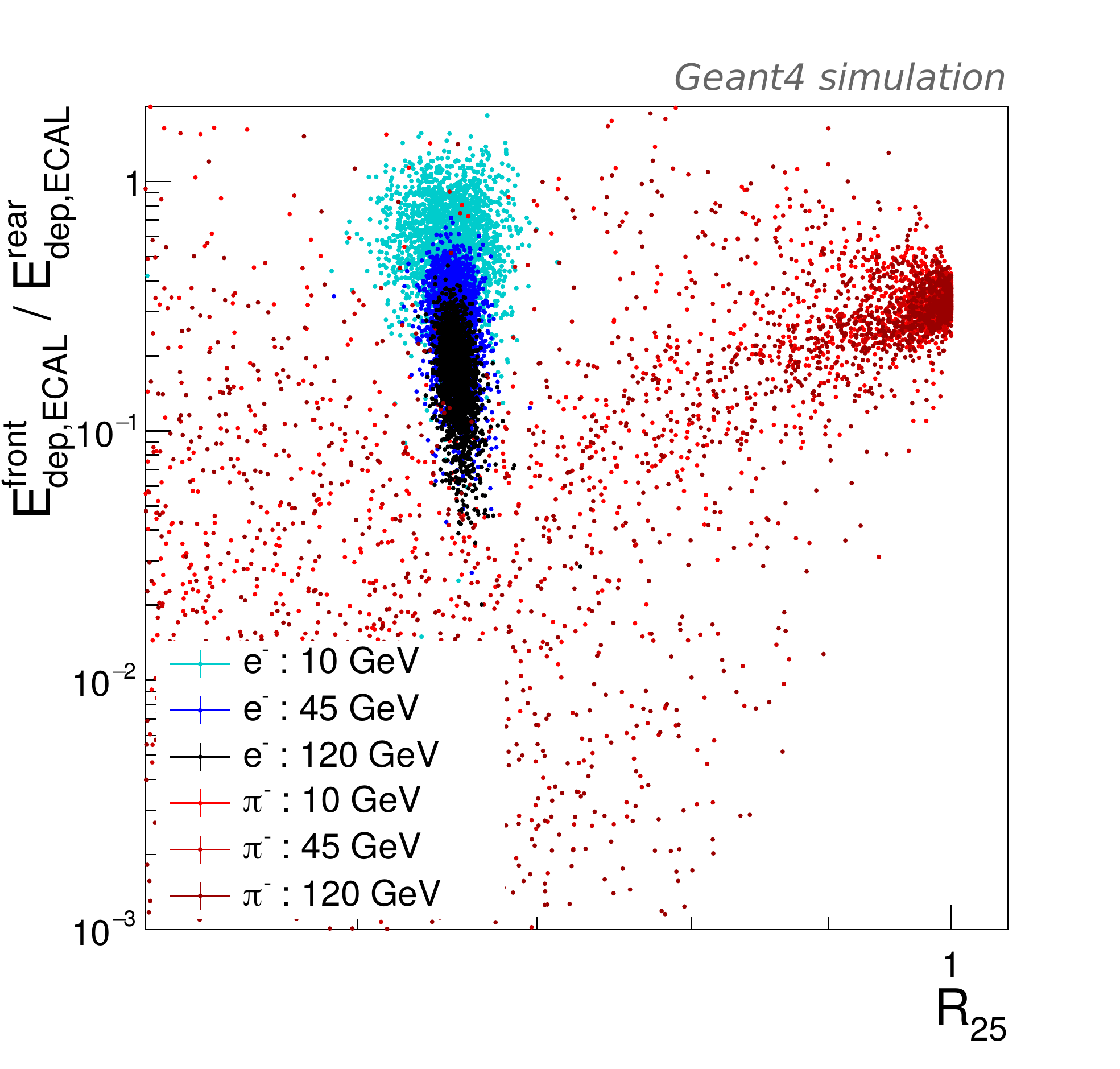}
%\caption{\label{fig:ecal_epi_separation} Energy deposits in different ECAL layers for 45 GeV pions (left) and ratio of energy deposit in the two ECAL layers as a function of the total energy deposit in the ECAL for both 45 GeV electrons and pions, combined on the same plot. It is clear the potential of discrimination between electrons and pions that convert in the ECAL based on the longitudinal segmentation.}
\caption{\label{fig:ecal_epi_separation} Left: ratio of the energy deposit in the front and rear ECAL layer as a function of the ECAL energy deposit divided by the beam energy. Red (blue) scatter plots are for pions (electrons) of 10, 45, 120 GeV energy. Right: ratio of the energy deposit in the front and rear ECAL layer as a function of R$_{25}$ defined as the ratio of energy in the central crystal (particle impact point) and the total energy deposited in a $5\times5$ crystal matrix around the central crystal.}
\end{figure}

Regardless of their energy, pions will mainly interact as MIPs in the thin timing layers (T1,T2) while electrons are more likely to shower and be detected as multiple MIPs, with on-average a larger signal in T2.
Similarly, in the ECAL segments, the pions will mostly behave as MIPs unless a shower is initiated. However, in both cases the ratio of the energy deposit in the rear (E2) with respect to the front (E1) segment will show a clear footprint that can be clearly distinguished from that of an electron regardless of their energy.
This is shown in Figure~\ref{fig:ecal_epi_separation} where events from both electrons and pions are displayed. A simple two-dimensional discrimination based on estimators of transverse and longitudinal shower profiles can yield an electron efficiency of 99\% at 99\% pion rejection.
More sophisticated combination of the information and the use of a convolutional neural network could be used to further improve the discrimination power. 

The addition of dual-readout capabilities to the SCEPCal would further improve particle identification, as discussed later in Section~\ref{sec:scepcal_dro}.
Furthermore, differences in the measured signal pulse shape could also be exploited in some cases to provide further insights on the secondary particle composition of calorimeter clusters (e.g. $\gamma$ and $K^0_L$), as recently investigated for the Belle II CsI crystal-based calorimeter \cite{BelleII_PID}.

%\newpage

\subsection{Combining a crystal ECAL with a dual-Readout HCAL}
To achieve the desired resolution for jets, as discussed in Section~\ref{sec:pfa_drivers}, a calorimeter with resolution for neutral hadrons better than $45\%/\sqrt{E}$ is required.
To meet such a requirement, it is necessary to reduce the contribution to the energy resolution from fluctuations of the electromagnetic component of the hadron shower (driven by the number of $\pi^{0}$s) as discussed in \cite{DRO_calorimetry_Ferrari2109}.
%Comment on Compensating calorimeters are a possibility to d
A calorimeter capable of measuring the EM fraction of the shower on an event-by-event basis should thus be used. 

Several ideas have been put forward, in this context, exploiting differences in the calorimeter response to the hadron and EM components. For instance, a correlation between the EM fraction of the shower and its time development has been proposed in \cite{Benaglia_TICAL_4D} (dual-gate approach).
A more common and broadly studied approach is the so-called \emph{dual-readout} (DRO) in which the simultaneous measurement of the Cherenkov and scintillation light signals is performed, and their ratio is used as an indicator of EM fraction \cite{DRO_calorimetry_Ferrari2109}. 
This exploits the fact that the EM component of the hadron shower is dominated by light charged particles which are more relativistic and produce a larger Cherenkov signal.

The possibility to develop a crystal-based hadron calorimeter with dual-readout capability has a long history, and several designs have been proposed \cite{CrystalFibersDRO_Mavromanolakis, LuAG_FibersTB_Lucchini_2013, LuAG_FibersTB_Benaglia_2016}. The performance of a fully homogeneous hadronic calorimeter made of crystals where dual-readout capabilities are combined with PFA was also simulated, indicating that an outstanding energy resolution to hadrons could be achieved \cite{Magill_2012}. However, the actual implementation of such an option is strongly disfavored due to the high cost related to the large crystal volume required.

A hadron calorimeter made of several longitudinal layers that alternate an absorber material with two or more sensitive materials (scintillators, Cherenkov radiators, neutron-sensitive elements) optimized for detection of different components of the hadronic shower has also been proposed recently. % \cite{YONEL_DWINN}.
Such a design could present several advantages but, to achieve sufficient energy resolution, would require a high sampling fraction, i.e. a large number of active layers and of readout channels. 
%We explored this option by simulating sandwich structure made of a Cu layer, and two active layers of 5 mm thickness each consisting of a plastic scintillator and quartz tiles.
%The thickness of the Cu layer was then varied from XXX to YYY to change the sampling fraction of the calorimeter. As shown in Figure~\ref{}, to achieve a resolution of ZZZ a Cu thickness smaller than AAA is required which would require about NNN channels/m$^2$ assuming a transverse granularity of VVV~mm.

Another approach, which has been extensively investigated by the DREAM collaboration in the last decades, consists of a sampling calorimeter where two different active media in the shape of fibers are embedded in an absorber structure \cite{DRO_calorimetry_Ferrari2109}.
Some challenges in the practical implementation of such a design exist nonetheless. In this context, a recent proposal has been made to build the calorimeter out of sensitive fibers encased in brass tubes, which are then assembled and glued together. 
%\cite{}
%[I have contacted the speaker to have a reference about his talk as I could not find any peer reviewed reference for this: \url{https://agenda.infn.it/event/19360/contributions/95824/attachments/64280/77719/Capillary_tube_RBI_Proposal.pdf}]
%
Based on this latter approach, the calorimeter proposed for the IDEA detector at future lepton colliders is designed to work for both electromagnetic and hadron showers detection with a performance of about $15\%/\sqrt{E}$ and $25\%/\sqrt{E}$ respectively.

\begin{figure}[!tbp]
\centering % \begin{center}/\end{center} takes some additional vertical space
\includegraphics[width=0.495\textwidth]{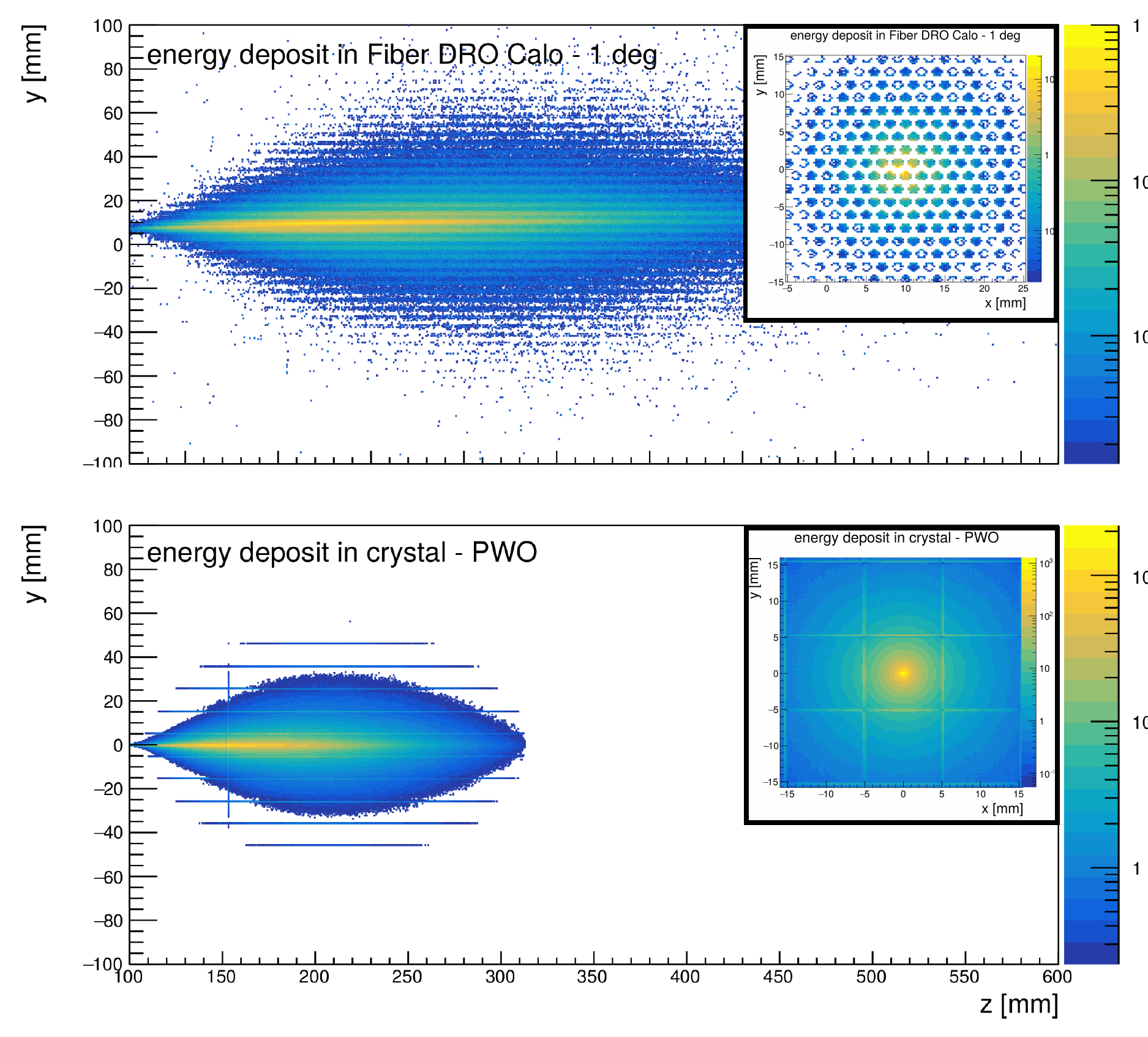}
\includegraphics[width=0.495\textwidth]{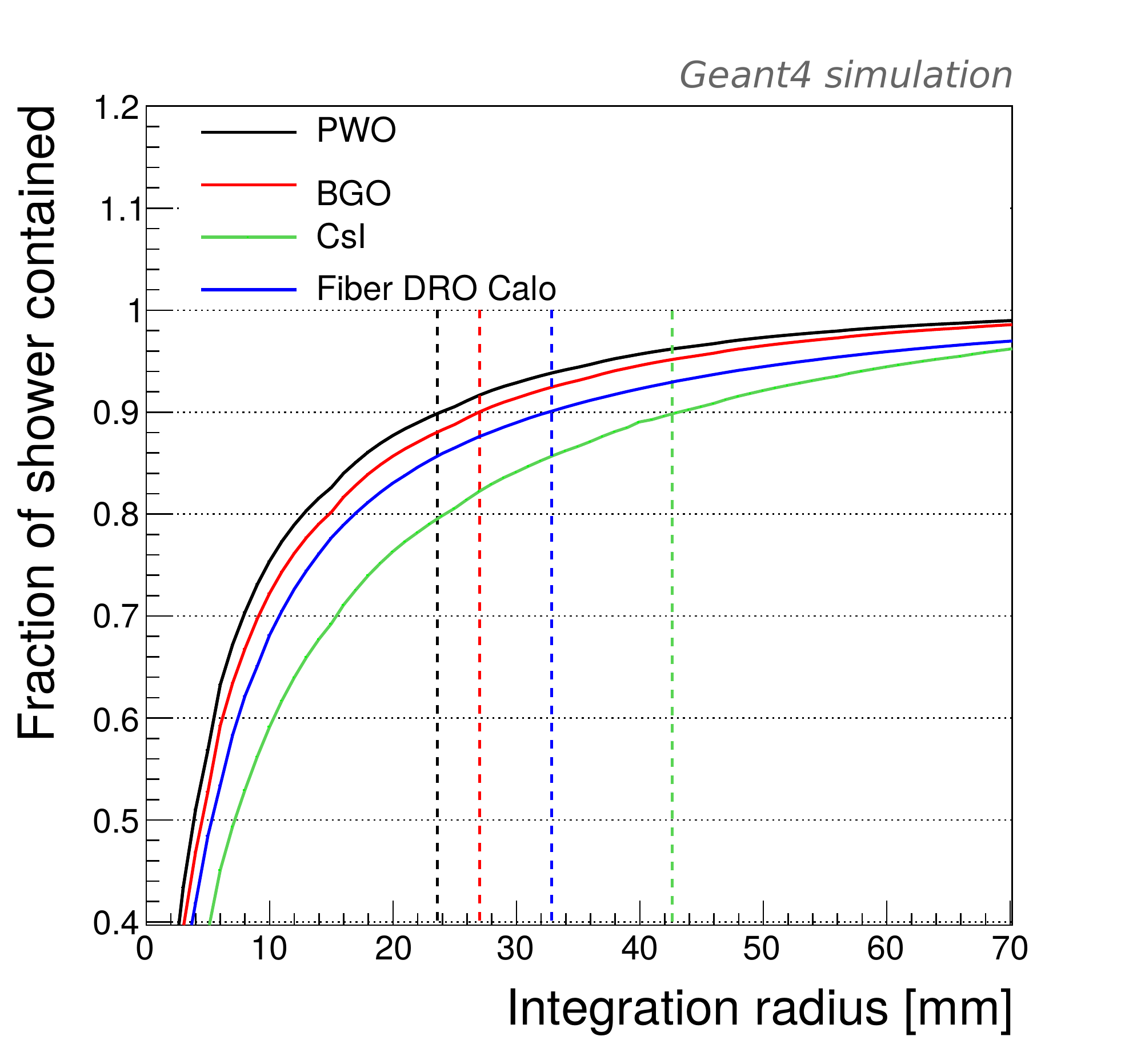}
\caption{\label{fig:moliere_radii} Left: longitudinal shower profile for 60 GeV electrons in a fiber DRO HCAL compared to that of a PbWO$_4$-based SCEPCal (the inset in the plot shows the transverse shower profile). Right: comparison of Moli\`ere radii between homogeneous crystal calorimeters (PbWO$_4$, BGO, CsI) and the fiber DRO HCAL.}
\end{figure}

While the performance for hadron showers detection is outstanding, the performance for EM showers is limited.
%
%For particles impacting at normal incidence in a detector where fibers point towards the interaction point, a spatial non-uniformity in the response leads to a substantial constant term. In addition, channeling effects appear due to particles traveling through the air gaps between brass tubes or through the active fiber which has a much smaller density than brass and leads to showers that develop later in the calorimeter. 
%These effects can be mitigated by a non-fully-pointing geometry where fibers are oriented at a tilt angle with respect to the interaction point, which, however, partially increases the effective Moli\`ere radius since the shower spreads over a larger number of fibers.
Figure~\ref{fig:moliere_radii} shows a comparison of electron shower profiles and Moli\`ere radii between homogeneous crystals and the fiber-based calorimeter described above and illustrates how the former option could yield a smaller Moli\`ere radius.
In addition, to maintain a resolution for EM particles better than $15\%/\sqrt{E}$, the design of the calorimeter in terms of sampling fraction becomes very demanding and requires a large number of sensitive fibers with sizable impact on the detector cost.
%
%Provided that a sufficient hadron resolution is maintained and that thus the dual-readout method is still applicable, the introduction of a dedicated electromagnetic calorimeter based on homogeneous crystals could thus bring a global performance advantage to the calorimeter system, easing the design requirements of a pure fiber-based calorimeter.
However, the introduction of a crystal electromagnetic calorimeter with dual readout capabilities can solve these problems, easing the design requirements of a pure fiber-based caloroimeter, while maintaining excellent hadron resolution (see Section~\ref{sec:hcal_opt}).

Such possibility was first explored by the DREAM collaboration through a set of beam tests where a BGO crystal ECAL module was placed in front of a DRO hadron calorimeter tower \cite{AKCHURIN2009488, Fiber_and_crystals_DRO}. The applicability of dual-readout for several crystals was verified (BGO, BSO, PbWO$_4$ with and without Molybdenum codoping). It was shown that the DRO method can still be applied for those events in which the hadron shower starts in the ECAL segment. However, the overall results obtained were limited by a poor initial performance of the ECAL calorimeter due to a suboptimal configuration of the photodetectors. In particular, the need to detect both Cherenkov and scintillation light using the same photodetector required to quench the scintillation signal to a level that allowed the Cherenkov contribution to be distinguished from the measured waveforms but lead to a very poor photostatistics dominating the stochastic term of energy resolution.

In the following, we investigate further the potential of combining the dual-readout information from a crystal-based ECAL with that from the HCAL and demonstrate that the two systems can be combined to provide an efficient correction for fluctuations of the EM shower component, f$_{\rm EM}$.
In particular, we combine the SCEPCal calorimeter described in Section~\ref{sec:ECAL_part} with a DRO HCAL made of fibers inserted into brass tubes as described in Figure~\ref{fig:overall_layout}, each with a substantially better performance than the calorimeters tested in \cite{AKCHURIN2009488}.

\subparagraph{Dual-readout method in a hybrid calorimeter}

The response of such a calorimeter to single electrons and neutral kaons ($K^{0}_{L}$) in the energy range (5-300 GeV) was studied using a standalone \textsc{Geant4} simulation. 
The dual-readout method relies on the simultaneous measurement of a scintillation (S) and Cherenkov (C) signal to correct for event-by-event fluctuations in the shower electrogmagnetic fraction, f$_{\rm EM}$. The energy, $E$, of the incoming particle is obtained by solving the following system of equations:
\begin{eqnarray}\label{eq:dualreadout}
\begin{cases}
S = E\left[{\rm f_{EM}} + \frac{1}{(e/h)|_S} (1-{\rm f_{EM}})\right]\\
C = E\left[{\rm f_{EM}} + \frac{1}{(e/h)|_C} (1-{\rm f_{EM}})\right]\\
\end{cases}
\end{eqnarray}
in which $(e/h)|_C$ ($(e/h)|_S$) is the ratio between the Cherenkov (scintillation) calorimeter response to the EM and non-EM shower component.
According to Equations~\ref{eq:dualreadout}, the electromagnetic fraction of the shower has a one-to-one correspondance with the measured C/S ratio which can thus be directly used to perform a correction on the measured signal.
The following steps were thus performed to reconstruct particle energies in the simulated hybrid calorimeter by exploiting dual-readout in both the ECAL and HCAL sections:
\begin{enumerate}
\itemsep0em 
\item Evaluate the C and S response of the hadronic section to electrons and calculate the corresponding sampling fractions, defined as the energy deposited in the active volume with respect to the particle input energy and calibrate them such that for a pure electromagnetic shower $\langle C\rangle = \langle S\rangle = E$;
\item Calculate the C/S HCAL correction using hadrons not interacting in the SCEPCal section;
\item Apply the HCAL DRO correction on the energy deposit in the HCAL;
\item Calculate the C/S ECAL correction using hadrons interacting in the SCEPCal section;
\item Apply the ECAL DRO correction on the energy deposit in the ECAL;
\item Sum the DRO-corrected and calibrated responses of the ECAL and the HCAL segments.
\end{enumerate}

%Show the energy resolution of the HCAL to electrons?
The sampling fraction of the SCEPCal, $\xi_{\rm ECAL}$, is close to 100\% by construction, since all the energy is deposited in the crystal except for shower leakages and the very small energy deposited in the services or reflective material between crystals. The sampling fraction for the HCAL scintillating (S) fibers (simulated as plastic scintillators of the BC type from Saint-Gobain), $\xi_{\rm HCAL}$, is about 3\%. Stochastic fluctuations due to photostatistics are neglected at first, and studied in detail in Section~\ref{sec:scepcal_dro}.

\begin{figure}[!tbp]
\centering % \begin{center}/\end{center} takes some additional vertical space
\includegraphics[width=0.495\textwidth]{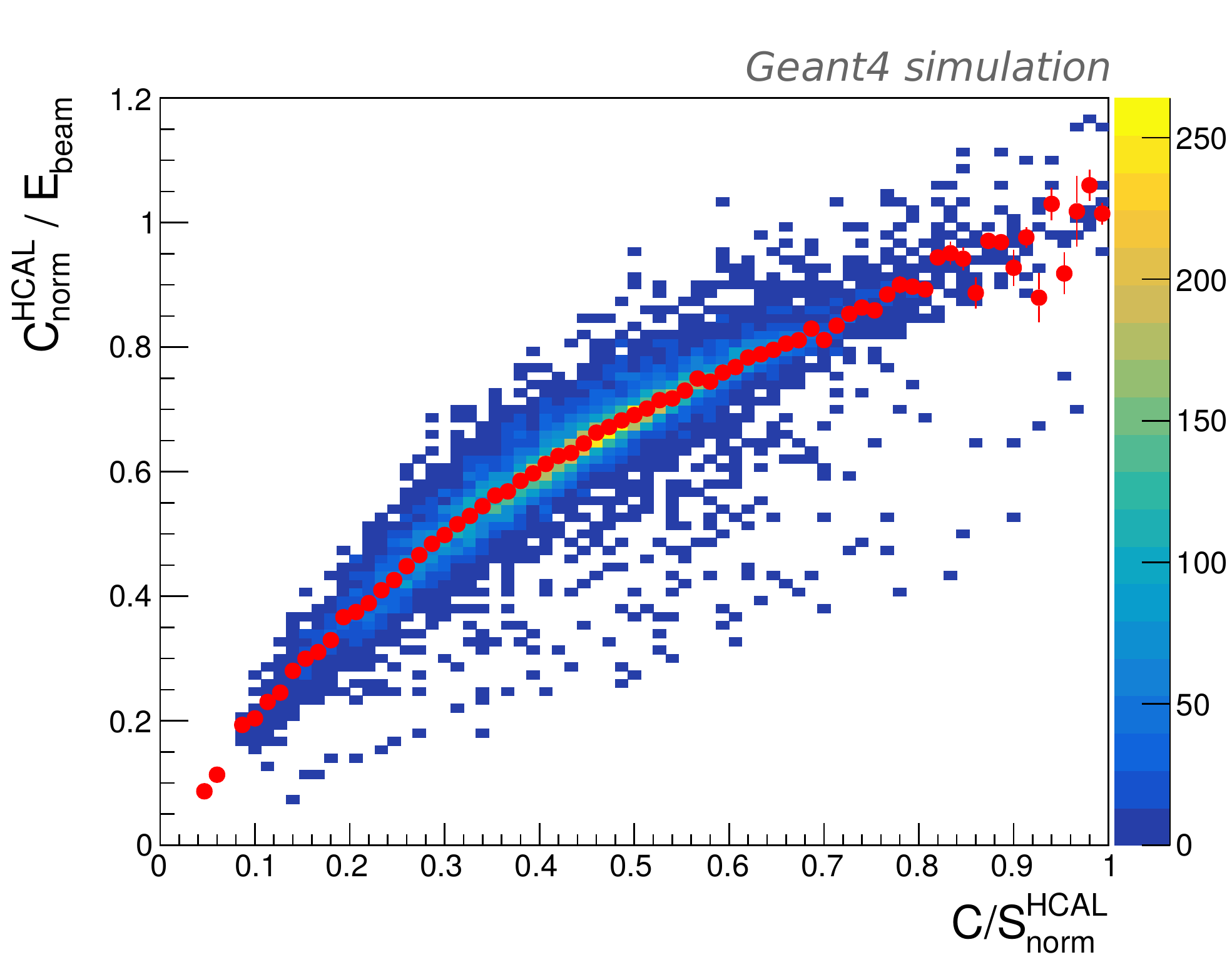}
\includegraphics[width=0.495\textwidth]{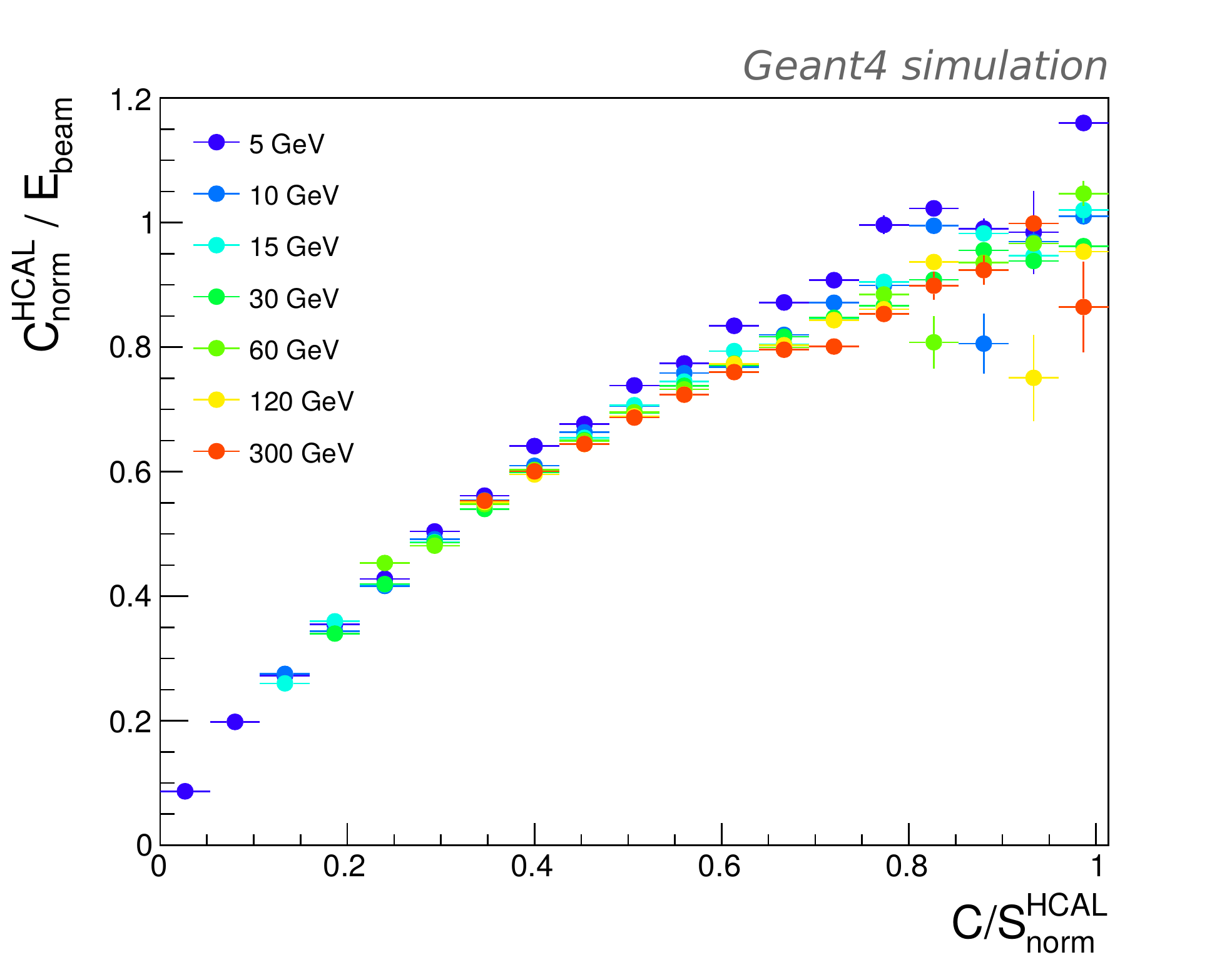}
\caption{\label{fig:hcal_dro_correction} Left: correlation between the normalized Cherenkov signal, $C^{HCAL}_{norm}$, and the $C/S^{HCAL}$ ratio measured in the HCAL segment for events that do not interact in the SCEPCal. Right: DRO correction curves for the HCAL segment for different hadron energies.}
\end{figure}

\begin{figure}[!tbp]
\centering % \begin{center}/\end{center} takes some additional vertical space
\includegraphics[width=0.495\textwidth]{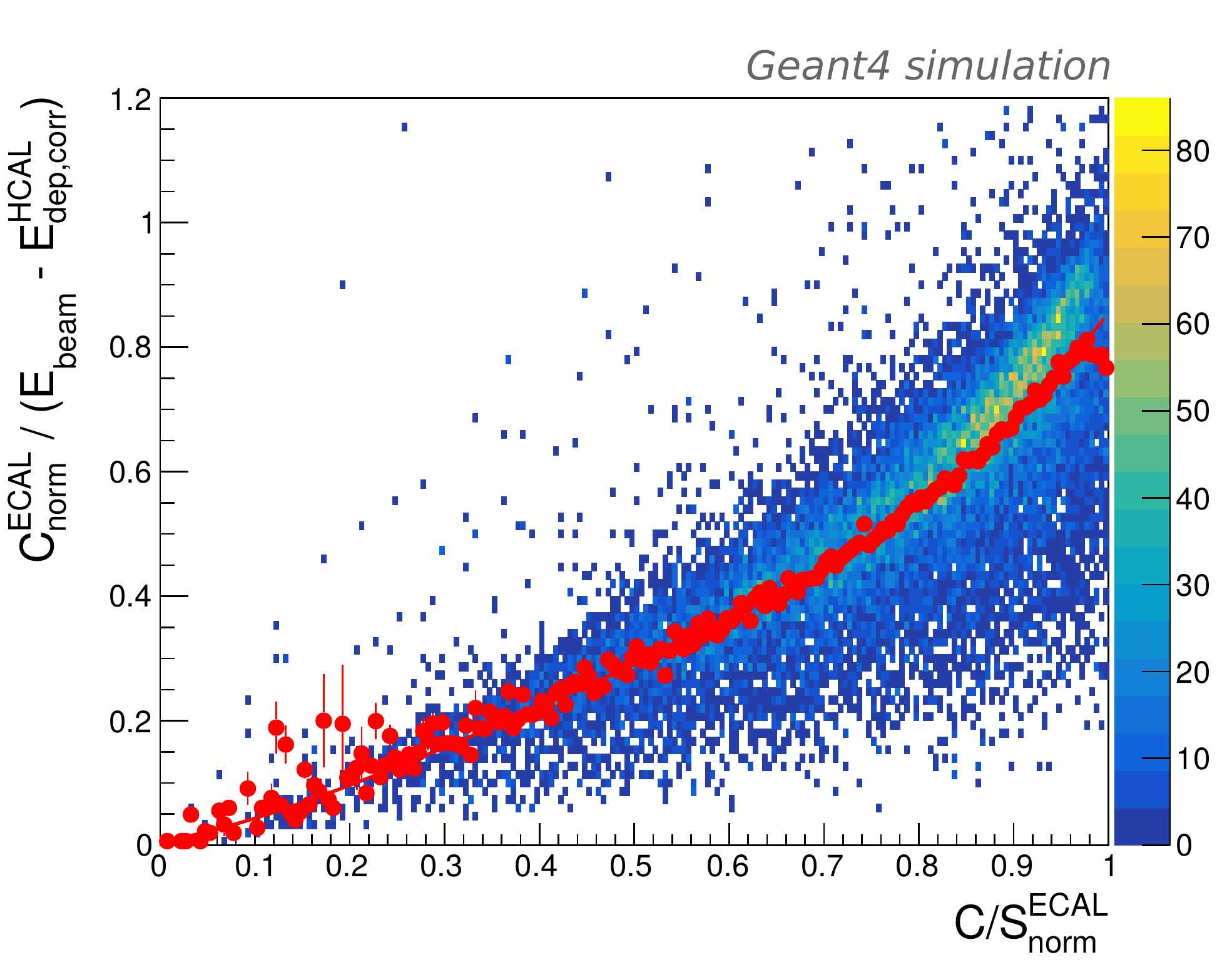}
\includegraphics[width=0.495\textwidth]{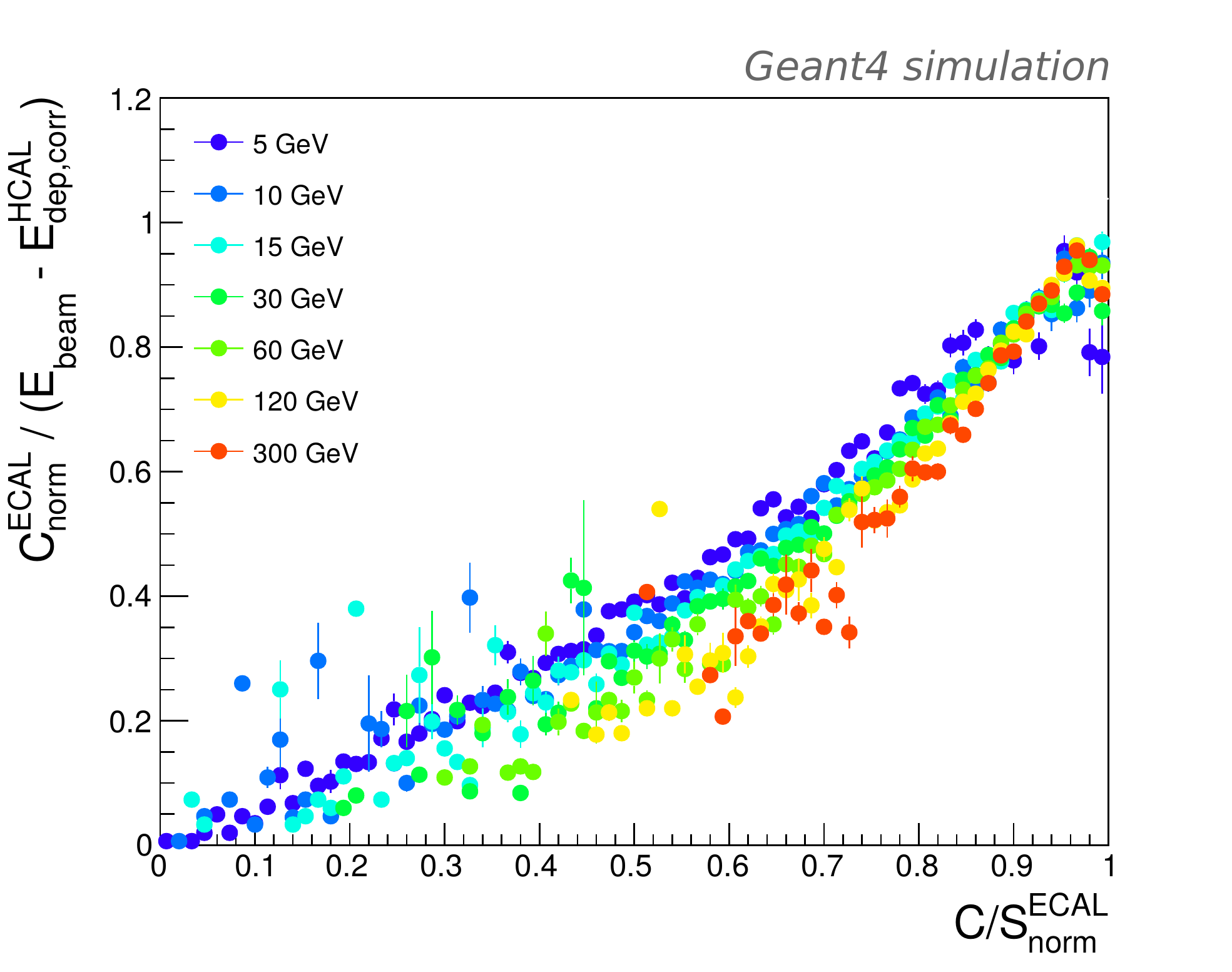}
\caption{\label{fig:ecal_dro_correction} Left: correlation between the normalized Cherenkov signal, $C^{ECAL}_{norm}$, and the $C/S^{ECAL}$ ratio measured in the SCEPCal for events that deposit a fraction of energy in the SCEPCal. Right: DRO correction curves for the SCEPCal for different hadron energies.}
\end{figure}

Charged hadrons passing through the ECAL deposit at least a minimum energy corresponding to that of a MIP, about 10~MeV/cm in PbWO$_4$ (9.2~MeV/cm in BGO).
Conversely, neutral hadrons do not produce ionization in the SCEPCal unless they start showering, producing secondary charged particles.
For the simulated beam of neutral kaons, we consider at first events that do not deposit energy in the SCEPCal section and define a DRO correction based on the ratio of the Cherenkov and scintillation signals (C/S) measured in the HCAL, f$_{(\rm C/S)_{HCAL}}$. The obtained curve is, as expected, independent of the beam energy as shown in Figure~\ref{fig:hcal_dro_correction}.
For electron showers about 6000 Cherenkov photons per unit energy are produced in the quartz fibers within the range of wavelengths from 300 to 1000~nm.
For neutral kaon showers, the Cherenkov signal per beam energy increases as expected when C/S approaches one, i.e. when the hadron shower is dominated by the EM component, f$_{\rm EM}$.
In Figures~\ref{fig:hcal_dro_correction} and \ref{fig:ecal_dro_correction}, the Cherenkov signal, C, is normalized to the mean number of Cherenkov photons produced per GeV when f$_{\rm EM} = 1$, and are indicated as $C^{HCAL}_{norm}$ and $C^{ECAL}_{norm}$ for the HCAL and ECAL segment respectively.

The obtained function can then be used to correct, event-by-event, the HCAL energy deposit for all events, including those with energy deposited in the SCEPCal, improving the hadron energy resolution.
Nonetheless, the fraction of the hadronic energy deposited in the SCEPCal is still subject to fluctuations of the EM component of the shower developed in the ECAL section and thus degrades the overall energy resolution of the combined SCEPCal+HCAL calorimeter.

To correct for these fluctuations in the ECAL, we look at the ratio of the C and S signals measured by the SCEPCal. 
These signals are estimated respectively as the number of the Cherenkov photons produced in the crystal in the range of wavelengths between 300 and 1000 nm (C) and the energy deposited in the crystal through ionization processes (S). The effect of a finite number of detected photons, responsible for photostatistics fluctuations, is discussed in the following section.
No clear correlation is observed between the C/S in the ECAL and the C/S in the HCAL indicating that the way the hadron shower develops in the HCAL, (e.g. number of $\pi^{0}$'s) is independent of the f$_{\rm EM}$ in the ECAL. Thus it is not possible to correct for f$_{\rm EM}$ fluctuations in the ECAL based on the C/S measured in the HCAL.

As expected, correlation is instead observed between the C/S in the ECAL and the Cherenkov (or scintillation) signal measured in the ECAL, as shown in Figure~\ref{fig:ecal_dro_correction}. Such a correlation is mostly independent of the beam energy and can thus be used to correct for the f$_{\rm EM}$ fluctuations.

\begin{figure}[!tbp]
\centering 
\includegraphics[width=0.495\textwidth]{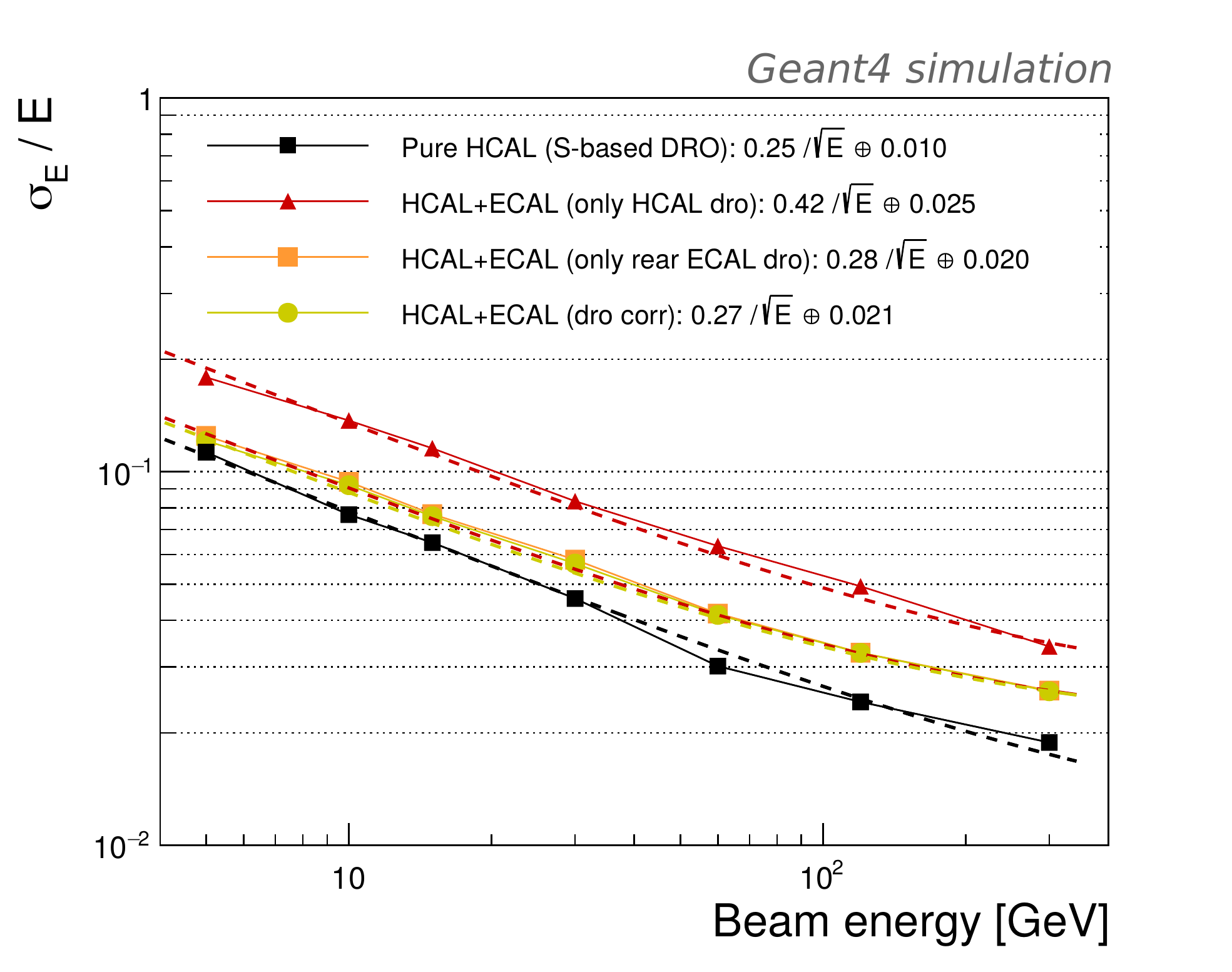}
\includegraphics[width=0.495\textwidth]{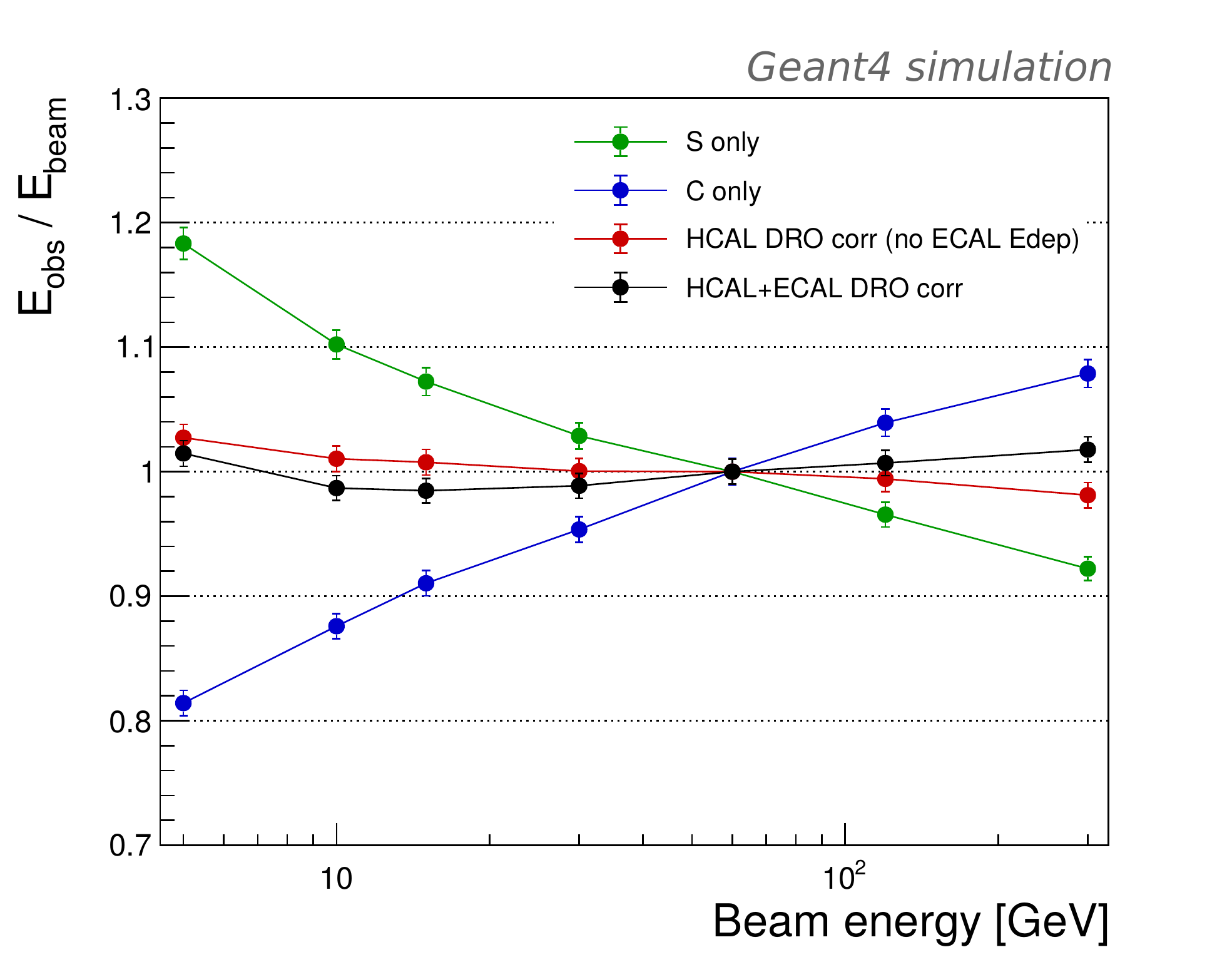}
\caption{\label{fig:combined_hadron_performance} Response to neutral kaons in terms of energy resolution (left) and linearity (right) for a dual-readout hybrid calorimeter consisting of the SCEPCal followed by an ultrathin-bore solenoid and a dual-readout fiber HCAL. The dashed lines in the right plot are the results of a fit to the data with a $\sigma_{E}/E = A/\sqrt{E} \oplus C$ function, which has been used to estimate the stochastic (A) and constant (C) terms of the energy resolution reported in the legend.}
\end{figure}

Exploiting this second correction, f$_{(\rm C/S)_{ECAL}}$, the total reconstructed energy can be written as:
\begin{equation}
\text{E}_{\rm tot}^{\rm corr} = \frac{\rm E_{\rm HCAL}}{\xi_{\rm HCAL}}\cdot \rm f_{(\rm C/S)_{\rm HCAL}} + \frac{\rm E_{\rm ECAL}}{\xi_{\rm ECAL}}\cdot {\rm f}_{(\rm C/S)_{\rm ECAL}} 
\end{equation}
This equation provides an energy independent method for correcting the f$_{\rm EM}$ fluctuations and leads to Gaussian and narrower energy distributions, achieving an overall energy resolution at the level of $28\% \oplus 2\%$ and a linearity within 2\% over the 5-300~GeV energy range, as shown in Figure~\ref{fig:combined_hadron_performance}.
The calorimetric performance in measuring hadrons achieved with the hybrid dual-readout system, including the ultrathin-bore solenoid between the ECAL and the HCAL segment, is close to that of a pure dual-readout HCAL. In particular, the stochastic term of the energy resolution, which drives the contribution from low momentum neutral hadrons to the jet resolution, is almost identical. A slightly larger constant term is observed and is attributed to the intrinsic limitation of a system that combines segments with different e/h ratios, and to the material budget from the ECAL services and the solenoid.
Nonetheless, the difference in resolution remains small and within the target performance discussed in Section~\ref{sec:pfa_drivers}.

\subparagraph{Implementation of dual-readout in the SCEPCal}\label{sec:scepcal_dro}

While for the hadronic segment the dual-readout method can be implemented by reading separately the scintillating and quartz fibers, a different approach should be used for the crystal electromagnetic section.
Scintillation and Cherenkov photons are simultaneously and abundantly generated inside crystals due to relativistic charged particles produced in electromagnetic showers. For example, the high density and refractive index of lead tungstate ($n=2.1$) is such that for each MeV of deposited energy, about 56 Cherenkov photons are produced with a $1/\lambda^{2}$ distribution in the range of wavelengths from 300 to 1000~nm.
To measure accurately the C and S signal in the SCEPCal three possible methods can be considered: 
\begin{enumerate}
\itemsep0em 
\item \textbf{\emph{Dual-SiPM} method}: Use of two separate SiPMs (with wavelength filters) coupled to the far side of the rear crystals and optimized to detect mainly the S and C photons as discussed below;
\item \textbf{Time constants}: Exploitation of the different time constants of C photons (emitted promptly) and S photons (emitted accordingly to the scintillation decay time);
\item \textbf{Dedicated Cherenkov radiator}: Introduction of specific C-sensitive elements (e.g. quartz or undoped crystals) nested within the scintillating crystal matrix, e.g. at the corners of the crystals.
\end{enumerate}

\begin{figure}[!tbp]
\centering 
\includegraphics[width=0.495\textwidth]{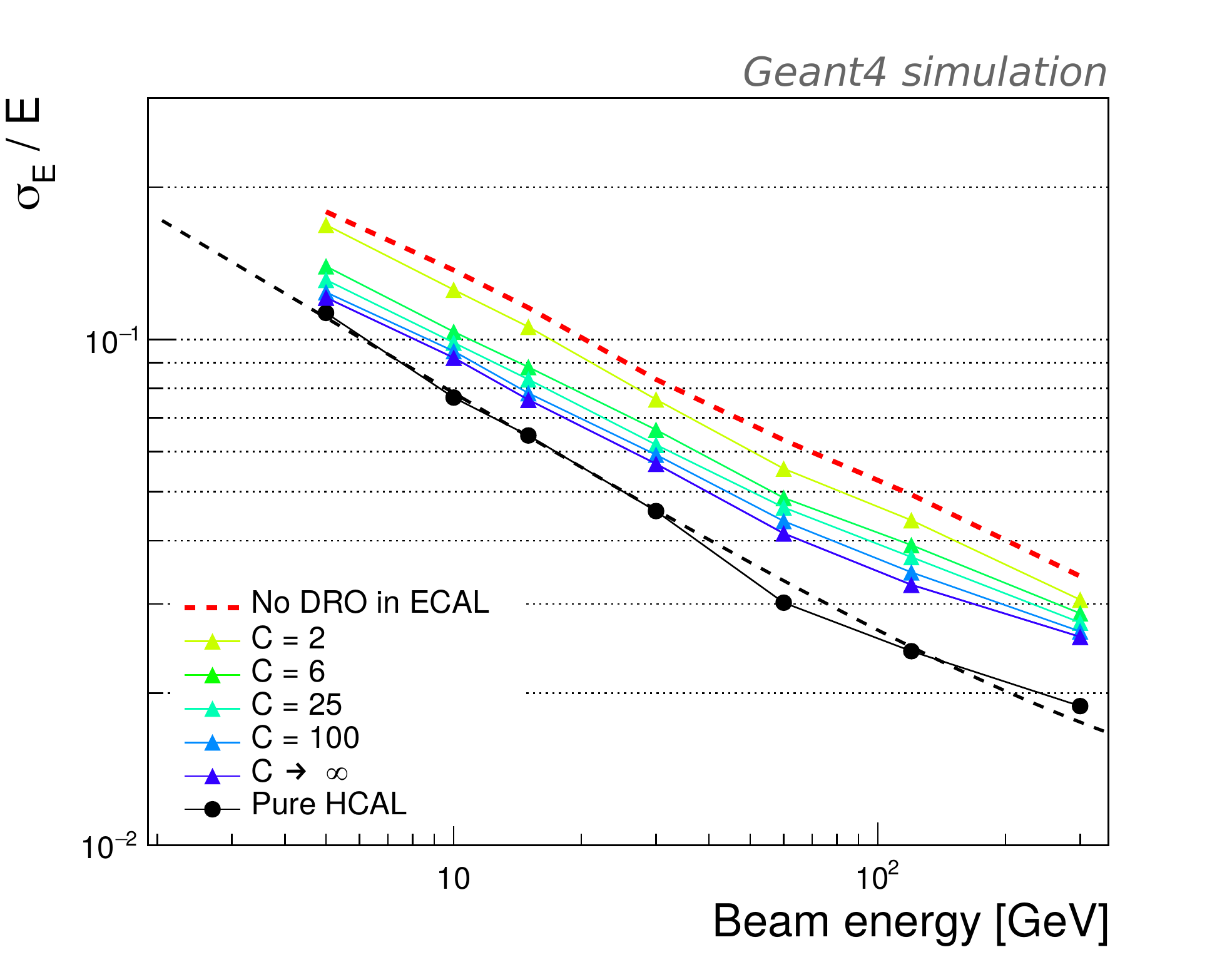}
\includegraphics[width=0.495\textwidth]{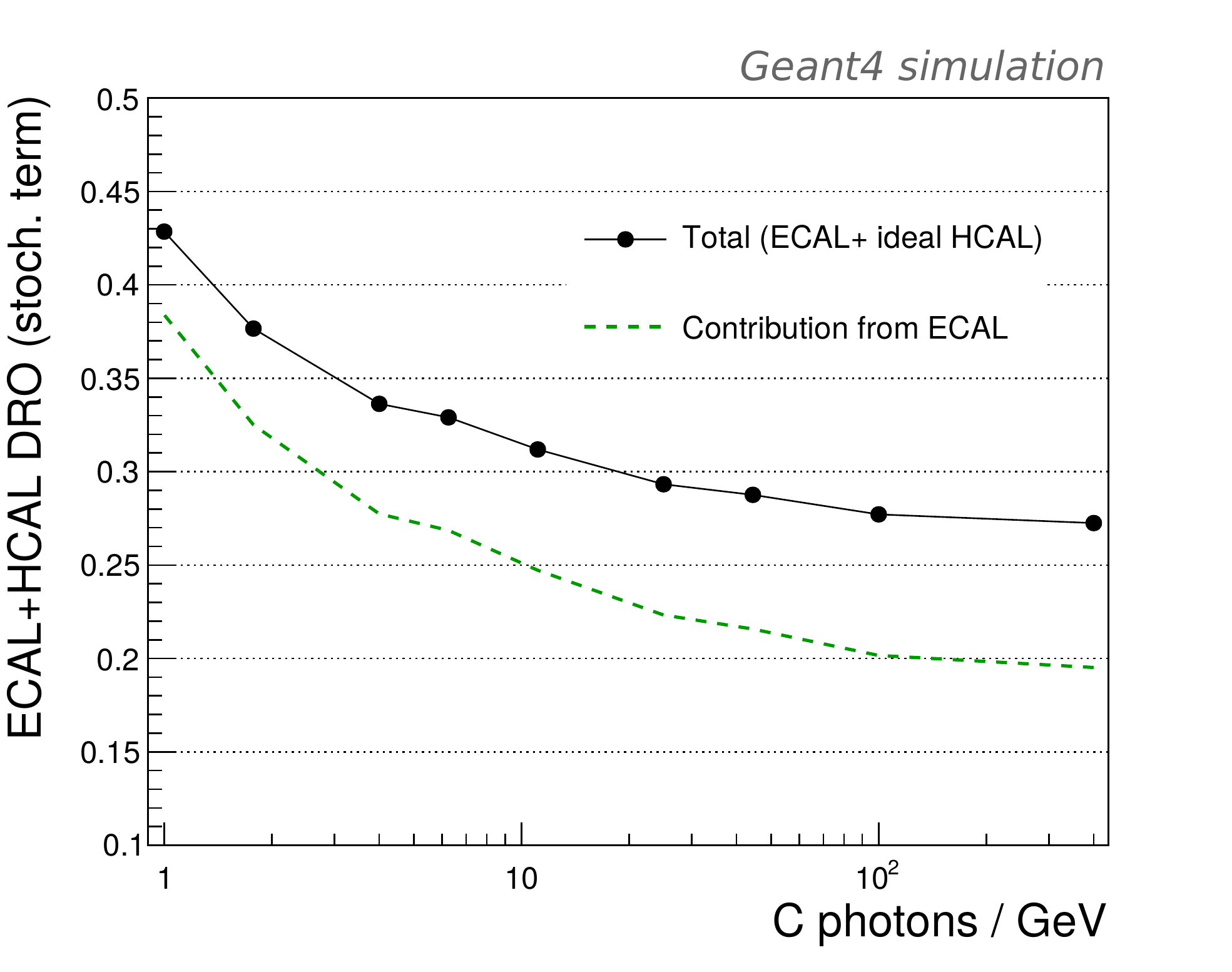}
\caption{\label{fig:scepcal_dro_stochastic} Left: energy resolution as a function of particle energy under different assumption of Cherenkov photon yield (C), from 1 phe/GeV to more than 1000 phe/GeV compared with the energy resolution for a pure HCAL (black) and for a hybrid SCEPCal+HCAL calorimeter where no dual-readout in the SCEPCal is present. Right: impact of photostatistics from the Cherenkov signal on the stochastic term of the hadron energy resolution for the combined calorimeter system (black) and difference in quadrature of the stochastic term with respect to the fiber-based calorimeter only in the absence of the crystal segment (green dashed).}
\end{figure}

For all methods, it is important to detect a sufficient number of Cherenkov photons (C) such that the stochastic fluctuations due to photostatistics are not a limitation to the precision of the dual-readout correction.
A smearing to the C signal from the \textsc{Geant4} simulation was applied, corresponding to Poisson fluctuations for different assumptions for the Cherenkov light detection efficiency, and the impact on the overall hadronic resolution was evaluated for each case. The results are shown in Figure~\ref{fig:scepcal_dro_stochastic}.
We estimate that a minimum of 50 C-photoelectrons/GeV is needed to maintain the stochastic term of the calorimeter energy resolution below $28\%/\sqrt{E}$. The constant term instead is only marginally affected since at higher energies, even for a small yield of C-photons, the signal is sufficiently large that Poisson fluctuations are negligible.

%Both the first and second methods have been previously explored in \cite{AKCHURIN2009488} where for instance a number of 55 Cherenkov photons per GeV was measured in BGO and it was similarly concluded that such number was sufficient.

%
Since the hadrons interacting in the SCEPCal deposit energy mostly in the rear ECAL segment, it is not necessary to instrument the front segment with dual-readout capabilities. The performance achieved by applying a dual-readout correction based on the C/S measured in the rear segment only is the same within statistical fluctuations to that of a full SCEPCal dual-readout as shown in Figure~\ref{fig:combined_hadron_performance}. This simplification reduces by a factor of two the number of channels required for the readout of the C-component.

\begin{figure}[!tbp]
\centering 
\includegraphics[width=0.495\textwidth]{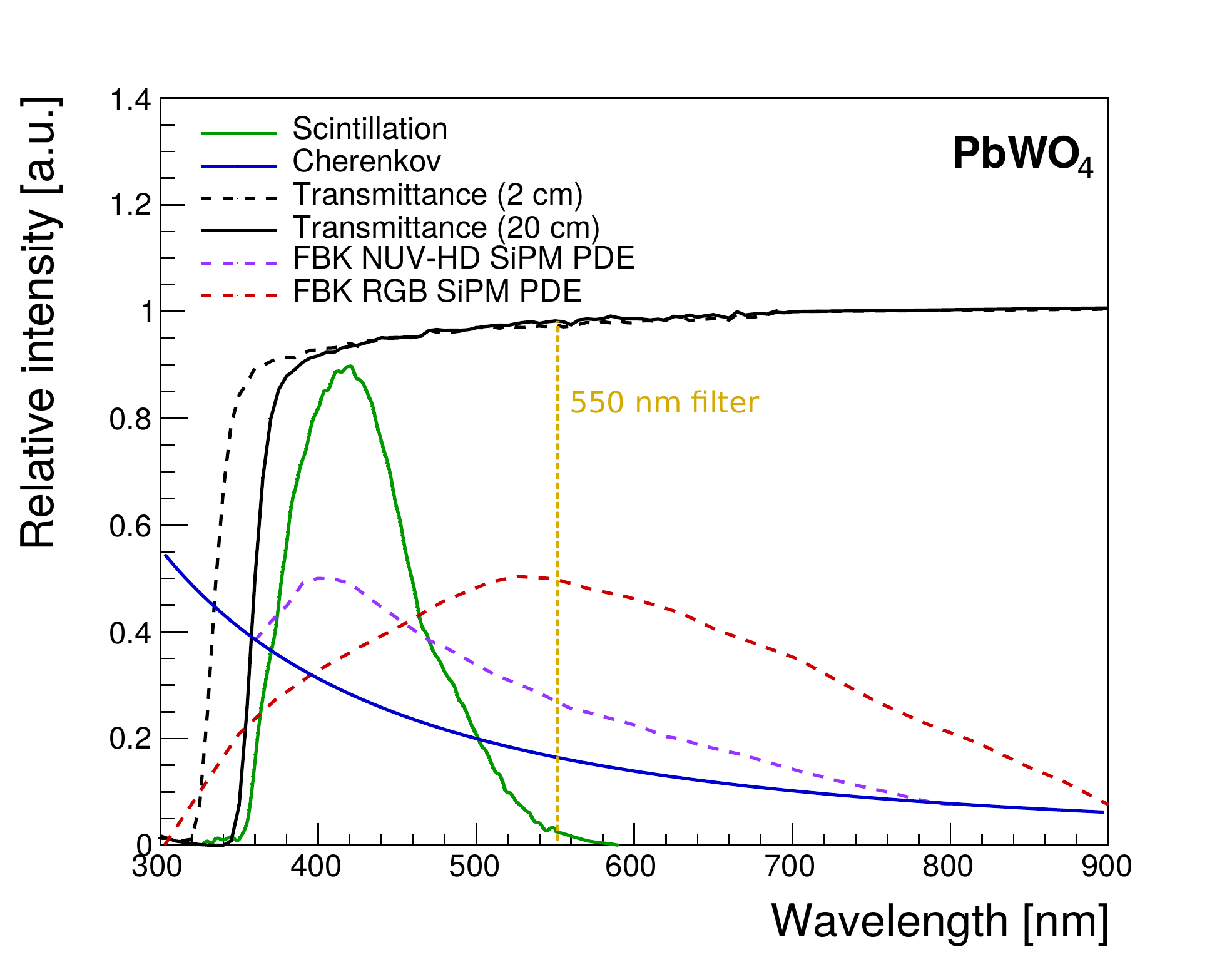}
\includegraphics[width=0.495\textwidth]{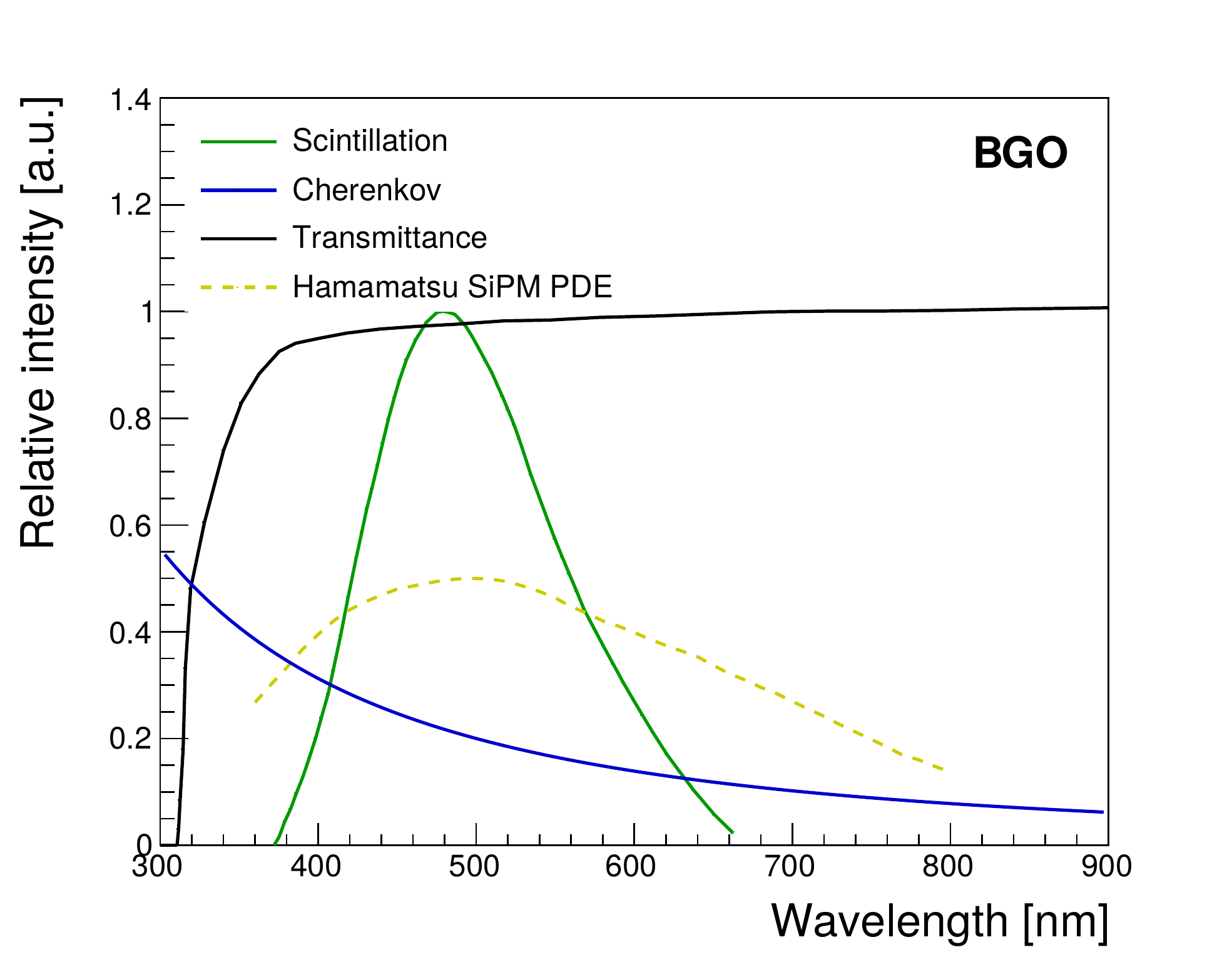}
\caption{\label{fig:scepcal_dro_options} The transmission curves of the crystals (black) are compared with the wavelength spectrum for scintillation (green) and Cherenkov light (blue) and with the photon detection efficiency of commercially available SiPMs from FBK and Hamamatsu. A comparison of PbWO$_4$ (left) and BGO (right) crystals shows how different spectral regions of the Cherenkov light could be used for better separation from the scintillation component. }
\end{figure}

In the following, we discuss the possibility to measure the S and C components using the \emph{Dual-SiPM} method.
As shown in Figure~\ref{fig:scepcal_dro_options}, different crystals may require a different optimization of the readout scheme for optimal separation of the detected S and C signals.
For instance, BGO crystals show a larger Stokes shift with respect to PbWO$_4$, providing a wider transparency window for Cherenkov photons in the 300-400~nm range. For this reason, the BGO potential for simultaneous Cherenkov and scintillation detection has also recently triggered new interest because of its possible exploitation in Time-Of-Flight Positron Emission Tomography scanners \cite{Kratochwil_BGO}.

On the other hand, the scintillation emission peak for PbWO$_4$ is closer to the UV (420~nm), and leaves a larger window for detection of Cherenkov photons in the region between 550 and 1000~nm. A further advantage of exploiting large-wavelength Cherenkov photons is the better transparency of the crystal in this region, which leads to a more uniform and linear response of the C component.
Other variants of these crystals can be explored and further developed, for instance, bismuth silicate (Bi$_4$Si$_3$O$_{12}$, BSO) features a smaller light yield and faster decay time than BGO \cite{ISHII_Nikl_BSO, AKCHURIN_BGO_BSO} and PbWO$_4$ doped with Molybdenum shifts the emission peak towards 500~nm \cite{ANNENKOV_PMWO}.

\begin{figure}[!tbp]
\centering 
\includegraphics[width=0.495\textwidth]{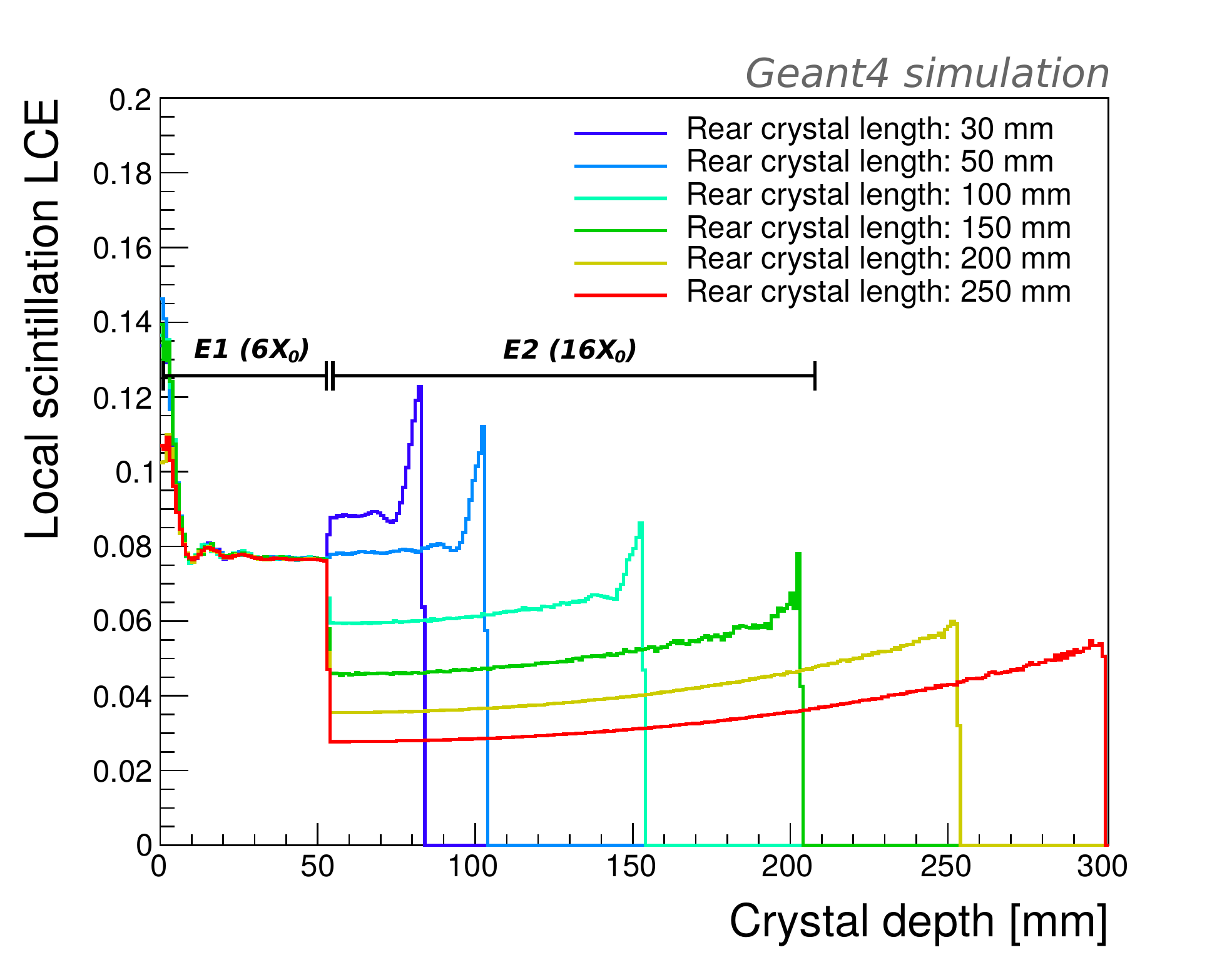}
\includegraphics[width=0.495\textwidth]{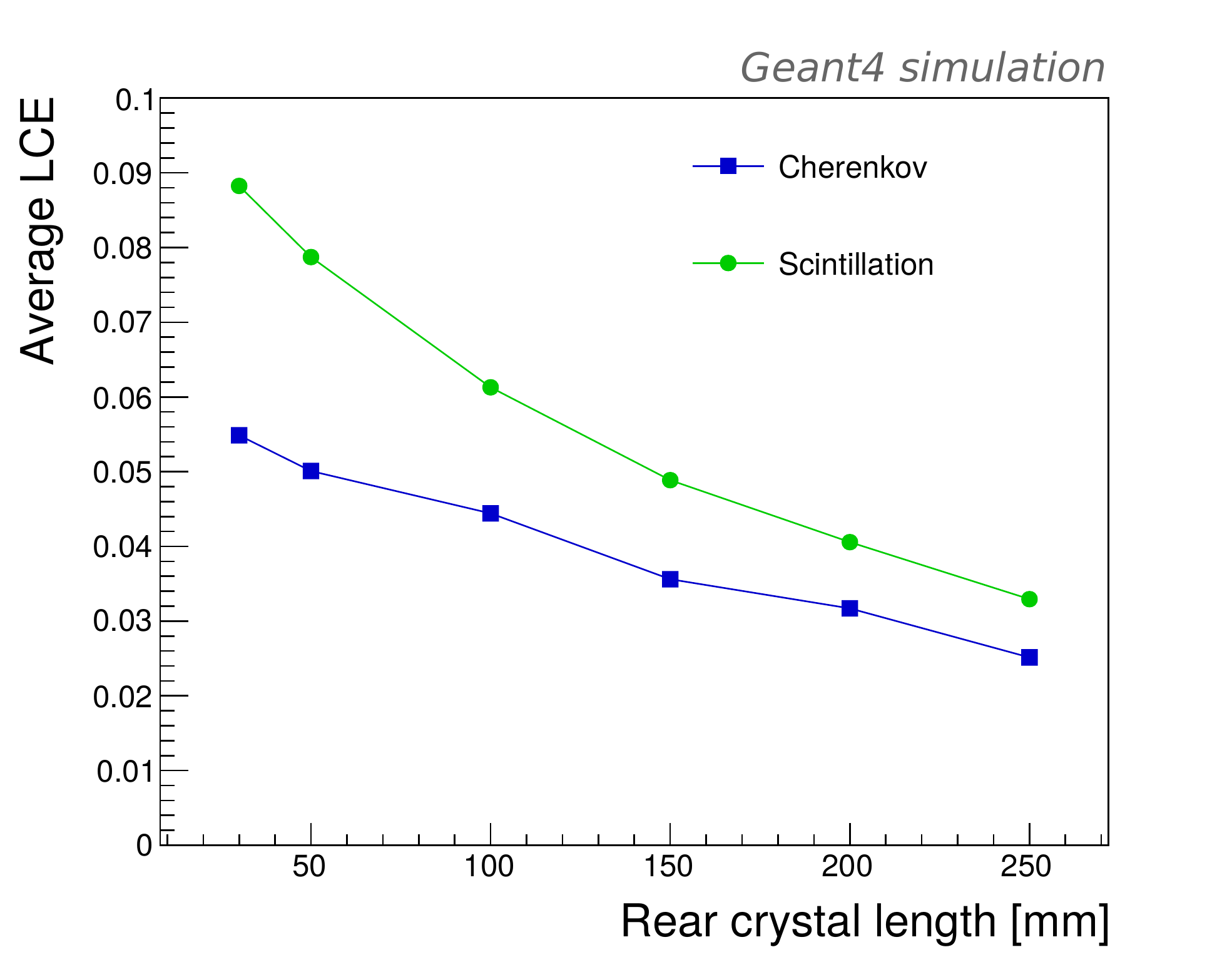}
\caption{\label{fig:scepcal_lce_optimization} Left: local light collection efficiency for scintillation light along the SCEPCal for both the front and rear segments when varying the length of the rear crystal segment (E2). Right: average light collection efficiency for scintillation and Cherenkov light for the rear segment (E2) as a function of crystal length.}
\end{figure}

Being a highly segmented calorimeter, the SCEPCal features shorter and smaller crystals with respect to previous crystal calorimeters (e.g. CMS ECAL), leading to an improved light collection efficiency. This effect, quantified using a \textsc{Geant4}-based ray-tracing simulation, is shown in Figure~\ref{fig:scepcal_lce_optimization}, where the light collection efficiency for both scintillation and Cherenkov photons is shown as a function of the crystal length.

Based on recent developments in photodetector technologies, in particular in SiPMs, we propose to instrument the rear crystals with two different SiPM technologies, featuring peak sensitivities (PDE) in different wavelength regions.
For example the FBK-NUV-HD technology \cite{FBK_NUV_SiPMs} with maximum PDE at around 420~nm would be well-suited for the detection of S photons in PbWO$_4$ while the FBK-RGB-HD technology \cite{FBK_RGB_UHD} with peak PDE at 550~nm (extending up to 1000~nm) can better detect the C photons above 550~nm.
A SiPM cell size of $7.5-10$~\SI{}{\um} would be optimal for the readout of the abundant scintillation photons and provide a large dynamic range. A larger cell size on the order of $15-25$~\SI{}{\um}, would instead be optimal to readout the C component that requires a larger photon detection efficiency.
The use of a filter to remove wavelengths above 550~nm on the SiPM optimized for S-light detection would minimize the contamination of Cherenkov photons to the measured light signal $P_1$.
Similarly, by applying on the window of the SiPM optimized for C-light detection a high pass optical filter, removing wavelengths below 550~nm, the scintillation photons could be largely excluded from the measured signal $P_2$.
Our simulation estimates that about 160 C-photoelectrons per GeV can be measured from the rear SCEPCal segment (E2) with negligible S contamination. A summary of the estimated photon yields for Cherenkov and scintillation photons for such a configuration (PbWO$_4$ crystal with two SiPMs and optical filters) is reported in Table~\ref{tab:SCEPCal_S_C_yields}.

Exploiting this \emph{dual-SiPM} method, the S and C components can then be extracted from the simultaneous light measurements from the pair of SiPMs, $P_1$ and $P_2$, according to the equations:
\begin{eqnarray}\label{eq:dualsipm}
\begin{cases}
P_1 = f_{S,1}\cdot S + f_{C,1}\cdot C\\
P_2 = f_{S,2}\cdot S + f_{C,2}\cdot C
\end{cases}
\end{eqnarray}
where $f_{S,1}$ ($f_{S,2}$) and $f_{C,1}$($f_{C,2}$) are respectively the fraction of S and C components measured by the first and second SiPM. As long as a sufficient contrast between the S and C signal is maintained in each SiPM (i.e. $(f_{C,1}\cdot C) /(f_{S,1}\cdot S)\lesssim 0.1$ and $(f_{S,2}\cdot S) /(f_{C,2}\cdot C)\lesssim0.1$), as is the case for the numbers reported in Table~\ref{tab:SCEPCal_S_C_yields}, the solution of the system~\ref{eq:dualsipm} provides an accurate estimate of the pure C and S components that can be used for dual readout corrections. 
\begin{table}[!tbp]
\centering
\caption{\label{tab:SCEPCal_S_C_yields} Photon yield for both Cherenkov and scintillation light in response to a 45 GeV electron shower in the rear SCEPCal segment assuming a PbWO$_4$ crystal and the SiPM spectral sensitivity shown in Figure~\ref{fig:scepcal_dro_options} (left).}
\smallskip
\begin{tabular}{|l|c|c|c|c|}
\hline
		  						 & Scintillation  	& $f_{S}$ & Cherenkov    & $f_{C}$ 		\\
		  						 & [photons/GeV] 	&  [\%]	  &	[photons/GeV]& [\%]		 	\\	     
\hline
\hline
Generated 			 			 &  200000			& 100     & 56000		 & 100 		\\
Collected 			 			 &  10000			& 5.0	  & 2130		 & 3.8		\\
\hline
Detected by NUV SiPM \#1  ($\lambda<550$ nm)	 	 &  2000		& 1.0     & 140			 & 0.25		\\
Detected by RGB SiPM \#2 ($\lambda>550$ nm)  	 &  $<20$		& $<0.01$&	160			 & 0.3		\\
\hline
\end{tabular}
\end{table}
The scintillation light yield from PbWO$_4$ is small enough that it would contribute only about 12\% to the measured C-signal with an RGB SiPM and the 550~nm filter. The possibility to exploit the different time constants of S and C could further reduce such contamination.

In addition to dual-readout corrections, the measurement of C/S in the SCEPCal will also enhance its particle identification capabilities discussed in Section~\ref{sec:scepcal_pid}. In particular, a requirement on the C/S signal to be compatible with that from electron showers would improve the discrimination of charged pions that start showering in the SCEPCal from electrons. With such a cut the efficiency for charged pion rejection improves from 99.0\% to 99.4\% for the same electron identification efficiency.

\subparagraph{Cost optimization of the dual-readout HCAL}\label{sec:hcal_opt}

\begin{figure}[!tbp]
\centering 
\includegraphics[width=0.495\textwidth]{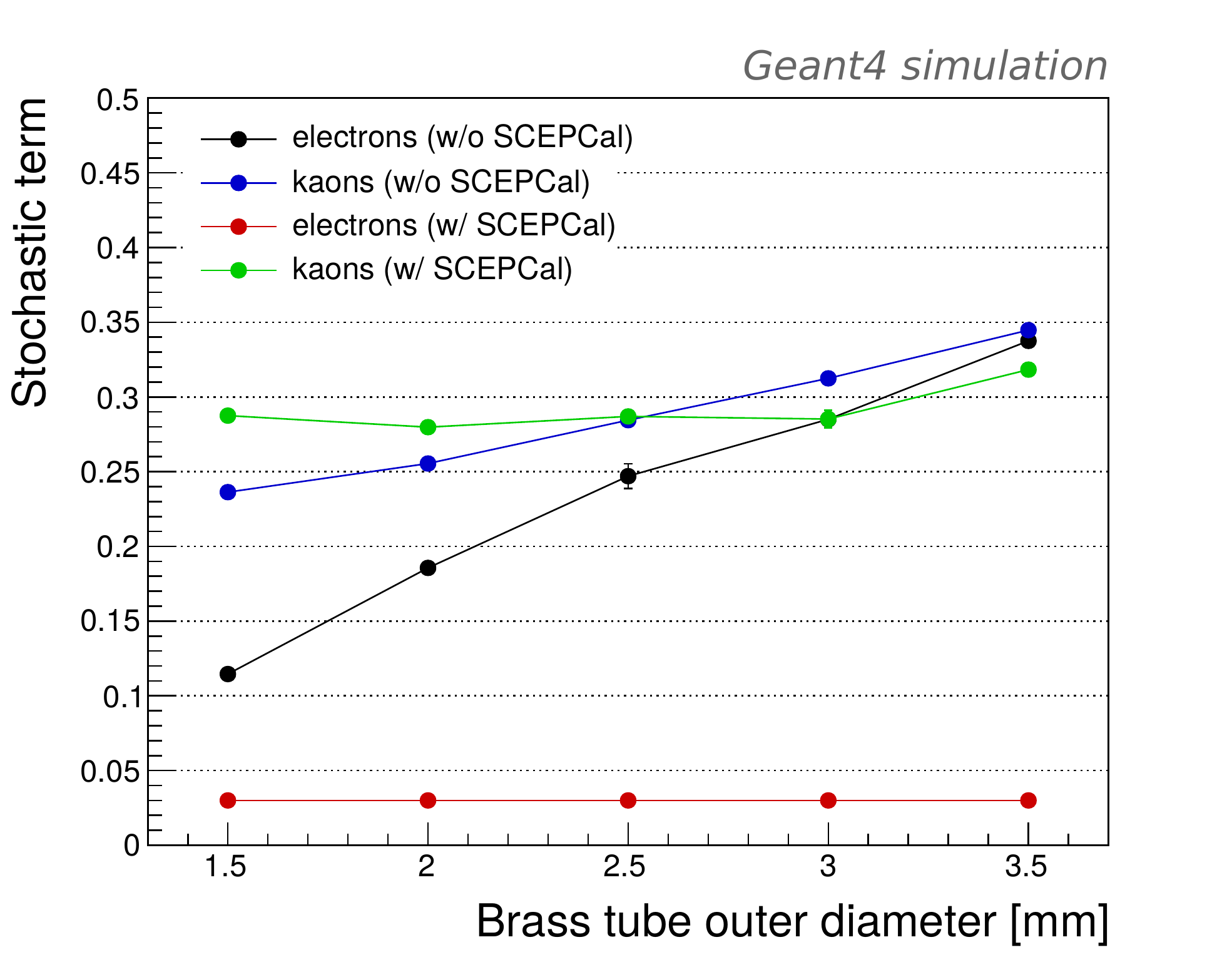}
\includegraphics[width=0.495\textwidth]{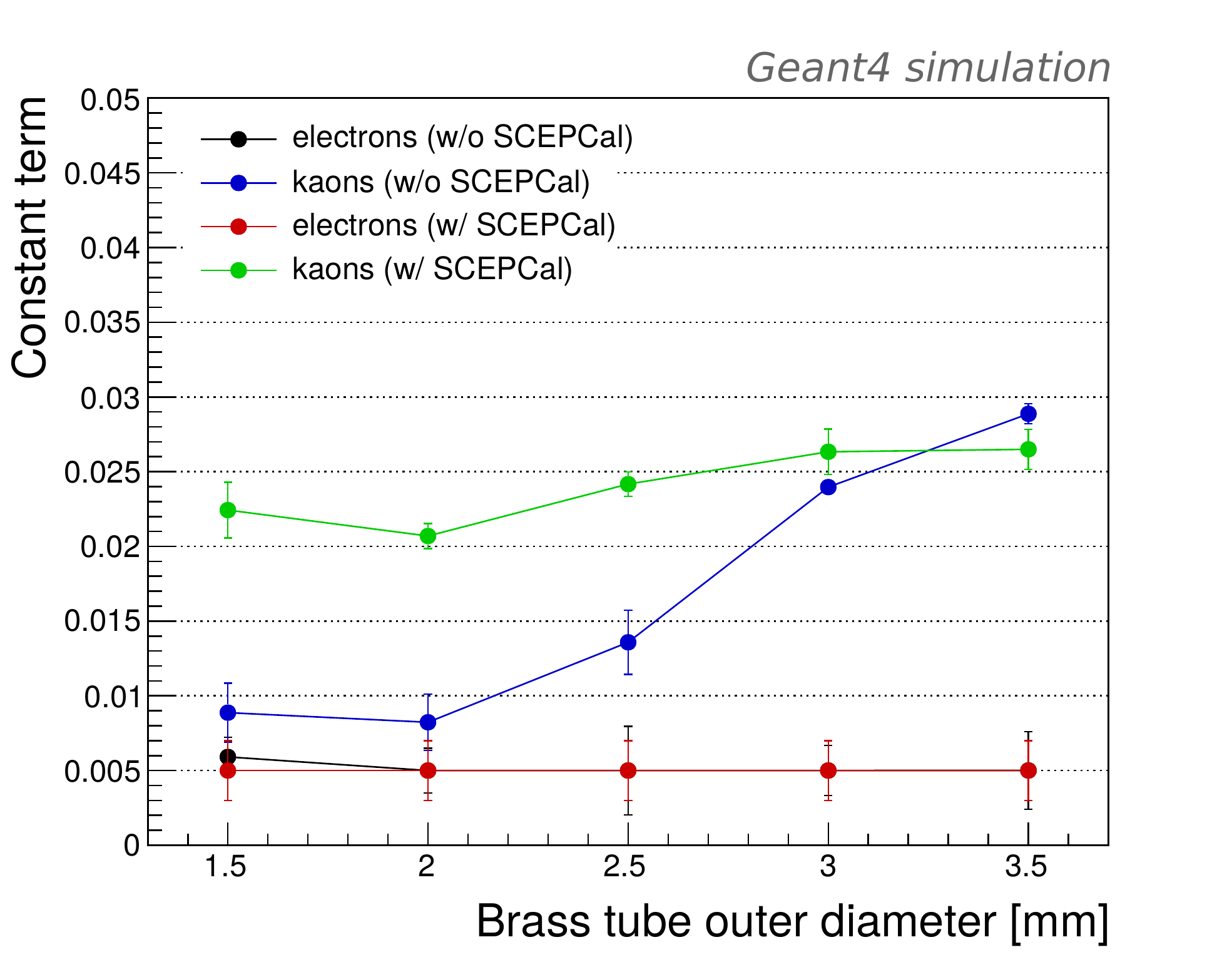}
\caption{\label{fig:hcal_optimization} Performance of the calorimeter system in terms of stochastic (left) and constant (right) term of the energy resolution to both electrons and neutral kaons, with and without the SCEPCal and by varying the dimension of the outer brass tube diameter in the fiber HCAL.}
\end{figure}

In the combined hybrid calorimeter system presented above, excellent resolution, as well as high transverse segmentation for EM showers, are provided by the SCEPCal segment. The cost of the system can be reduced by having a coarser segmentation for the transverse granularity and sampling fraction of the HCAL. 
%Without such tight requirements for EM particle detection, the design of the HCAL segment can be further optimized in terms of cost/performance. In particular, its sampling fraction and transverse granularity can be relaxed.
%
We have studied this possibility by varying the outer diameter (OD) of the brass capillaries while keeping the fiber and inner tube diameters unchanged. In this way, the sampling fraction and number of channels decrease with larger outer diameter of the brass tubes. As shown in Figure~\ref{fig:hcal_optimization}, if the electromagnetic resolution requirement of dual-readout fiber calorimeter is relaxed, the brass tube thickness can be increased to 3--3.5 mm with marginal impact on the hadron resolution but a relative channel reduction and cost decrease scaling approximately with $1/\text{OD}^2$.
In particular, because of the presence of the crystal ECAL, which maintains an excellent EM resolution independently of the brass tube size, the hadronic resolution of the combined hybrid calorimeter is comparable or better than a pure HCAL for outer diameters larger than 3~mm. This is attributed to the fraction of hadron energy deposited in the SCEPCal which is always measured accurately regardless of the sampling fraction in the HCAL segment.

\subsection{Discussion}
%We showe
The performance of future detectors for $e^+e^-$ colliders should provide sufficient resolution to measure jet energy at the level of 3\% at 50~GeV. The event reconstruction will likely exploit all the information collected from different sub-detectors as in the particle flow approach. 
In this context, the key to improved jet energy resolution was assumed to be mainly the calorimeter granularity. 
We showed, however, that a calorimeter capable of providing excellent energy resolution for the neutral component of the jet is crucial. In particular, in events characterized by a dense jet topology, the possibility of reconstructing $\pi^0$'s from the multitude of photons represents an excellent tool to enhance the correct assignment of photons to the corresponding jet, further boosting the performance of Particle Flow algorithms.

To achieve such performance, we propose a hybrid calorimeter design consisting of a high-resolution electromagnetic calorimeter made of segmented crystals placed in front of a fiber-based dual-readout hadronic calorimeter.
The combination of these two technologies has the following features:
\begin{itemize}
\itemsep0em 
\item Excellent EM resolution (3\%/$\sqrt{E}$ instead of 10-15\%/$\sqrt{E}$) to enable efficient recovery of bremsstrahlung photons and pre-clustering of $\pi^0$ photons to enhance the performance of jet clustering algorithms;
\item More compact EM showers: smaller Moli\`ere radius and radiation length;
\item Longitudinal segmentation to enhance particle identification and PFA performance;
\item Possibility to incorporate a thin solenoid after the ECAL with no degradation of the ECAL resolution;
\item Fewer constraints on the design of the HCAL section allowing a cost reduction (e.g. smaller sampling fractions, thus fewer readout channels);
\item Energy resolution to neutral hadrons better than $30\%/\sqrt{E}$ to maintain the contribution to 50 GeV jet resolution below 1.5\%.
\end{itemize}

\section{Summary}
\label{sec:summary}

We have explored in this work major innovations in collider detector performance that can be achieved with crystal calorimetry when high granularity segmentation and dual-readout capabilities are combined with a new high EM resolution approach to PFA in multi-jet events such as Higgs factory ZH all-hadronic final-states. 
In particular, a calorimeter with EM resolution at the level of $3\%/\sqrt{E}$ can effectively improve the resolution of the recoil mass of the Z boson decay into electron pairs in ZH events to 80\% of that for muon pairs by improving the electron momentum resolution through the recovery of bremsstrahlung photons. 
In addition, we show that such a calorimeter has sufficient resolution to efficiently cluster the multitude of photons from $\pi^0$ decays in multi-jet events. Exploiting a graph-based approach through a Blossom V algorithm we demonstrate that the fraction of photon pairs correctly clustered into $\pi^0$ over those wrongly clustered is a factor of five better for a $3\%/\sqrt{E}$ EM resolution with respect to a $30\%/\sqrt{E}$ case.
With such clustering, applied in advance of the jet clustering algorithms, the efficiency in correctly assigning photons to the corresponding jet is improved by a factor of three for the worst jet in 6-jets event topologies with potential beneficial impact on both the jet angular and energy resolution.
We present the design and optimization of a segmented crystal electromagnetic calorimeter with $3\%/\sqrt{E}$ energy resolution and demonstrate how it can be combined with a dual-readout HCAL to achieve a resolution to neutral hadrons better than $30\%/\sqrt{E}$. Such a hybrid system provides room for additional independent optimization of the two calorimeter segments in terms of cost and achieves the best performance on both photons and neutral hadrons while maintaining a high-level of timing, segmentation and particle identification for PFA.
The calorimeter concept discussed in this paper, can represent a forward evolution of the use of crystal calorimeters at future lepton collider experiments and open new perspectives for the exploitation of high-resolution EM calorimeters within the PFA approach.

\acknowledgments
We thank Sergei Chekanov for his help and guidance in exploiting the HepSim tools and repository that have been used in this work. We are grateful to Franco Bedeschi and Gabriella Gaudio for the useful discussions on their previous studies on dual-readout with crystals and optimization of a fiber-based dual-readout HCAL layout.

%\paragraph{Note added.} This is also a good position for notes added after the paper has been written.

%\section*{References}

\bibliography{mybibfile}

%\begin{thebibliography}{99}
%\bibitem{a}
%Author, \emph{Title}, \emph{J. Abbrev.} {\bf vol} (year) pg.

% Please avoid comments such as "For a review'', "For some examples",
% "and references therein" or move them in the text. In general,
% please leave only references in the bibliography and move all
% accessory text in footnotes.

% Also, please have only one work for each \bibitem.

%\end{thebibliography}
\end{document}